\documentclass{aa}
\usepackage{url}
\titlerunning{short title}
\usepackage[utf8]{inputenc}
\DeclareUnicodeCharacter{2212}{\textminus}
\usepackage{CJKutf8}
\usepackage{hyperref}
\hypersetup
{
    colorlinks= true,
    linkcolor = blue,
    filecolor = blue,      
    urlcolor  = blue,
    citecolor = blue
}
\usepackage{xcolor}
\usepackage{caption}

\usepackage{graphicx}
\usepackage{tikz}
\usetikzlibrary{arrows.meta}
\usepackage{txfonts}
\usepackage{natbib}
\usepackage{url}
\bibpunct{(}{)}{;}{a}{}{,} 
\begin{document}

\title{Physical and Chemical Conditions of Molecular Gas in NGC~1068}
\subtitle{The nuclear feedback in the circumnuclear disk and starburst ring}

\author{Bin Jia(\begin{CJK*}{UTF8}{gbsn}贾彬\end{CJK*})\inst{\ref{STRW}}\fnmsep\thanks{E-mail: \url{jia@strw.leidenuniv.nl}}, Serena Viti\inst{\ref{STRW},\ref{TRA},\ref{UCL}}, Erica Behrens\inst{\ref{UVA}}, Yun-Hao Zhang(\begin{CJK*}{UTF8}{gbsn}张云皓\end{CJK*})\inst{\ref{STRW},\ref{UOE}}}

\institute{
Leiden Observatory, Leiden University, P.O. Box 9513, 2300 RA Leiden, The Netherlands\label{STRW}
\and Transdisciplinary Research Area (TRA) ‘Matter’/Argelander-Institut für Astronomie, University of Bonn, Bonn, Germany\label{TRA}
\and Department of Physics and Astronomy, University College London, Gower Street, London, UK\label{UCL}
\and Department of Astronomy, University of Virginia, P. O. Box 400325, 530 McCormick Road, Charlottesville, VA 22904-4325, USA\label{UVA}
\and Institute for Astronomy, University of Edinburgh, Royal Observatory, Blackford Hill, Edinburgh, EH9 3HJ, The United Kingdom\label{UOE}
}

    \abstract
    {Molecular gas in galaxies is shaped by both star formation and active galactic nuclei. In NGC~1068, the circumnuclear disk and the starburst ring offer a nearby case to study these effects with many molecular tracers. Earlier work has shown strong outflow activity and complex chemistry, which motivates the use of methods that combine radiative transfer with time-dependent chemistry.}
    {Our aim is to map the physical conditions across the circumnuclear disk and the starburst ring of NGC~1068 and to test whether the nuclear outflow influences the molecular gas in the ring. We also examine whether the heating or the quiescent cloud scenario better matches the observations.}
    {We use archival ALMA observations obtained in Bands 3, 4, and 5, covering molecular species including HCN, HCO$^{+}$, HNC, CS, CN and C$_2$H. All data cubes are convolved to a common resolution of 0\farcs8 and are sampled into 56~pc hexagons with a signal-to-noise threshold of three. We perform hierarchical Bayesian inference that links a non-LTE radiative transfer module (\texttt{SpectralRadex}) with chemical modelling. To make the analysis efficient, we replace direct \texttt{UCLCHEM} calculations with a neural network emulator trained on a large model grid. Sampling is done with \texttt{Nautilus}. We also compare our results with previous studies that used RADEX and \texttt{UCLCHEM} for selected regions.}
    {The emulator reproduces the \texttt{UCLCHEM} abundances with low error and allows inference at modest computational cost. We find clear radial and azimuthal variations in gas density, temperature, column density, and cosmic-ray ionization rate. Toward the inner edge of the starburst ring, both physical properties and CO line profiles suggest that the gas there is highly disturbed.}
    {By coupling chemistry and radiative transfer within a hierarchical framework, and by using a neural network to speed up the chemical calculations, we obtain spatially resolved constraints on the physical conditions in NGC~1068. The radial gradients and the non-Gaussian CO profiles indicate that the nuclear outflow influences the inner part of the starburst ring. Higher resolution observations will help better constrain the gas properties.}

    \maketitle
    \nolinenumbers
%

\section{Introduction}

Galaxy evolution is strongly influenced by feedback processes that regulate star formation through the activity of active galactic nuclei (AGNs). The energy released by AGNs can suppress star formation by removing gas from host galaxies (negative feedback; e.g., \citealt{Fabian2012,Harrison2017}). Conversely, AGN-driven outflows can compress gas and potentially enhance star formation (positive feedback; e.g., \citealt{Silk2013,Zubovas2013}). Stellar feedback, which is driven by winds and supernovae, plays a similarly important role in this cycle \citep{Koudamani2019Fast,Koudamani2022Two,Leung2019The,Murthy2022Cold}. As a direct probe of these activities, submillimeter molecular tracers serve as a crucial tool for studying the physical conditions in galaxy centers, as most of the gas is in molecular form and is often obscured by dust. Different molecules trace distinct environments: HCO, HCO$^{+}$, and C$_{2}$H are prominent in photon-dominated regions (PDRs; e.g., \citealt{Garcia2002,Gerin2009,Martin2009,Garcia2017,Holdship2021}); HCN and CS trace dense gas clumps (e.g., \citealt{Gao2004,Aladro2011,Scourfield2020,Li2021Dense,Josh2022}); and CH$_{3}$OH, HNCO, and SiO are enhanced in shocked gas (e.g., \citealt{kelly2017,huang2022,huang2023NGC253}). This correspondence is not unique, however, since the same species can also trace cosmic-ray dominated regions (CRDRs), where a high ionization rate produces PDR-like chemistry deep within clouds at high visual extinction \citep{Holdship2021}.

NGC~1068 represents an archetypal system for studying the interplay between AGN and starburst activity. Located at a distance of 14~Mpc (\citealt{Bland1997}; 1\arcsec $\sim$ 70~pc), its proximity enables sufficient spatial resolution to distinguish between star formation and AGN-driven processes. Interferometric CO mapping by \cite{Schinnerer2000} revealed three principal molecular gas components: an extended starburst ring at a galactocentric radius of $\sim$1.5~kpc, a compact circumnuclear disk (CND) extending to $\sim$200~pc from the nucleus, and an elongated stellar bar spanning approximately 2~kpc with a position angle of $\sim$48\degr{} \citep{Scoville1988}. The galaxy hosts a biconical ionized and molecular gas outflow driven by the central AGN (e.g., \citealt{Das2007,Garcia2014,Saito2022}). Kinematic evidence demonstrates that molecular gas has been displaced out of the galactic plane into a three-dimensional outflow geometry \citep{Garcia2019}. Detailed CO outflow mass rates within the CND have been calculated by \cite{Zhang2025}, and molecular outflow signatures have been detected in bow-shock arc kinematics \citep{Garcia2014,Garcia2019,Sanchez2022}.

Despite the detailed characterization of the central outflow, the broader connection between AGN-driven outflows and star formation in the starburst ring of NGC~1068 remains unresolved. \cite{Sanchez2022} investigated the relationship between star formation rate and dense molecular gas using CO, HCN, and HCO$^{+}$ from ALMA, together with Pa$\alpha$ emission from \textit{HST}. By assessing gas boundedness, they attributed the enhanced dense gas fractions in the bar-ring interaction region mainly to molecular gas compression caused by intense cloud-cloud collisions. However, because this interface also overlaps with the potential outflow impact zone, the contribution from AGN-driven outflows could not be fully ruled out. The limited spatial resolution of their observations further restricts the ability to distinguish between these processes. Therefore, determining whether AGN-driven outflows reach the starburst ring and influence its star formation activity remains a key open question.

Resolving this question requires looking beyond kinematics to the underlying physical conditions of the gas, where molecular chemistry offers a complementary probe. Previous surveys have already established a rich chemical baseline for the nuclear region of NGC~1068. Notably, \cite{viti2014} combined ALMA and Plateau de Bure Interferometer data to characterize five chemically distinct regions within the circumnuclear disk. Their analysis revealed a multiphase interstellar medium and demonstrated that the starburst ring exhibits lower molecular column densities than the CND, despite similarities in chemical composition. Subsequent multi-transition studies further explored these differences using dense gas and shock tracers \citep[e.g.,][]{Scourfield2020,huang2022,Josh2022}. A common limitation in these studies is the strong degeneracy between gas density and temperature when radiative transfer or chemical modeling is applied in isolation. This highlights the need for modeling frameworks that self-consistently couple non-LTE excitation with chemistry to robustly constrain the physical conditions of molecular gas.

To overcome the degeneracies inherent to isolated radiative transfer or chemical modeling, previous studies have demonstrated that coupling non-LTE radiative transfer calculations with time-dependent chemistry provides significantly tighter constraints on gas density and temperature \citep{Viti2017,Harada2019}. However, the computational cost of such coupled modeling is substantial. To mitigate this limitation, \cite{mijolla2019} developed the first neural-network emulators designed to accelerate the inference process. Building on this approach, \cite{Berhens2024} recently utilized these emulators to enable robust constraints on the cosmic-ray ionization rate in the nearby starburst galaxy NGC\,253. In parallel, advances in statistical methodologies have shown that hierarchical Bayesian frameworks can effectively suppress spurious parameter correlations, as demonstrated by \cite{tanaka2024} for molecular gas in NGC~253.

Building on this foundation, we apply this advanced modeling framework to NGC~1068. We examine how AGN-driven outflows reshape the molecular content along their paths and evaluate whether they ultimately regulate star formation in the starburst ring.

The paper is organized as follows: Section \ref{section:Observations} describes the observations and data reduction. We present the Bayesian method and neural network workflow in Section \ref{section:BayesianandNN}. Section \ref{section:Results} presents the model favorability, corner plots, and parameter distribution maps. In Section \ref{section:Discussion}, we compare our results with previous studies and assess whether AGN-launched outflows reach the starburst ring. We summarize our findings in Section \ref{section:conclusion}.


\section{Observations}
\label{section:Observations}
    \subsection{ALMA  data}

    We use archival data from three ALMA programs: 2018.1.01506.S (PI: S.\,Viti) with Band~3 and Band~4, 2018.1.01684.S (PI: T.\,Tosaki) with Band~3, and 2018.1.00321.S (PI: D.\,Pesce) with Band~5. Table~\ref{table:lines} lists all molecular transitions, project IDs, and synthesized beam sizes. The native beam sizes range from $0\farcs3$ ($\sim20$ pc) to $0\farcs8$ ($\sim60$ pc) at the distance of NGC~1068, which is comparable to the size of a giant molecular cloud \citep{Leroy2015}. All data were calibrated and imaged using CASA \citep{CASA}.
    
    Since our science goal requires maps that cover both the full starburst ring and the CND with comparable resolution and sensitivity, we  focus on low-$J$ transitions of common tracers well detected across both regions, including HCN, HNC, HCO$^{+}$, CS, CN, and C$_{2}$H. These lines provide wide spatial coverage and strong signal-to-noise, suitable for Bayesian modeling based on real detections. Although the archive includes many additional transitions, most are only bright in limited areas and are weak or undetected elsewhere. Including them would introduce numerous $3\sigma$ upper limits with little benefit for our Bayesian analysis. Thus, we restrict our study to the lines listed in Table~\ref{table:lines}.
    
    The rest frequencies were determined using the systemic velocity from \citet{Garcia2019}, with $v_{sys}(HEL) \simeq$ 1130 km/s. The phase tracking center was $\alpha_{2000} = 02^{\mathrm{h}}42^{\mathrm{m}}40.771^{\mathrm{s}}, \delta_{2000} = -00\degr00\arcmin47.84\arcsec$, which is the galaxy's center according to SIMBAD, taken from the Two Micron All Sky Survey (2MASS; \citet{Skru}).

    \begin{table*}
    \caption{Observational details}
    \label{table:lines}
    \centering
    \footnotesize
    \begin{tabular}{lcccccccc}
    \hline\hline
    Molecule & Transition & Rest Frequency & $E_{\rm u}/k$ & Spatial Resolution & $N_{Regions}$ $>$3$\sigma$ & Project ID&MRS&RMS \\
             &            & (GHz)          & (K)           &                    &                                   &           & (\farcs)  &  (mJy/beam)  \\
    \hline
    C$_2$H      & $N=1$--0                & 87.3169  & 4.2  & 0.$^{\arcsec}$66 $\times$ 0.$^{\arcsec}$53 & 64 & 2018.1.01506.S &6.6& 0.42\\
    HCN         & 1--0                    & 88.6316  & 4.3  & 0.$^{\arcsec}$34 $\times$ 0.$^{\arcsec}$32 & 263 & 2018.1.01506.S &6.6&0.42\\
    HNC         & 1--0                    & 90.6636  & 4.4  & 0.$^{\arcsec}$34 $\times$ 0.$^{\arcsec}$32 & 214 & 2018.1.01506.S &6.6&0.42\\
    HCO$^{+}$   & 1--0                    & 89.1885  & 4.3  & 0.$^{\arcsec}$34 $\times$ 0.$^{\arcsec}$32 & 241 & 2018.1.01506.S &6.6&0.42\\
    CS          & 2--1                    & 97.9809  & 7.1  & 0.$^{\arcsec}$78 $\times$ 0.$^{\arcsec}$45 & 308 & 2018.1.01684.S &9.8&0.82\\
    CN          & $N=1$--0, $J=1/2$--1/2  & 113.1441 & 5.4  & 0.$^{\arcsec}$43 $\times$ 0.$^{\arcsec}$36 & 152 & 2018.1.01684.S &9.8&0.46\\
    CN          & $N=1$--0, $J=3/2$--1/2  & 113.4910 & 5.4  & 0.$^{\arcsec}$43 $\times$ 0.$^{\arcsec}$36 & 257 & 2018.1.01684.S &9.8&0.46\\
    HNC         & 2--1                    & 181.3248 & 13.1 & 0.$^{\arcsec}$34 $\times$ 0.$^{\arcsec}$30 & 210 & 2018.1.00321.S &6.0&0.41\\
    \hline
    \end{tabular}
    \tablefoot{
    Molecular transitions observed as part of the ALMA projects 2018.1.01506.S, 2018.1.01684.S, and 2018.1.00321.S.
    Rest frequencies and upper-level energies are taken from the Cologne Database for Molecular Spectroscopy (CDMS; \citealt{e}). 
    The column ``Regions $>$3$\sigma$'' lists the number of hexagons where the integrated intensity exceeds a signal-to-noise ratio of 3. The MRS corresponds to the largest angular scale recoverable by the observations. The RMS shows the channel noise level measured after convolving the cubes to a resolution of 0$\farcs$8.}
\end{table*}

    \subsection{Region selection and Moment-0 Maps}
    \label{sec:regionselection}
    We first convolved all image cubes to 0.$^{\arcsec}8$, which is the largest beam size among the datasets. The line intensity maps were integrated over selected velocity ranges to capture significant emission from both rotational and outflow motions in NGC 1068. For most transitions, we integrated fluxes within $|v-v_{sys}| \leq 230\ \mathrm{km\ s}^{-1}$ to encompass these motions \citep{Garcia2014}. However, some transitions required modified velocity windows: The C${_2}$H (N = 1−0) line, which will be denoted as C${_2}$H (1−0) throughout this work, shows a complex spectral pattern consisting of six hyperfine transitions that can be separated into two distinct fine structure groups. Since these components appear blended at our velocity resolution, we employed a broader integration window of $v-v_{sys} \in [-510, 230]\ \mathrm{km\ s}^{-1}$ to include the complete C$_{2}$H (1−0) transition \citep{Garcia2017}.

    We used the HExagonal Region Averager (HERA)\footnote{https://github.com/ebehrens97/HERA} code \citep{erica_behrens_2024_13839853} to partition the integrated intensity maps into adjoining hexagonal regions, each with a characteristic size of 56\,pc, equivalent to the $\sim0\farcs8$ beam size. The code retains only those regions that satisfy a specified signal-to-noise ratio threshold, in our case S/N $\geq$ 3. The noise on the moment 0 map, $\sigma_{\mathrm{mom}0}$ (in Jy beam$^{-1}$ km s$^{-1}$), is calculated using the noise in one line-free channel, $\sigma_{\mathrm{chan}}$ (in Jy beam$^{-1}$), the number of channels included for each pixel in the moment 0 map, $N_{\mathrm{chan}}$, and the velocity resolution, $\Delta V$ (in km s$^{-1}$):$\sigma_{\mathrm{mom}0} = \sigma_{\mathrm{chan}} \sqrt{N_{\mathrm{chan}}}\Delta V$.

    Although this approach leaves some hexagons with measurements for only a few species, it captures genuine spatial variations in molecular content across NGC~1068. We therefore analyze only hexagonal regions with more than five detected transitions (see Section ~\ref{subsection:HB-model}).

    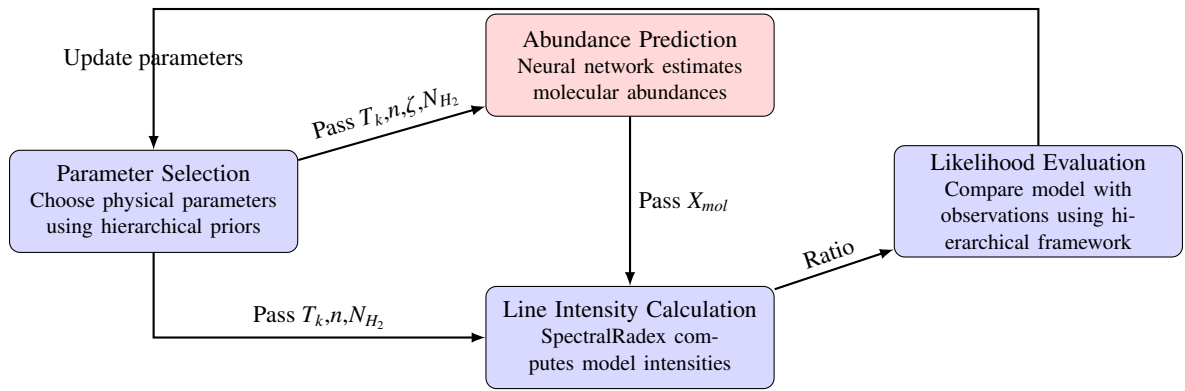
\begin{figure*}
    \centering
    \begin{tikzpicture}[scale=0.9, transform shape,
        box/.style={rectangle, rounded corners, draw=black, minimum width=4cm, minimum height=1.5cm, text centered, text width=4cm, align=center},
        arrow/.style={->, >=latex, thick},
        node distance=3cm
    ]
    
    \node[box, fill=blue!15] (A) at (0,0) {Parameter Selection \\ \small{Choose physical parameters using hierarchical priors}};
    
    \node[box, fill=red!15] (B) at (7,2) {Abundance Prediction \\ \small{Neural network estimates molecular abundances}};
    
    \node[box, fill=blue!15] (C) at (7,-2) {Line Intensity Calculation \\ \small{SpectralRadex computes model intensities}};
    
    \node[box, fill=blue!15] (D) at (13,0) {Likelihood Evaluation \\ \small{Compare model with observations using hierarchical framework}};
    
    \draw[arrow] (A) -- node[above, sloped] {Pass $T_{k}$,$n$,$\zeta$,$N_{H_{2}}$} (B);
    \draw[arrow] (A) |- node[above, near end] {Pass $T_{k}$,$n$,$N_{H_{2}}$} (C);
    \draw[arrow] (B) -- node[right] {Pass $X_{mol}$} (C);
    \draw[arrow] (C) -- node[above, sloped] {Ratio} (D);
    
    \draw[arrow] (D.north) -- ++(0,2) -| node[above, near end] {Update parameters} (A.north);
    
    \end{tikzpicture}
    \caption{Workflow of our hierarchical Bayesian analysis. The neural network component (red) accelerates the analysis by replacing computationally intensive chemical modeling. Each box shows a key step in the process, with arrows indicating data flow between steps. The feedback loop allows continuous refinement of parameter estimates until convergence is achieved.}
    \label{fig:workflow}
    \end{figure*}

\section{Bayesian and Neural Network Workflow}
\label{section:BayesianandNN}
   
    To characterize the global molecular gas properties and to investigate whether the outflow launched from the AGN reaches the starburst ring, we developed a modeling framework that combines chemical evolution, radiative transfer, and Bayesian inference. The chemical abundances are computed with \texttt{UCLCHEM} \citep{holdship2017}\footnote{https://github.com/uclchem/UCLCHEM}, a gas grain chemical modeling code that uses user-defined chemical networks and physical modules to simulate a wide range of conditions. In this work, we employ two types of chemical models: a quiescent cloud model and a heating model. The two models differ only in their temperature evolution, where the heating model allows the temperature to rise gradually to mimic the effect of stellar or AGN activities, while the quiescent model holds the temperature fixed. In the quiescent model, the gas temperature is held at the input value throughout the chemical evolution, whereas in the heating model the gas starts at 10 K and increases gradually to the specified final temperature. Because the chemical network is time-dependent, this difference in thermal history produces distinct molecular abundances even when both models reach the same final temperature. We performed radiative transfer calculations using \texttt{SpectralRadex}\footnote{Our analysis confirms that all transitions remain optically thin at the best-fit parameters in all regions, particularly for $C_{2}H$, which consists of hyperfine structures.} \citep{Holdship2021}\footnote{https://spectralradex.readthedocs.io}. This code, based on \texttt{RADEX} \citep{radex}, predicts molecular line intensities based on collisional rate coefficients obtained from the LAMBDA database \citep{Lambda2005}\footnote{https://home.strw.leidenuniv.nl/~moldata/}. We adopted linewidths of 150, 100, and 50\,km\,s$^{-1}$ for the CND, stellar bar, and starburst ring, respectively \citep{Scourfield2020}.
    
    Bayesian inference is carried out with \texttt{Nautilus} \citep{nautilus}, which combines neural-network accelerated transport, importance sampling, and ultrafast likelihood approximations. This accelerated sampling strategy enables efficient exploration of high-dimensional posteriors and provides robust constraints on the physical conditions across the galaxy with reduced computational cost.

    However, performing full chemical network calculations within the Bayesian inference process would be computationally prohibitive for analysing the entire galaxy. To overcome this limitation, we implemented a neural network that can efficiently emulate the chemical evolution calculations, following the methods of \cite{mijolla2019}, replacing \texttt{UCLCHEM} in our analysis pipeline (see Fig.~\ref{fig:workflow}). The neural network architecture is described in Sect.~\ref{subsection:NN-Architrcture}, and the Bayesian inference methodology is detailed in Sect.~\ref{subsection:HB-model}.

\subsection{Neural Network Architecture}
\label{subsection:NN-Architrcture}

    We constructed a neural network using the TensorFlow package to emulate the chemical modeling calculations. The network takes four inputs, corresponding to the chemical model parameters: kinetic temperature ($T_{\rm K}$), H$_2$ volume density ($n$), cosmic-ray ionization rate ($\zeta$), and H$_2$ column density ($N_{\rm H_2}$). The architecture includes four hidden layers with 2048 nodes each. The output layer returns the predicted molecular abundances for the species considered in this study.

    During training, we use a mean-squared-error loss function to quantify the difference between the neural network predictions and the abundances from the training data. This loss is used to update the network weights and biases throughout each epoch, where an epoch corresponds to one full pass through the training set. A separate validation set, not used during training, is evaluated at the end of each epoch to monitor the model’s performance on unseen data.

    To prevent overfitting, we implemented an early stopping criterion that halts training when the validation loss has not improved over 20 consecutive epochs. The model then reverts to using the weights from the epoch with the lowest validation loss. This approach ensures the model generalizes well to new parameter combinations rather than simply memorizing the training data.

\subsection{Training Data Set}
\label{subsection:Trainingdataset}

    Our neural network model was trained to predict molecular abundances using input parameters generated by \texttt{UCLCHEM} (see Table \ref{table:priors}). \texttt{UCLCHEM} employs a two-stage modeling approach. In stage 1, a freefall collapse model simulates the conversion of atomic/ionic gas to molecular form under fixed physical conditions: initial density of 100 cm$^{-3}$, temperature of 10 K, with the standard cosmic ray ionization rate for molecular hydrogen of $\zeta_{0}$ = 1.3 x 10$^{-17} s^{-1}$\citep[e.g.,][]{indriolo2015}. This stage tracks chemical evolution over 6 Myr, with the final molecular abundances serving as initial conditions for stage 2. The second stage incorporates four of the parameters listed in Table \ref{table:priors} to model chemical evolution for an additional 1 Myr, by which time most reactions have reached chemical equilibrium. In this stage, we adopt a visual extinction of $A_v$ = 10 mag for all models. At this extinction, the external UV radiation field is fully attenuated and UV driven photochemistry does not contribute to the chemical evolution \citep{Scourfield2020,Holdship2021}. This choice is supported by previous studies of NGC\,1068, which infer molecular column densities throughout the CND and the starburst ring that imply visual extinctions well in excess of 10\,mag \citep{Scourfield2020,Josh2022}. The column densities we derive (See Fig.~\ref{fig:physical_parameters}) are consistent with this picture: adopting the conversion relation between $N_{H_2}$ and $A_V$ \citep{Dyson1997}, all analyzed regions remain comfortably in the $A_V > 10$\,mag regime. The final abundances from stage 2 constitute our neural network training dataset.

    Table \ref{table:priors} presents the parameter ranges explored in our models, specifies the assumed prior distributions for each parameter, and indicates the number of sampled values within each range. This sampling scheme yielded 481,988 unique parameter combinations for our chemical modeling.

    \texttt{UCLCHEM} solves coupled ordinary differential equations, and convergence problems can occur when the system is stiff. As noted by \citet{Heyl2023}, such problems can produce unrealistically low molecular abundances that come from numerical error rather than real chemistry. To keep only physically meaningful results in our training set, we set a lower limit of $10^{-15}$ for fractional abundances and removed any outputs below this value. This choice is practical because abundances below $10^{-15}$ are far below our observational sensitivity.

    We divided the \texttt{UCLCHEM} output into three subsets: a training set containing 80\% of the data ($\sim$360,000 points), a validation set with 10\% ($\sim$40,000 points), and a test set with the remaining 10\% ($\sim$40,000 points). The parameter values listed in Table \ref{table:priors} serve as inputs to the first layer of the neural network, which are then propagated through subsequent layers to generate molecular abundance predictions. Since our chemical abundances span eleven orders of magnitude ($10^{-15}$ to $10^{-4}$), normalization to $(-1, 1)$ prevents gradient vanishing for small values and centers the data around zero, which improves the performance of ReLU activation functions and accelerates convergence.
    
    Throughout each training epoch, we continuously monitor the mean-squared error loss between the network predictions and the actual training data. The validation loss, calculated using the validation dataset, is evaluated only after each epoch to assess the network's generalization capability. Once the stopping criterion is met (see Sec \ref{subsection:NN-Architrcture}), we evaluate the trained model's performance using the test set. We can quantitatively evaluate the model's predictive accuracy by comparing the network's abundance predictions for the test set against the corresponding \texttt{UCLCHEM} calculations. 

\begin{table*}
\caption{Prior Distributions and Training Data}
\label{table:priors}
\centering
\begin{tabular}{lcccc}
\hline\hline
Parameter & Description & Range & Distribution Type & No. Points  \\
\hline
$T_{\mathrm{K}}$\tablefootmark{a} & Temperature (K) & 30-300 & Log & 26 \\
$n_{\mathrm{H}_2}$ & Volume Density (cm$^{-3}$) & 10$^3$--10$^7$ & Log & 23 \\
$\zeta$ & Cosmic-ray Ionization Rate ($\zeta_0$)\tablefootmark{b} & 1--10$^{3}$ & Log & 31 \\
$N_{\mathrm{H}_2}$ & H$_2$ Column Density (cm$^{-2}$) & 10$^{22}$--10$^{25}$  & Log & 26 \\
\hline
\end{tabular}
\tablefoot{
\tablefootmark{a} Maximium temperature in the heating model\\
\tablefoottext{b}{$\zeta_0 = 1.36 \times 10^{-17}$ s$^{-1}$}\\
}
\end{table*}

\subsection{Traning Results}
\label{subsection:Trainingresults}

\begin{figure}
   \centering
   \includegraphics[width=.9\hsize]{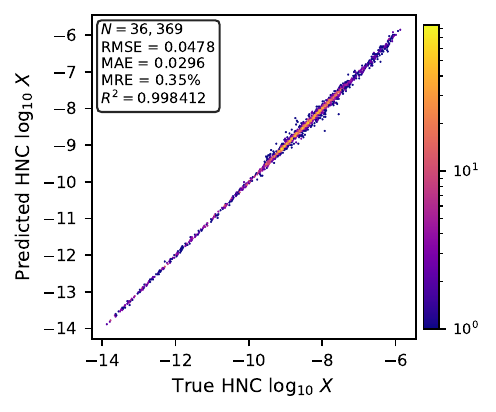}
   \caption{Comparison between the HNC fractional abundances predicted by the neural network emulator and those computed by \texttt{UCLCHEM} for the test dataset containing 36,369 chemical models. The emulator accurately reproduces the chemical model predictions over approximately eight orders of magnitude in abundance ($10^{-10}$ to $10^{-6}$). The performance metrics show a root mean square error of 4.78$\times10^{-2}$ dex, a mean absolute error of 2.96$\times10^{-2}$ dex, and a mean relative error of 0.35\%, with  $R^2 = 0.9984$. These results demonstrate that the neural network provides a reliable surrogate for the chemical model across the explored parameter space.}
   \label{fig:NN_results_HCN}
\end{figure}

    Fig.~\ref{fig:NN_results_HCN} shows the performance of the neural network model using HNC as an example. The network reproduces the \texttt{UCLCHEM} predictions with a mean relative error of 0.35\% and accurately recovers abundance variations spanning approximately eight orders of magnitude. A small number of outliers deviate from the true abundances by several orders of magnitude, and these cases occur mainly when the input volume density ($n_{\mathrm{H_2}}$) or column density ($N_{\mathrm{H_2}}$) lies close to the boundaries of the parameter space \citep{Berhens2024}. Although \cite{Berhens2024} introduced additional data points outside the target parameter ranges to mitigate this behaviour, we do not extend the parameter space with buffer points because the current model performance is sufficient for our analysis and previous studies of NGC~1068 show that the molecular gas rarely reaches these boundary values \citep{viti2014, Scourfield2020, huang2022, Josh2022}. Similar performance results for other molecular species are shown in Fig.~\ref{fig:nn_performance}.

\subsection{Hierarchical Bayesian model}
\label{subsection:HB-model}

We infer the physical conditions of molecular gas using a hierarchical Bayesian (HB) framework following \citet{Tanaka2018} and \citet{tanaka2024}. The HB approach enables a self-consistent treatment of measurement uncertainties and additional non-statistical scatter, while constraining the population distribution of physical parameters across all regions.

For each region $i$, we define a parameter vector $ \boldsymbol{\theta}_i =
(\log n_{\rm H_2},\, \log T_{\rm K},\, \log N_{\rm H_2},\, \log \zeta)_i$. These parameters describe the local physical state inferred from our non-LTE radiative transfer coupled with chemistry modelling. The ensemble of $\boldsymbol{\theta}_i$ is modeled with a multivariate Student-$t$ distribution with location parameter $\boldsymbol{\mu}$ and covariance matrix $\boldsymbol{\Sigma}$, which describes the intrinsic dispersion and correlations of the physical parameters across all regions.

In contrast to previous intensity-based implementations, we use line ratios as observables. For region $i$, the observed data vector is defined as $\mathbf{y}_i^{\rm obs} = \ln(\mathbf{r}_i)$, where $\mathbf{r}_i$ contains ratios constructed from all available transitions with positive detections. To avoid over-weighting specific transitions, we construct $m = n - 1$ ratios in a chain form, where $n$ is the number of transitions detected above 3$\sigma$ in a given region. These ratios take the form $r_k = I_k / I_{k+1}$ for $k = 1,\dots,n-1$, after sorting the transitions. Because adjacent ratios share common intensities, the elements of $\mathbf{y}_i^{\rm obs}$ are correlated. We therefore adopt a multivariate normal likelihood: $\mathbf{y}_i^{\rm obs} \sim \mathcal{N}(\mathbf{y}_i^{\rm mod},\, \mathbf{C}_i)$, where $\mathbf{y}_i^{\rm mod}$ is the model prediction and $\mathbf{C}_i$ is the covariance matrix derived from measurement uncertainties and ratio propagation.

To account for residual non-statistical effects such as calibration systematics and modelling imperfections, we introduce an additive term in log-ratio space:
$\mathbf{y}_i^{\rm obs} =
\mathbf{y}_i^{\rm mod} + \boldsymbol{\epsilon}_i$, where $\boldsymbol{\epsilon}_i$ follows a normal distribution with zero mean and global scale parameter $\sigma_{\rm error}$. This corresponds to a multiplicative factor in ratio space, consistent with the formulation in \citet{tanaka2024}.
\begin{table}
\footnotesize
\caption{Hyperparameters and prior ranges adopted in the hierarchical Bayesian analysis}
\label{table:hyperparams}
\centering
\begin{tabular}{lc}
\hline\hline
Parameter & Prior range \\
\hline
\multicolumn{2}{c}{Location parameters ($\boldsymbol{\mu}$)} \\
$\mu_{n}$ & (3.0, 6.00) \\
$\mu_{T}$ & (1.55, 2.30) \\
$\mu_{N}$ & (22.5, 24.0) \\
$\mu_{\zeta}$ & (0.0, 2.0) \\
\hline
\multicolumn{2}{c}{Scale parameters ($\mathbf{S}$)} \\
$\sigma_{n}$ & (0.1, 1.0) \\
$\sigma_{T}$ & (0.1, 0.8) \\
$\sigma_{N}$ & (0.1, 1.0) \\
$\sigma_{\zeta}$ & (0.1, 0.5) \\
\hline
\multicolumn{2}{c}{Correlation coefficients ($\mathbf{R}$)} \\
$R_{nT}$ & (0.0, ...) \\
$R_{nN}$ & (0.0, ...) \\
$R_{T\zeta}$ & (0.0, ...) \\
$R_{N\zeta}$ & (...,0.0) \\
\hline
\multicolumn{2}{c}{Non-statistical scatter} \\
$\sigma_{\rm error}$ & (0.02, 0.5) \\
\hline
\end{tabular}
\tablefoot{All physical parameters are defined in logarithmic scale. 
The population covariance matrix is written as $\boldsymbol{\Sigma} = \mathbf{S}\mathbf{R}\mathbf{S}$. 
The correlation matrix $\mathbf{R}$ is constructed to be positive definite, and the listed ranges are implemented as soft bounds.}
\end{table}

The population covariance matrix is decomposed as $\boldsymbol{\Sigma} = \mathbf{S}\mathbf{R}\mathbf{S}$, where $\mathbf{S}$ is a diagonal matrix that sets the scale of each parameter, and $\mathbf{R}$ is the correlation matrix. This decomposition separates the marginal dispersions from the parameter correlations and improves numerical stability during sampling. The correlation matrix is parameterized through a lower-triangular form to guarantee positive definiteness. We impose soft constraints on selected correlation coefficients in order to suppress unphysical degeneracies between strongly coupled parameters. In particular, we allow positive correlations between $N_{\rm H_2}$ and $n_{\rm H_2}$, and between $T_{\rm K}$ and $\zeta$, while a negative correlation is favored between $N_{\rm H_2}$ and $\zeta$, consistent with  \citet{padovani2018}. Although the correlation between $n_{\rm H_2}$ and $T_{\rm K}$ is not immediately intuitive, it has been reported for the Galactic center \citep{Nagai2007, Tanaka2018} and in extragalactic hierarchical Bayesian studies \citep{tanaka2024}. We note, however, that in our analysis the final posterior distributions do not show a significant change whether this correlation is imposed or not. The adopted hyperparameters and their prior ranges are summarized in Table~\ref{table:hyperparams}.

Under this hierarchical structure, the joint posterior distribution is written as
\begin{equation}
P(\boldsymbol{\Theta}, \boldsymbol{\epsilon}, \boldsymbol{\eta}\mid \mathbf{Y})
\propto
P(\mathbf{Y}\mid \boldsymbol{\Theta}, \boldsymbol{\epsilon})\,
P(\boldsymbol{\Theta}, \boldsymbol{\epsilon}\mid \boldsymbol{\eta})\,
P(\boldsymbol{\eta}),
\end{equation}
Here $\mathbf{Y}=\{\mathbf{y}_i^{\rm obs}\}$ is the set of observed log-ratio vectors,
$\boldsymbol{\Theta}=\{\boldsymbol{\theta}_i\}$ are the region parameters,
$\boldsymbol{\epsilon}=\{\boldsymbol{\epsilon}_i\}$ are the non-statistical error terms,
and $\boldsymbol{\eta}=\{\boldsymbol{\mu},\boldsymbol{\Sigma},\sigma_{\rm error}\}$ denotes the hyperparameters.

The HB inference yields posterior probability distributions of 
$T_{\rm K}$, $n_{\rm H_2}$, $N_{\rm H_2}$, and $\zeta$ for each 56\,pc region, together with 68\% credible intervals.

\section{Results}
\label{section:Results}
  
    In this section, we present the results of our hierarchical Bayesian analysis of the molecular gas in NGC~1068. The framework yields posterior constraints on the physical conditions of the gas, including density, temperature, cosmic-ray ionization rate, and column density. These parameters enable us to construct detailed physical parameter maps across NGC~1068, revealing spatial gradients that trace the signatures of AGN-driven outflow through the galaxy.

\subsection{Posterior Results and Model Preference}
\begin{figure}
       \centering
       \includegraphics[width=0.9\hsize]{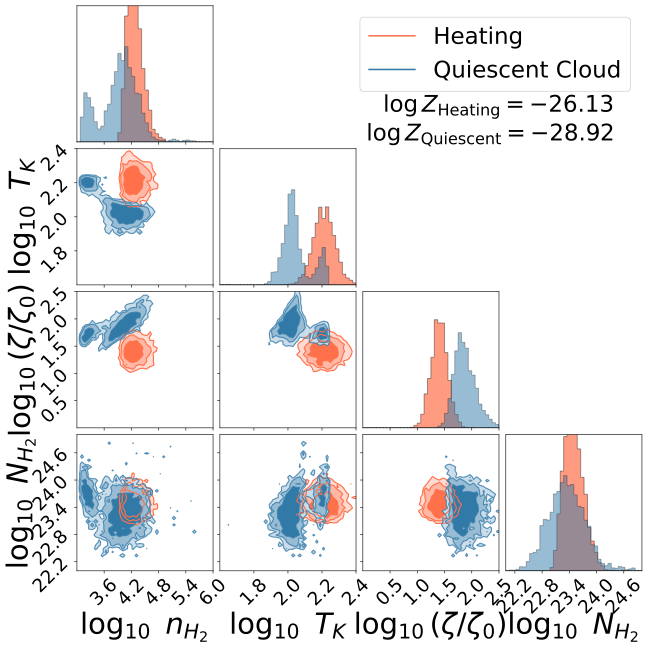}
       \caption{Corner plots showing the Bayesian inference results for a molecular cloud near the CND-S. Orange and blue correspond to the heating and quiescent cloud models. Contours show the 50, 80 and 95 per cent credible intervals. The log-evidence values indicate that the heating model is preferred.}
       \label{fig:corner_plots}
    \end{figure}

    We identify 198 regions where more than five molecular transitions are detected above the three-sigma level, ensuring that the number of observational constraints exceeds the number of free parameters in the Bayesian inference. Figure~\ref{fig:corner_plots} presents an example of the posterior distributions for one of the sub-regions listed in Table~\ref{tab:subregions}, located near the CND-S. The posteriors are shown for both the heating and quiescent cloud models. The Bayesian evidence (log $\mathcal{Z}$) provides an objective criterion for model selection, as it measures the marginal likelihood averaged over the full parameter space. A higher value indicates stronger support when both fit quality and model complexity are taken into account. In this representative region, the evidence values are $-26.13$ for the heating model and $-28.92$ for the quiescent model. The difference of $2.8$ in log-evidence corresponds to a Bayes factor of $\exp(2.8)\approx 16$, which provides strong support for the heating model.

\begin{figure*}[htbp]
        \centering
        \includegraphics[width=0.8\hsize]{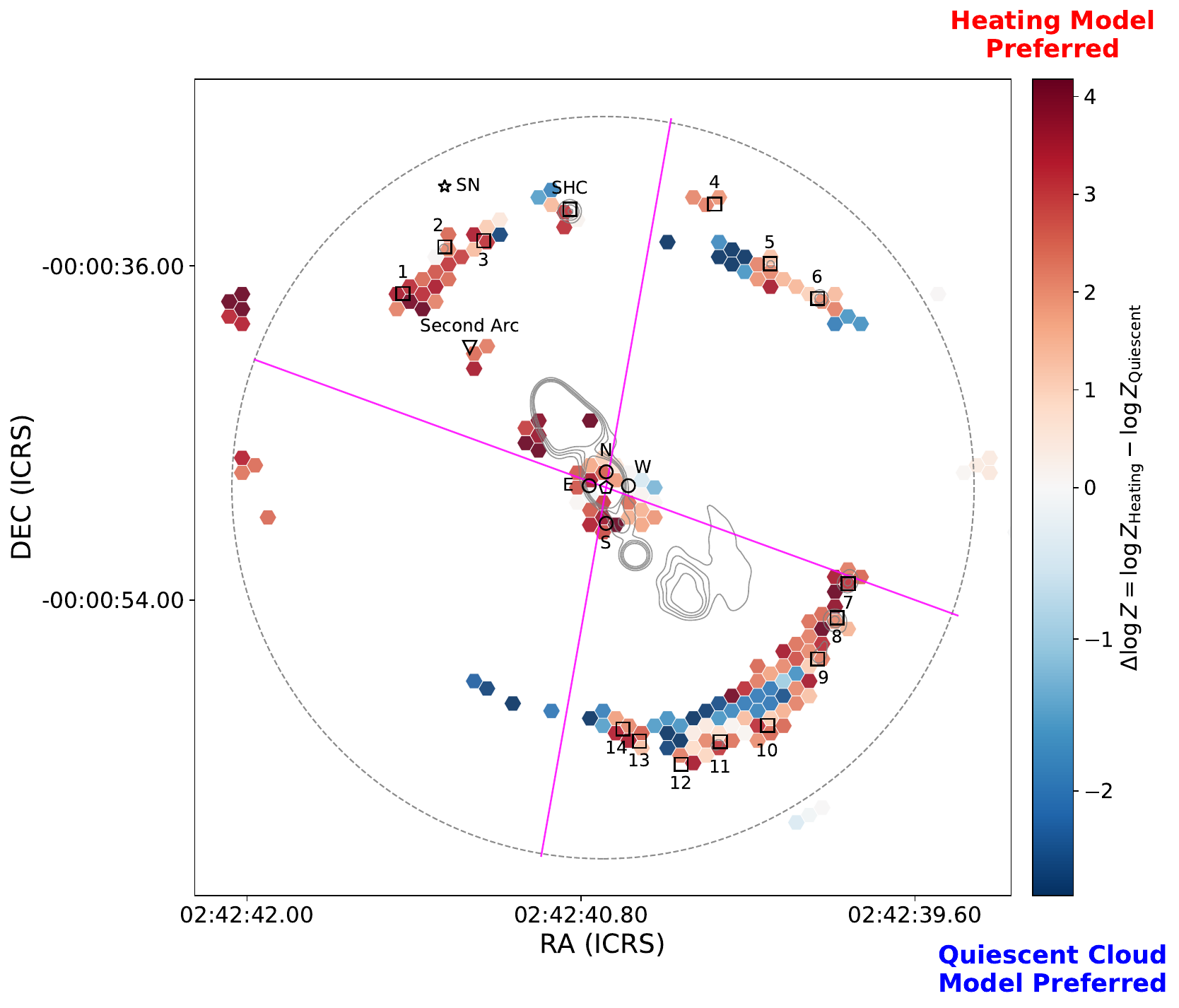}
        \caption{Spatial distribution of the difference in log-evidence between the heating and quiescent models across the CND and the starburst ring. Red colours indicate support for the heating model and blue colours indicate support for the quiescent model. Contours show the 100\,GHz continuum map and symbols mark regions of interest. Labels denote the corresponding SSCs and sub-regions in the CND. The magenta line outlines the region possibly influenced by the AGN wind described by \citet{Das2007}.}
        \label{fig:model_evidence}
    \end{figure*}

    We then construct a spatial model-selection map using the difference in Bayesian evidence (Fig.~\ref{fig:model_evidence}). Regions of particular interest are marked with different symbols. Circles denote sub-regions in the CND and the AGN, squares indicate superstar clusters (SSCs) in the starburst ring, and a pentagram marks the position of Type II SN 2018ivc \citep{bostroem2020}. The SSC positions are adopted from \citet{Villas2021}, where Pa$\alpha$ emission was used to trace compact young stellar clusters. The coordinates of all marked sub-regions are provided in Table~\ref{tab:subregions}. Regions within the CND along the AGN wind direction show the strongest preference for the heating model, consistent with previous studies that identified outflow-driven disturbances in these areas \citep{Garcia2014,Garcia2019}. Some regions outside the ionized bicone also favor heating, in agreement with non-rotational CO(2--1) motions extending beyond the bicone\citep{Garcia2014,Zhang2025}. In contrast, regions located further from the AGN, on the eastern side of the East Knot and the western side of the West Knot, preferentially support the quiescent model, consistent with the absence of clear outflow signatures in those areas.

    Several regions along the jet direction on the eastern side correspond to the arc features seen by \cite{Garcia2014}. High-sensitivity \textit{e-}MERLIN and VLA images from \cite{mutie2024} show that the NE arc has a spectral index of $-1.1\pm0.1$, typical of optically thin synchrotron emission, indicating strong jet–ISM interaction. These regions also favor the heating model, supporting this interpretation. A shock model would likely better describe these conditions, but since shock-based emulators are still in development, we focus here on the heating–quiescent comparison.

    \begin{table}
    \footnotesize
    \centering
    \caption{Sub-region coordinates in NGC 1068}
    \begin{tabular}{l l l}
    \hline\hline
    Sub-region & RA (J2000) & Dec (J2000) \\
    \hline
    E Knot & 02$^h$42$^m$40.771$^s$ & $-$00$^\circ$00$'$47.84$''$ \\
    W Knot & 02$^h$42$^m$40.630$^s$ & $-$00$^\circ$00$'$47.84$''$ \\
    AGN & 02$^h$42$^m$40.710$^s$ & $-$00$^\circ$00$'$47.94$''$ \\
    CND-N & 02$^h$42$^m$40.710$^s$ & $-$00$^\circ$00$'$47.09$''$ \\
    CND-S & 02$^h$42$^m$40.710$^s$ & $-$00$^\circ$00$'$49.87$''$ \\
    SHC & 02$^h$42$^m$40.84$^s$ & $-$00$^\circ$00$'$32.94$''$ \\
    SSC 1 & 02$^h$42$^m$41.44$^s$ & $-$00$^\circ$00$'$37.48$''$ \\
    SSC 2 & 02$^h$42$^m$41.29$^s$ & $-$00$^\circ$00$'$34.97$''$ \\
    SSC 3 & 02$^h$42$^m$41.15$^s$ & $-$00$^\circ$00$'$34.64$''$ \\
    SSC 4 & 02$^h$42$^m$40.32$^s$ & $-$00$^\circ$00$'$32.66$''$ \\
    SSC 5 & 02$^h$42$^m$40.12$^s$ & $-$00$^\circ$00$'$35.87$''$ \\
    SSC 6 & 02$^h$42$^m$39.95$^s$ & $-$00$^\circ$00$'$37.75$''$ \\
    SSC 7 & 02$^h$42$^m$39.84$^s$ & $-$00$^\circ$00$'$53.12$''$ \\
    SSC 8 & 02$^h$42$^m$39.88$^s$ & $-$00$^\circ$00$'$54.96$''$ \\
    SSC 9 & 02$^h$42$^m$39.95$^s$ & $-$00$^\circ$00$'$57.88$''$ \\
    SSC 10 & 02$^h$42$^m$40.13$^s$ & $-$00$^\circ$01$'$00.76$''$ \\
    SSC 11 & 02$^h$42$^m$40.30$^s$ & $-$00$^\circ$01$'$01.64$''$ \\
    SSC 12 & 02$^h$42$^m$40.44$^s$ & $-$00$^\circ$01$'$02.85$''$ \\
    SSC 13 & 02$^h$42$^m$40.59$^s$ & $-$00$^\circ$01$'$01.61$''$ \\
    SSC 14 & 02$^h$42$^m$40.65$^s$ & $-$00$^\circ$01$'$00.95$''$ \\
    SN2018ivc & 02$^h$42$^m$41.29$^s$ & $-$00$^\circ$00$'$31.71$''$ \\
    \hline
    \end{tabular}
    \label{tab:subregions}
    \end{table}
    
    In the northern starburst ring, regions near the SSCs, compact young stellar clusters associated with intense star formation, and at the bar–ring interface favor the heating model, consistent with strong cloud compression. The average log-evidence difference exceeds 3. The northwestern starburst ring, however, shows mixed behavior: areas outside the SSCs favor the quiescent model, suggesting that molecular clouds at $\sim$50\,pc resolution remain largely unperturbed. The southern starburst ring shows a layered pattern. The bar–ring interface again favors heating, while regions slightly farther out lean toward the quiescent model, implying weaker external influence. At the outer edge, the heating model dominates once more. \cite{Villas2021} showed that gas responds to the spiral inside corotation by entering the spiral arm, where accumulation and/or compression triggers star formation. Our results align with this picture, as the outer starburst ring edge appears to host gas undergoing active heating and star formation.

\subsection{Physical Properties of the Cloud}
\label{subsection:parametermaps}

    \begin{figure*}
    \centering
    \includegraphics[width=0.84\textwidth]{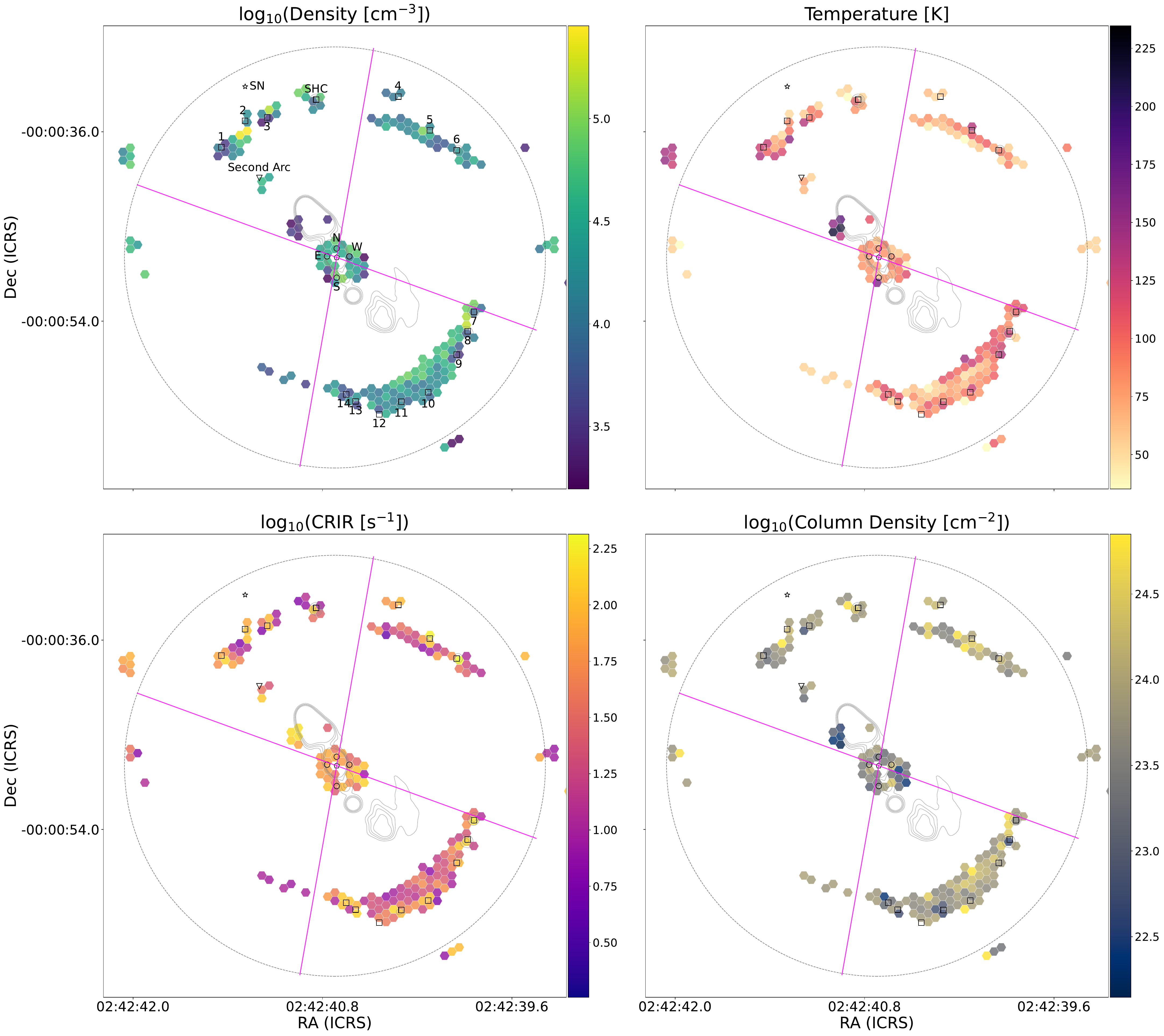}
    \caption{Physical parameters derived in NGC 1068. The panels show the spatial distribution of molecular gas properties: volume density (upper left), kinetic temperature (upper right), cosmic ray ionization rate (lower left), and column density (lower right). Labels denote the corresponding SSCs and sub-regions in the CND, and are listed in the density panel.}
    \label{fig:physical_parameters}
    \end{figure*}

    Fig.~\ref{fig:physical_parameters} presents the physical parameters derived from the HB model. We analyzed the parameter statistics across the three main regions of NGC 1068 (CND, stellar bar, and starburst ring), revealing distinct physical conditions in each environment. The CND exhibits the highest molecular gas density ($n_{\rm H_2} \approx 10^{4.42 \pm 0.31}$ cm$^{-3}$), denser than the starburst ring ($n_{\rm H_2} \approx 10^{4.17 \pm 0.32}$ cm$^{-3}$) and the stellar bar ($n_{\rm H_2} \approx 10^{3.78 \pm 0.29}$ cm$^{-3}$). This trend is consistent with previous studies \citep{viti2014,Scourfield2020,Josh2022}, although we note that the exact values might not match those from previous studies, and such differences will be discussed in Section \ref{subsection:Comparsion}. Similar trends are observed in the H$_2$ column density, with the CND showing the highest values ($N_{\rm H_2} \approx 10^{23.52 \pm 0.34}$ cm$^{-2}$), followed by the starburst ring ($N_{\rm H_2} \approx 10^{23.23 \pm 0.33}$ cm$^{-2}$) and the stellar bar ($N_{\rm H_2} \approx 10^{22.84 \pm 0.31}$ cm$^{-2}$).

    The kinetic temperature distribution reveals that the stellar bar exhibits the highest mean temperature ($T_{\rm kin} = 140.8 \pm 36.6$ K), substantially exceeding both the CND ($T_{\rm kin} = 82.2 \pm 28.3$ K) and the starburst ring ($T_{\rm kin} = 76.2 \pm 28.1$ K). This temperature enhancement in the stellar bar is consistent with expectations for regions proximate to the jet \citep{mutie2024}, where mechanical heating and cosmic ray heating contribute to elevated temperatures. The southern starburst ring displays higher temperatures than its northern counterpart, which may be attributed to the presence of younger SSCs~\citep{Villas2021}. A similar temperature gradient linked to SSC evolution has been observed in NGC~4945: \citet{Bellocchi2023} found that the youngest proto-SSCs have rotational temperatures of about 65\,K, much higher than the 25\,K measured in more evolved SSCs. In early stages, strong protostellar activity and feedback heat the surrounding molecular gas. As SSCs evolve, heating becomes confined to H\,II regions while the larger molecular cloud cools, producing the observed decline in temperature.

    The cosmic ray ionization rate (CRIR) remains uniformly high across the central kiloparsec, with only a moderate decline toward the outer regions. The CND shows the highest mean value, $\log(\zeta/\zeta_{0}) \approx 1.95 \pm 0.21$, followed by the stellar bar ($1.91 \pm 0.23$) and the SSCs in the starburst ring ($1.88 \pm 0.24$). The molecular clouds not associated with SSCs show a more pronounced decrease, with an average of $1.03 \pm 0.21$. The elevated CRIR in the CND likely originates from particle acceleration near the AGN or within AGN-driven outflows \citep[e.g.][]{Rodrigues2020Active}. The similar values found in the bar suggest that jet–ISM interactions or large-scale shocks maintain a high level of cosmic ray activity there, consistent with observations of cosmic ray production in jet systems such as Centaurus~A \citep{Collaboration2020Resolving}. The slightly lower but still enhanced CRIR in the starburst ring is consistent with cosmic ray generation by clustered supernovae and massive stellar winds in active star-forming regions \citep{Holdship2022,Berhens2024}. The overall distribution therefore points to a broadly elevated cosmic ray field across the CND, bar, and SSCs, with a clear reduction only in the more quiescent molecular clouds.

\section{Discussion}
\label{section:Discussion}

    In this section, we first compare our results with previous work to establish consistency with earlier studies of NGC~1068. We then investigate whether the AGN-driven outflow affects the starburst ring by analyzing physical parameter gradients along the outflow trajectory and across the ring, with particular focus on regions at the inner edges of the northern and southern starburst rings. Finally, we present a preliminary analysis of CO(1--0) line profiles, demonstrating that single-Gaussian fits inadequately describe the emission in the starburst rings, which provides additional evidence for outflow activity.

\subsection{Comparisons with Previous Studies}
\label{subsection:Comparsion}

\begin{figure*}
    \centering
    \includegraphics[width=0.8\textwidth]{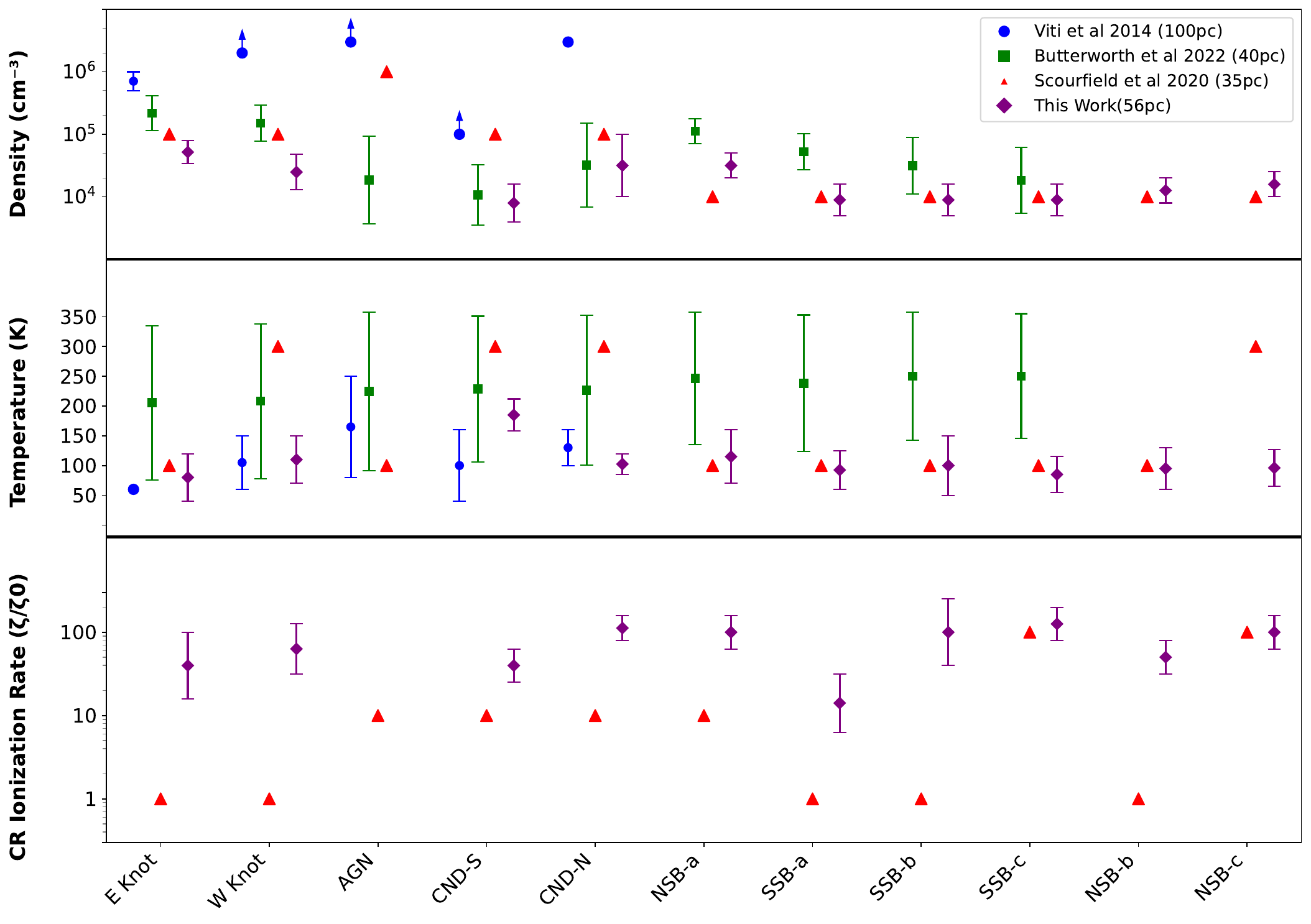}
    \caption{Comparison of H$_2$ volume density, kinetic temperature, and cosmic ray ionization rate across different regions in NGC~1068. Values are shown for the E-knot, W-knot, CND-N, CND-S, and selected starburst ring regions. Results from \citet{viti2014} at 100 pc resolution, \citet{Josh2022} at 40 pc resolution, \citet{Scourfield2020} at 35 pc resolution, and this work at 56 pc resolution are shown for reference.}
    \label{fig:parameter_comparison}
    \end{figure*}

    We compare our results with previous multi molecular studies of NGC~1068 in Fig.~\ref{fig:parameter_comparison}, focusing on \citet{Josh2022}, \citet{viti2014}, and \citet{Scourfield2020}. For \citet{viti2014} and \citet{Scourfield2020}, we adopt the parameters derived from their best fit \texttt{UCLCHEM} models to remain consistent with the coupled chemical and radiative transfer framework used in this work. Although \citet{Josh2022} did not include chemical modelling, the overlap in molecular tracers makes their results suitable for comparison. More details of those previous studies are summarized in Appendix.~\ref{sec:appxE}.
    
    The largest discrepancies among studies are found in the derived gas densities within the CND, particularly in the E-knot and W-knot, where our values are systematically lower than those reported previously. This difference likely reflects a combination of spatial sampling and tracer selection. These regions may not be fully sampled by our hexagonal grid, and the area-weighted averaging of overlapping hexagons may incorporate lower density surroundings. More importantly, our analysis relies on low-J transitions, which are less sensitive to dense gas than the high-J transitions used in earlier work, consistent with \citet{Josh2022}, who showed that low-J HCN is not an effective tracer of high-density material. Due to the limited number of available molecular constraints in the innermost CND, likely related to the strong ionized outflow \citep{Garcia2014,huang2022,vollmer2022}, we do not derive physical parameters for the AGN region. In the CND-N and CND-S regions, our density estimates agree more closely with \citet{Josh2022}, while the larger offsets relative to \citet{viti2014} and \citet{Scourfield2020} can be attributed to differences in chemical model sampling, since their parameter grids were limited in size and unlikely to capture the strong chemical differentiation across the CND.
    
    In the starburst ring, discrepancies in densities are mainly confined to NSB-a and SSB-a. These regions host very young SSCs with ages of order $10^{5}$ years \citep{Villas2021}, and  previous studies have shown that clusters at this stage can disperse nearby gas \citep[e.g.][]{Miura2018,Mills2021,Sun2024}. The differences we find likely arise because our $\sim50$\,pc resolution beam contains multiple gas components. Young SSCs create low-density environments in their immediate surroundings, yet dense clumps can still exist within the beam. Different molecules therefore trace different phases that cannot be separated at this scale. This naturally leads to lower inferred densities compared to studies that rely on higher excitation transitions, such as \citet{Josh2022}, and indicates that different data sets probe different components of the interstellar medium rather than inconsistent physical conditions.
    
    For kinetic temperature (the second row in Fig.~\ref{fig:parameter_comparison}), the main discrepancies arise when comparing our results with \citet{Scourfield2020}. In particular, our temperature estimates for the W-knot, CND-N, and CND-S are higher than those reported in their study. This difference likely reflects their reliance on CS transitions alone and the use of a sparse temperature grid, limited to 100 and 300 K, which restricts the ability to capture the full range of conditions present within the CND. In contrast, the temperature we derive within the CND is broadly consistent with \citet{viti2014}, although their lower spatial resolution of about 100 pc smooths over the temperature variations that we resolve, naturally leading their estimates toward lower values.

    The cosmic ray ionization rate (the third row in Fig.~\ref{fig:parameter_comparison}) shows large variation among studies, particularly when compared with the CS-based analysis of \citet{Scourfield2020}. This difference reflects the limitation of using a single molecular species to trace ionization level. Although our results disagree with \citet{Scourfield2020} for several sub-regions, both studies indicate that the ionization rate in these regions remains below about $100\,\zeta_{0}$, a regime where the CS abundance becomes strongly non-linear (see Fig.~\ref{fig:CS_abundance}). This behavior makes the CRIR derived from CS alone unreliable compared with our multi-species approach.
    
\subsection{The outflow impact}

    In this section, we examine whether the AGN-driven outflow in NGC 1068 influences star formation activity within the starburst ring. To address this question, we analyze the spatial distribution of physical parameters (Sec.~\ref{subsubsection:physicalparametersacross}), search for signatures of gas compression caused by the outflow and investigate CO(1–0) line profiles across the galaxy (Sec.~\ref{subsec:COlineprofile}).

\subsubsection{Gas Density Enhancement}
\label{subsubsection:physicalparametersacross}
    In Fig.~\ref{fig:physical_parameters}, we presented the physical parameters throughout the galaxy. Given that the aim of this work is to investigate the impact of the outflow on the starburst ring, we focus our analysis on regions beyond the CND. The arc regions (See Fig.\ref{fig:COline-profile}) show clear signatures of jet influence, exhibiting relatively low densities ($\sim10^{3}$ cm$^{-3}$). Such reduced densities are consistent with strong jet–ISM interaction scenario, which can disturb or partially remove molecular gas through enhanced turbulence and cloud disruption, as shown in numerical simulations \citep{lauvikas2024} and in observational studies (e.g., \citealt{quillen2005,Arce2010}). The arc regions also exhibit the highest gas temperatures measured in our study, indicating the extreme physical conditions present.

    Moving outward from the CND toward the starburst ring along the jet propagation direction reveals distinct patterns in the gas properties. The second arc regions (Green triangle in Fig.~\ref{fig:physical_parameters}), situated between the starburst ring and the primary arc structures, display significantly enhanced gas densities ($\sim$10$^{5}$ cm$^{-3}$) and lower gas temperatures ($\sim$60 K). These physical properties, together with the characteristic V-shaped morphology traced by the CO emission (Fig.~\ref{fig:COline-profile}), suggest that although the region is not directly impacted by the jet, compression associated with jet-driven outflows likely contributes to the observed molecular gas structure and influences the local physical conditions.

    The northeastern starburst ring shows spatial variations in gas density that point to two different feedback effects. In regions that coincide with the SSCs, the gas density is systematically lower than in the surrounding regions, consistent with the removal or dispersal of gas by stellar winds from the young clusters \citep[e.g.,][]{pabst2019, rosen2022, Bonne2022}. In contrast, nearby regions that do not overlap with any SSC exhibit density enhancements above 10$^{4}$ cm$^{-3}$. When these values are compared with the northwestern ring, where densities remain below 10$^{4}$ cm$^{-3}$, the enhanced densities are consistent with a scenario in which the outflow compresses the gas in these locations, although other mechanisms such as cloud collisions cannot be excluded. The observed pattern therefore suggests that stellar winds reduce the gas content around the SSCs, while the outflow may contributes to gas compression in adjacent clouds.
    
    The southwestern starburst ring displays less fragmentation than its northern counterpart. Similar density patterns emerge near the bar-ring interface region, where decreased densities within SSCs are accompanied by elevated densities in surrounding molecular material. \cite{Sanchez2022} attributed the significant compression of molecular gas at the bar-ring interface to intense cloud-cloud collisions. In their analysis, the degree of gravitational boundedness of the dense gas was quantified through the parameter $b \equiv \Sigma_{\text{dense}}/\sigma^2$, which is proportional to the inverse of the virial parameter and measures the balance between self-gravity and kinetic energy of the gas, following the methodology of \cite{Leroy2017}. A higher value of $b$ indicates gas that is more gravitationally bound and therefore more likely to form stars efficiently. Our results are consistent with this scenario; however, we note that the boundedness parameter method has important limitations. As demonstrated in their study, the boundedness parameter shows significant scatter at the native resolution of $\sim$56 pc, with correlations only becoming statistically significant when averaged over larger scales ($\geq$100 pc). Furthermore, the parameter assumes virial equilibrium, which may not hold in regions experiencing strong dynamical perturbations from outflows or cloud-cloud collisions. Additionally, we identify density enhancements exceeding 10$^{4}$ cm$^{-3}$ accompanied by elevated temperatures ($>$120 K) at the inner edge of the southwestern starburst ring. The spatial coincidence of enhanced density and temperature in this region, combined with its location relative to the outflow axis, suggests that outflow-driven compression rather than cloud-cloud collisions may be the dominant mechanism in this particular zone.

    We note that while the combined density and temperature measurements provide valuable insights into the physical conditions across the starburst ring, these diagnostics alone cannot definitively distinguish between the various compression mechanisms at play. The observed enhancements are consistent with both jet-driven compression through ram pressure (e.g., \citealt{wagner2012}) and shock compression from cloud-cloud collisions at the bar-ring interface (e.g., \citealt{choi2023,maity2024,maeda2025}), as moderate-strength shocks from different physical origins produce similar signatures in our observed parameter range. Furthermore, projection effects and beam averaging may blend regions experiencing different mechanisms, particularly where the outflow influence zone overlaps with the bar-ring interface. Detailed kinematic analysis, which can reveal shock velocities and systematic velocity patterns associated with different acceleration mechanisms, will be required to disentangle these possibilities and will be addressed in future work. For the present analysis, we proceed with the understanding that significant compression is occurring throughout the starburst ring, with the specific driving mechanisms likely varying spatially depending on proximity to the jet axis and the bar-ring interface.
        
\subsubsection{CO Line Profiles}
\label{subsec:COlineprofile}

    In the previous sections, we investigated the potential effects of the outflow on the molecular gas properties. Here, we examine more direct evidence for gas outflow motion through analysis of CO line profiles. While detailed line profile analysis and kinematic modeling are beyond the scope of this work, we present a preliminary examination of outflow signatures in the CO line profiles across the galaxy.

    \begin{figure*}[htbp]
        \centering
        \includegraphics[width=.95\textwidth]{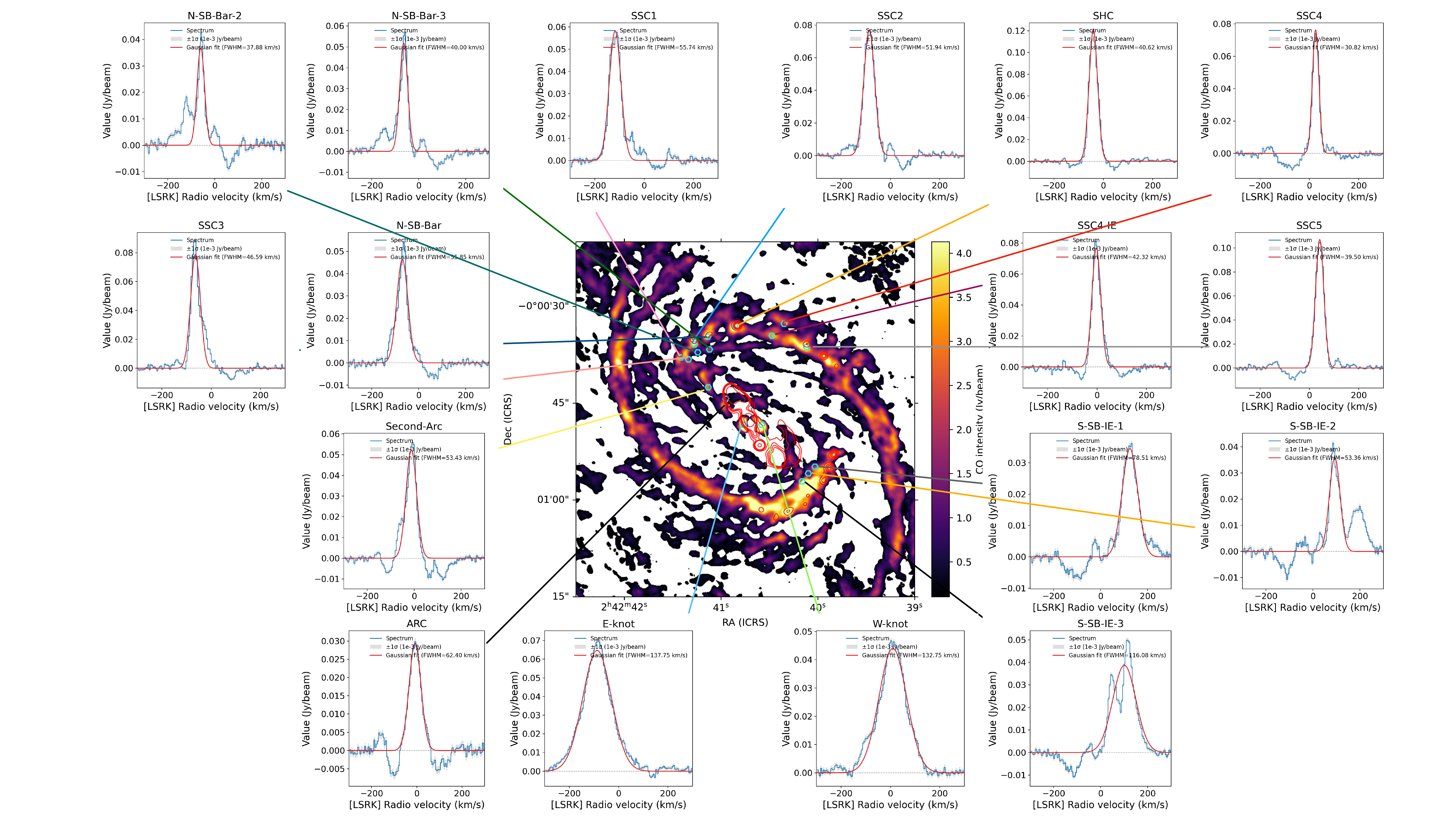}
        \caption{CO(1--0) line profiles across NGC~1068. The color map shows the CO(1--0) integrated intensity with red contours representing the 100~GHz continuum emission at 5$\sigma$, 10$\sigma$, 15$\sigma$, 20$\sigma$, and 100$\sigma$ levels. Line profiles are extracted from hexagonal apertures of identical size. Each panel shows the observed spectrum (blue), the 1$\sigma$ noise level (grey), and a single-component Gaussian fit (red) to the emission above zero intensity. The hexagon positions and extraction regions are indicated on the map.}
        \label{fig:COline-profile}
    \end{figure*}

    Within the CND, we present CO line profiles from the E-knot and W-knot (middle-bottom subplots in Fig.~\ref{fig:COline-profile}). Single-component Gaussian fits clearly fail to reproduce the observed profiles. Indeed, \cite{Zhang2025} performed comprehensive line profile fitting in the CND regions and concluded that the E/W-knot require at least two velocity components to adequately fit the CO line profiles, providing evidence for molecular outflow.

    The Arc and second Arc regions (second column, third and fourth rows in Fig.~\ref{fig:COline-profile}) exhibit pronounced absorption features at both blue- and red-shifted velocities. While missing flux may contribute to these features, the single-component fit to the main emission is insufficient, and additional components with different central velocities would be required to adequately reproduce the observed profiles, which may indicate the presence of multiple kinematic components along the line of sight.

\begin{figure}[htbp]
    \centering
    \includegraphics[width=0.9\hsize]{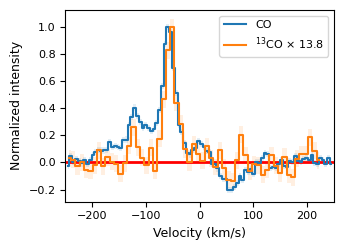}
    \caption{The CO(1--0) (blue) and $^{13}$CO(1--0) (scaled by a factor of 13.8, orange) line profiles extracted from the inner edge of the NE starburst ring. The secondary peak located at $-120$\,km\,s$^{-1}$ appears in both transitions. The error bars indicate the rms noise per velocity channel measured from line-free regions of the spectrum.}
     \label{fig:COvs13CO}
    \end{figure}

    We examine three CO(1--0) line profiles extracted from the inner edge of the northern SB ring (regions N\mbox{-}SB\mbox{-}Bar, N\mbox{-}SB\mbox{-}Bar\mbox{-}2, and N\mbox{-}SB\mbox{-}Bar\mbox{-}3; see the second column of the second row, the first row, and the first column of the first row in Fig.~\ref{fig:COline-profile}) and compare them with three spectra at nearby SSC positions (SSC1, SSC2, and SSC3). The SSC spectra are well described by single, nearly symmetric Gaussian components with FWHM values of $55.74$, $51.94$, and $46.59$\,km\,s$^{-1}$, while the N\mbox{-}SB\mbox{-}Bar spectra, although not systematically broader in FWHM (one is comparable at $55.85$\,km\,s$^{-1}$ and the other two are narrower at $37.88$ and $40.00$\,km\,s$^{-1}$), are markedly more \emph{non-Gaussian}: all three exhibit extended blue wings reaching $v \lesssim -150$\,km\,s$^{-1}$ and redward depressions/absorption-like dips around +20 to $+120$\,km\,s$^{-1}$, and in some cases an additional blue component is present. We present the spectra extracted from the N\mbox{-}SB\mbox{-}Bar\mbox{-}2 region, where both the CO(1--0) and $^{13}$CO(1--0) profiles display clear signatures of complex kinematics, as shown in Fig.~\ref{fig:COvs13CO}. The appearance of a secondary peak at $-120$\,km\,s$^{-1}$ in both transitions confirms this feature is intrinsic to the gas rather than an artifact of CO self absorption, while the extended blue wing in the CO emission suggests the presence of outflowing molecular gas across this region.

    At the inner edge of the southern starburst ring, we examine three regions, S-SB-IE-1, S-SB-IE-2, and S-SB-IE-3 (fifth column third row, sixth column third row, and fifth column fourth row in Fig.~\ref{fig:COline-profile}, respectively). The CO spectra reveal diverse kinematic structures. S-SB-IE-1 shows extended wing-like emission on both the redshifted and blueshifted sides of the main component, suggesting the presence of dynamically disturbed gas, potentially associated with outflowing motions. S-SB-IE-2 exhibits a more localized excess near $+200$\,km\,s$^{-1}$, with a stronger high-velocity contribution than in the surrounding regions. S-SB-IE-3 presents a pronounced double-peaked structure centered near $+100$\,km\,s$^{-1}$. We also inspected the corresponding \(^{13}\)CO(1--0) spectra to assess the robustness of these features. However, none of the regions displays a clear counterpart to the excess emission near $+200$\,km\,s$^{-1}$, and in particular the secondary contribution seen in CO in S-SB-IE-2 is not detected in \(^{13}\)CO. For S-SB-IE-3, the limited velocity resolution and modest signal-to-noise prevent us from confirming the presence of the weak feature near $+150$\,km\,s$^{-1}$. Taken together, these profiles likely trace a mixture of kinematic components, and their morphology may reflect CO self-absorption (e.g. \citealt{Combes2019, Rose2019, Rose2020}) or the superposition of multiple gas structures along the line of sight.

    While determining the precise origin of these complex CO line profiles is beyond the scope of this work, the comparison between regions along the proposed outflow trajectory and those in the starburst rings provides suggestive evidence for outflow activity. The observed absorption features at the line wings and the double-peaked profiles warrant detailed investigation in future work.

\section{Conclusions}
\label{section:conclusion}
We have combined high resolution ALMA observations with a hierarchical Bayesian framework that links non-LTE radiative transfer and chemical modeling to study the molecular gas in NGC~1068. By training a neural network emulator, we achieved efficient inference without losing predictive accuracy. This approach allows us to derive spatially resolved distributions of gas density, temperature, column density, and cosmic-ray ionization rate across both the CND and the starburst ring. Our main findings can be summarized as follows:
\begin{enumerate}
    \item The inferred physical parameters show clear radial and azimuthal variations. The CND displays higher temperatures and ionization rates than the starburst ring, while the ring hosts cooler and denser gas associated with star formation.
    \item Enhanced gas density and elevated temperatures are found along the inner edge of the starburst ring, suggesting that the nuclear outflow influences the local gas conditions beyond the CND.
    \item Line profiles of CO~(1--0) exhibit non-Gaussian features in regions near the outflow path, consistent with the presence of disturbed kinematics or multiple gas components.
    \item Comparison with previous single-pointing studies shows broad consistency, but our spatially resolved modeling highlights sub-structure that could not be captured by earlier analyses.
\end{enumerate}

These results demonstrate the capability of combining chemistry and radiative transfer calculation to interpret molecular emission on galactic scales. The method provides a practical way to investigate how feedback from an active nucleus reshapes the surrounding interstellar medium. Future extensions that include higher resolution observation, shock tracers, and time-dependent emulator will help to better disentangle heating, turbulence, and chemical effects in NGC~1068 and other nearby active galaxies.

\begin{acknowledgements}
We thank the anonymous referee for the constructive comments that improved the manuscript. BJ acknowledges support from the CSC (China Scholarship Council) scholarship program. The research of SV is funded by the European Research Council (ERC) Advanced Grant MOPPEX 833460.vii. BJ acknowledges assistance from Allegro, the European ALMA Regional Center node in the Netherlands.
\end{acknowledgements}

\bibliographystyle{aa}
\bibliography{main}

@ARTICLE{Berhens2024,
       author = {{Behrens}, Erica and {Mangum}, Jeffrey G. and {Viti}, Serena and {Holdship}, Jonathan and {Huang}, Ko-Yun and {Bouvier}, Mathilde and {Butterworth}, Joshua and {Eibensteiner}, Cosima and {Harada}, Nanase and {Mart{\'\i}n}, Sergio and {Sakamoto}, Kazushi and {Muller}, Sebastien and {Tanaka}, Kunihiko and {Colzi}, Laura and {Henkel}, Christian and {Meier}, David S. and {Rivilla}, V{\'\i}ctor M. and {van der Werf}, Paul P.},
        title = "{Neural Network Constraints on the Cosmic-Ray Ionization Rate and Other Physical Conditions in NGC 253 with ALCHEMI Measurements of HCN and HNC}",
      journal = {\apj},
         year = 2024,
        month = dec,
       volume = {977},
       number = {1},
          eid = {38},
        pages = {38},
          doi = {10.3847/1538-4357/ad85db},
archivePrefix = {arXiv},
       eprint = {2409.13821},
 primaryClass = {astro-ph.GA},
       adsurl = {https://ui.adsabs.harvard.edu/abs/2024ApJ...977...38B}
}

@ARTICLE{e,
       author = {{Endres}, Christian P. and {Schlemmer}, Stephan and {Schilke}, Peter and {Stutzki}, J{\"u}rgen and {M{\"u}ller}, Holger S.~P.},
        title = "{The Cologne Database for Molecular Spectroscopy, CDMS, in the Virtual Atomic and Molecular Data Centre, VAMDC}",
      journal = {Journal of Molecular Spectroscopy},
         year = 2016,
        month = sep,
       volume = {327},
        pages = {95-104},
          doi = {10.1016/j.jms.2016.03.005},
       adsurl = {https://ui.adsabs.harvard.edu/abs/2016JMoSp.327...95E}
}

@ARTICLE{Garcia2019,
       author = {{Garc{\'\i}a-Burillo}, S. and {Combes}, F. and {Ramos Almeida}, C. and {Usero}, A. and {Alonso-Herrero}, A. and {Hunt}, L.~K. and {Rouan}, D. and {Aalto}, S. and {Querejeta}, M. and {Viti}, S. and {van der Werf}, P.~P. and {Vives-Arias}, H. and {Fuente}, A. and {Colina}, L. and {Mart{\'\i}n-Pintado}, J. and {Henkel}, C. and {Mart{\'\i}n}, S. and {Krips}, M. and {Gratadour}, D. and {Neri}, R. and {Tacconi}, L.~J.},
        title = "{ALMA images the many faces of the <ASTROBJ>NGC 1068</ASTROBJ> torus and its surroundings}",
      journal = {\aap},
         year = 2019,
        month = dec,
       volume = {632},
          eid = {A61},
        pages = {A61},
          doi = {10.1051/0004-6361/201936606},
archivePrefix = {arXiv},
       eprint = {1909.00675},
 primaryClass = {astro-ph.GA},
       adsurl = {https://ui.adsabs.harvard.edu/abs/2019A&A...632A..61G}
}

@ARTICLE{Skru,
       author = {{Skrutskie}, M.~F. and {Cutri}, R.~M. and {Stiening}, R. and {Weinberg}, M.~D. and {Schneider}, S. and {Carpenter}, J.~M. and {Beichman}, C. and {Capps}, R. and {Chester}, T. and {Elias}, J. and {Huchra}, J. and {Liebert}, J. and {Lonsdale}, C. and {Monet}, D.~G. and {Price}, S. and {Seitzer}, P. and {Jarrett}, T. and {Kirkpatrick}, J.~D. and {Gizis}, J.~E. and {Howard}, E. and {Evans}, T. and {Fowler}, J. and {Fullmer}, L. and {Hurt}, R. and {Light}, R. and {Kopan}, E.~L. and {Marsh}, K.~A. and {McCallon}, H.~L. and {Tam}, R. and {Van Dyk}, S. and {Wheelock}, S.},
        title = "{The Two Micron All Sky Survey (2MASS)}",
      journal = {\aj},
         year = 2006,
        month = feb,
       volume = {131},
       number = {2},
        pages = {1163-1183},
          doi = {10.1086/498708},
       adsurl = {https://ui.adsabs.harvard.edu/abs/2006AJ....131.1163S}
}

@ARTICLE{CASA,
       author = {{CASA Team} and {Bean}, Ben and {Bhatnagar}, Sanjay and {Castro}, Sandra and {Donovan Meyer}, Jennifer and {Emonts}, Bjorn and {Garcia}, Enrique and {Garwood}, Robert and {Golap}, Kumar and {Gonzalez Villalba}, Justo and {Harris}, Pamela and {Hayashi}, Yohei and {Hoskins}, Josh and {Hsieh}, Mingyu and {Jagannathan}, Preshanth and {Kawasaki}, Wataru and {Keimpema}, Aard and {Kettenis}, Mark and {Lopez}, Jorge and {Marvil}, Joshua and {Masters}, Joseph and {McNichols}, Andrew and {Mehringer}, David and {Miel}, Renaud and {Moellenbrock}, George and {Montesino}, Federico and {Nakazato}, Takeshi and {Ott}, Juergen and {Petry}, Dirk and {Pokorny}, Martin and {Raba}, Ryan and {Rau}, Urvashi and {Schiebel}, Darrell and {Schweighart}, Neal and {Sekhar}, Srikrishna and {Shimada}, Kazuhiko and {Small}, Des and {Steeb}, Jan-Willem and {Sugimoto}, Kanako and {Suoranta}, Ville and {Tsutsumi}, Takahiro and {van Bemmel}, Ilse M. and {Verkouter}, Marjolein and {Wells}, Akeem and {Xiong}, Wei and {Szomoru}, Arpad and {Griffith}, Morgan and {Glendenning}, Brian and {Kern}, Jeff},
        title = "{CASA, the Common Astronomy Software Applications for Radio Astronomy}",
      journal = {\pasp},
         year = 2022,
        month = nov,
       volume = {134},
       number = {1041},
          eid = {114501},
        pages = {114501},
          doi = {10.1088/1538-3873/ac9642},
archivePrefix = {arXiv},
       eprint = {2210.02276},
 primaryClass = {astro-ph.IM},
       adsurl = {https://ui.adsabs.harvard.edu/abs/2022PASP..134k4501C}
}

@ARTICLE{Heyl2023,
       author = {{Heyl}, Johannes and {Butterworth}, Joshua and {Viti}, Serena},
        title = "{Understanding molecular abundances in star-forming regions using interpretable machine learning}",
      journal = {\mnras},
         year = 2023,
        month = nov,
       volume = {526},
       number = {1},
        pages = {404-422},
          doi = {10.1093/mnras/stad2814},
archivePrefix = {arXiv},
       eprint = {2309.06784},
 primaryClass = {astro-ph.GA},
       adsurl = {https://ui.adsabs.harvard.edu/abs/2023MNRAS.526..404H}
}

@ARTICLE{viti2014,
       author = {{Viti}, S. and {Garc{\'\i}a-Burillo}, S. and {Fuente}, A. and {Hunt}, L.~K. and {Usero}, A. and {Henkel}, C. and {Eckart}, A. and {Martin}, S. and {Spaans}, M. and {Muller}, S. and {Combes}, F. and {Krips}, M. and {Schinnerer}, E. and {Casasola}, V. and {Costagliola}, F. and {Marquez}, I. and {Planesas}, P. and {van der Werf}, P.~P. and {Aalto}, S. and {Baker}, A.~J. and {Boone}, F. and {Tacconi}, L.~J.},
        title = "{Molecular line emission in NGC 1068 imaged with ALMA. II. The chemistry of the dense molecular gas}",
      journal = {\aap},
         year = 2014,
        month = oct,
       volume = {570},
          eid = {A28},
        pages = {A28},
          doi = {10.1051/0004-6361/201424116},
archivePrefix = {arXiv},
       eprint = {1407.4940},
 primaryClass = {astro-ph.GA},
       adsurl = {https://ui.adsabs.harvard.edu/abs/2014A&A...570A..28V}
}

@ARTICLE{Josh2022,
       author = {{Butterworth}, J. and {Holdship}, J. and {Viti}, S. and {Garc{\'\i}a-Burillo}, S.},
        title = "{Understanding if molecular ratios can be used as diagnostics of AGN and starburst activity: The case of NGC 1068}",
      journal = {\aap},
         year = 2022,
        month = nov,
       volume = {667},
          eid = {A131},
        pages = {A131},
          doi = {10.1051/0004-6361/202244563},
archivePrefix = {arXiv},
       eprint = {2209.05928},
 primaryClass = {astro-ph.GA},
       adsurl = {https://ui.adsabs.harvard.edu/abs/2022A&A...667A.131B}
}

@ARTICLE{Scourfield2020,
       author = {{Scourfield}, M. and {Viti}, S. and {Garc{\'\i}a-Burillo}, S. and {Saintonge}, A. and {Combes}, F. and {Fuente}, A. and {Henkel}, C. and {Alonso-Herrero}, A. and {Harada}, N. and {Takano}, S. and {Nakajima}, T. and {Mart{\'\i}n}, S. and {Krips}, M. and {van der Werf}, P.~P. and {Aalto}, S. and {Usero}, A. and {Kohno}, K.},
        title = "{ALMA observations of CS in NGC 1068: chemistry and excitation}",
      journal = {\mnras},
         year = 2020,
        month = aug,
       volume = {496},
       number = {4},
        pages = {5308-5329},
          doi = {10.1093/mnras/staa1891},
archivePrefix = {arXiv},
       eprint = {2006.15041},
 primaryClass = {astro-ph.GA},
       adsurl = {https://ui.adsabs.harvard.edu/abs/2020MNRAS.496.5308S}
}

@ARTICLE{huang2022,
       author = {{Huang}, K. -Y. and {Viti}, S. and {Holdship}, J. and {Garc{\'\i}a-Burillo}, S. and {Kohno}, K. and {Taniguchi}, A. and {Martn}, S. and {Aladro}, R. and {Fuente}, A. and {S{\'a}nchez-Garc{\'\i}a}, M.},
        title = "{The chemical footprint of AGN feedback in the outflowing circumnuclear disk of NGC 1068}",
      journal = {\aap},
         year = 2022,
        month = oct,
       volume = {666},
          eid = {A102},
        pages = {A102},
          doi = {10.1051/0004-6361/202142831},
archivePrefix = {arXiv},
       eprint = {2202.05005},
 primaryClass = {astro-ph.GA},
       adsurl = {https://ui.adsabs.harvard.edu/abs/2022A&A...666A.102H}
}

@ARTICLE{tanaka2024,
       author = {{Tanaka}, Kunihiko and {Mangum}, Jeffrey G. and {Viti}, Serena and {Mart{\'\i}n}, Sergio and {Harada}, Nanase and {Sakamoto}, Kazushi and {Muller}, Sebastien and {Yoshimura}, Yuki and {Nakanishi}, Kouichiro and {Herrero-Illana}, Rub{\'e}n and {Emig}, Kimberly L. and {M{\"u}hle}, S. and {Kaneko}, Hiroyuki and {Tosaki}, Tomoka and {Behrens}, Erica and {Rivilla}, V{\'\i}ctor M. and {Colzi}, Laura and {Nishimura}, Yuri and {Humire}, P.~K. and {Bouvier}, Mathilde and {Huang}, Ko-Yun and {Butterworth}, Joshua and {Meier}, David S. and {van der Werf}, Paul P.},
        title = "{Volume Density Structure of the Central Molecular Zone NGC 253 through ALCHEMI Excitation Analysis}",
      journal = {\apj},
         year = 2024,
        month = jan,
       volume = {961},
       number = {1},
          eid = {18},
        pages = {18},
          doi = {10.3847/1538-4357/ad0e64},
archivePrefix = {arXiv},
       eprint = {2311.12106},
 primaryClass = {astro-ph.GA},
       adsurl = {https://ui.adsabs.harvard.edu/abs/2024ApJ...961...18T}
}

@ARTICLE{Garcia2014,
       author = {{Garc{\'\i}a-Burillo}, S. and {Combes}, F. and {Usero}, A. and {Aalto}, S. and {Krips}, M. and {Viti}, S. and {Alonso-Herrero}, A. and {Hunt}, L.~K. and {Schinnerer}, E. and {Baker}, A.~J. and {Boone}, F. and {Casasola}, V. and {Colina}, L. and {Costagliola}, F. and {Eckart}, A. and {Fuente}, A. and {Henkel}, C. and {Labiano}, A. and {Mart{\'\i}n}, S. and {M{\'a}rquez}, I. and {Muller}, S. and {Planesas}, P. and {Ramos Almeida}, C. and {Spaans}, M. and {Tacconi}, L.~J. and {van der Werf}, P.~P.},
        title = "{Molecular line emission in NGC 1068 imaged with ALMA. I. An AGN-driven outflow in the dense molecular gas}",
      journal = {\aap},
         year = 2014,
        month = jul,
       volume = {567},
          eid = {A125},
        pages = {A125},
          doi = {10.1051/0004-6361/201423843},
archivePrefix = {arXiv},
       eprint = {1405.7706},
 primaryClass = {astro-ph.GA},
       adsurl = {https://ui.adsabs.harvard.edu/abs/2014A&A...567A.125G}
}

@ARTICLE{Leroy2015,
       author = {{Leroy}, Adam K. and {Bolatto}, Alberto D. and {Ostriker}, Eve C. and {Rosolowsky}, Erik and {Walter}, Fabian and {Warren}, Steven R. and {Donovan Meyer}, Jennifer and {Hodge}, Jacqueline and {Meier}, David S. and {Ott}, J{\"u}rgen and {Sandstrom}, Karin and {Schruba}, Andreas and {Veilleux}, Sylvain and {Zwaan}, Martin},
        title = "{ALMA Reveals the Molecular Medium Fueling the Nearest Nuclear Starburst}",
      journal = {\apj},
         year = 2015,
        month = mar,
       volume = {801},
       number = {1},
          eid = {25},
        pages = {25},
          doi = {10.1088/0004-637X/801/1/25},
archivePrefix = {arXiv},
       eprint = {1411.2836},
 primaryClass = {astro-ph.GA},
       adsurl = {https://ui.adsabs.harvard.edu/abs/2015ApJ...801...25L}
}

@software{erica_behrens_2024_13839853,
	author = {Erica Behrens},
	doi = {10.5281/zenodo.13839853},
	month = sep,
	publisher = {Zenodo},
	title = {ebehrens97/HERA: ApJ Release},
	url = {https://doi.org/10.5281/zenodo.13839853},
	version = {v1.0},
	year = 2024}

@ARTICLE{kelly2017,
       author = {{Kelly}, G. and {Viti}, S. and {Garc{\'\i}a-Burillo}, S. and {Fuente}, A. and {Usero}, A. and {Krips}, M. and {Neri}, R.},
        title = "{Molecular shock tracers in NGC 1068: SiO and HNCO}",
      journal = {\aap},
         year = 2017,
        month = jan,
       volume = {597},
          eid = {A11},
        pages = {A11},
          doi = {10.1051/0004-6361/201628946},
archivePrefix = {arXiv},
       eprint = {1609.02023},
 primaryClass = {astro-ph.GA},
       adsurl = {https://ui.adsabs.harvard.edu/abs/2017A&A...597A..11K}
}

@ARTICLE{radex,
       author = {{van der Tak}, F.~F.~S. and {Black}, J.~H. and {Sch{\"o}ier}, F.~L. and {Jansen}, D.~J. and {van Dishoeck}, E.~F.},
        title = "{A computer program for fast non-LTE analysis of interstellar line spectra. With diagnostic plots to interpret observed line intensity ratios}",
      journal = {\aap},
         year = 2007,
        month = jun,
       volume = {468},
       number = {2},
        pages = {627-635},
          doi = {10.1051/0004-6361:20066820},
archivePrefix = {arXiv},
       eprint = {0704.0155},
 primaryClass = {astro-ph},
       adsurl = {https://ui.adsabs.harvard.edu/abs/2007A&A...468..627V}
}

@ARTICLE{nautilus,
       author = {{Lange}, Johannes U.},
        title = "{NAUTILUS: boosting Bayesian importance nested sampling with deep learning}",
      journal = {\mnras},
         year = 2023,
        month = oct,
       volume = {525},
       number = {2},
        pages = {3181-3194},
          doi = {10.1093/mnras/stad2441},
archivePrefix = {arXiv},
       eprint = {2306.16923},
 primaryClass = {astro-ph.IM},
       adsurl = {https://ui.adsabs.harvard.edu/abs/2023MNRAS.525.3181L}
}

@ARTICLE{Das2007,
       author = {{Das}, V. and {Crenshaw}, D.~M. and {Kraemer}, S.~B.},
        title = "{Dynamics of the Narrow-Line Region in the Seyfert 2 Galaxy NGC 1068}",
      journal = {\apj},
         year = 2007,
        month = feb,
       volume = {656},
       number = {2},
        pages = {699-708},
          doi = {10.1086/510580},
archivePrefix = {arXiv},
       eprint = {astro-ph/0611183},
 primaryClass = {astro-ph},
       adsurl = {https://ui.adsabs.harvard.edu/abs/2007ApJ...656..699D}
}

@ARTICLE{Villas2021,
       author = {{Rico-Villas}, F. and {Mart{\'\i}n-Pintado}, J. and {Gonz{\'a}lez-Alfonso}, E. and {Rivilla}, V.~M. and {Mart{\'\i}n}, S. and {Garc{\'\i}a-Burillo}, S. and {Jim{\'e}nez-Serra}, I. and {S{\'a}nchez-Garc{\'\i}a}, M.},
        title = "{Vibrationally excited HC$_{3}$N emission in NGC 1068: tracing the recent star formation in the starburst ring}",
      journal = {\mnras},
         year = 2021,
        month = apr,
       volume = {502},
       number = {2},
        pages = {3021-3034},
          doi = {10.1093/mnras/stab197},
archivePrefix = {arXiv},
       eprint = {2008.03693},
 primaryClass = {astro-ph.GA},
       adsurl = {https://ui.adsabs.harvard.edu/abs/2021MNRAS.502.3021R}
}

@ARTICLE{bostroem2020,
       author = {{Bostroem}, K.~A. and {Valenti}, S. and {Sand}, D.~J. and {Andrews}, J.~E. and {Van Dyk}, S.~D. and {Galbany}, L. and {Pooley}, D. and {Amaro}, R.~C. and {Smith}, N. and {Yang}, S. and {Anupama}, G.~C. and {Arcavi}, I. and {Baron}, E. and {Brown}, P.~J. and {Burke}, J. and {Cartier}, R. and {Hiramatsu}, D. and {Dastidar}, R. and {DerKacy}, J.~M. and {Dong}, Y. and {Egami}, E. and {Ertel}, S. and {Filippenko}, A.~V. and {Fox}, O.~D. and {Haislip}, J. and {Hosseinzadeh}, G. and {Howell}, D.~A. and {Gangopadhyay}, A. and {Jha}, S.~W. and {Kouprianov}, V. and {Kumar}, B. and {Lundquist}, M. and {Milisavljevic}, D. and {McCully}, C. and {Milne}, P. and {Misra}, K. and {Reichart}, D.~E. and {Sahu}, D.~K. and {Sai}, H. and {Singh}, A. and {Smith}, P.~S. and {Vinko}, J. and {Wang}, X. and {Wang}, Y. and {Wheeler}, J.~C. and {Williams}, G.~G. and {Wyatt}, S. and {Zhang}, J. and {Zhang}, X.},
        title = "{Discovery and Rapid Follow-up Observations of the Unusual Type II SN 2018ivc in NGC 1068}",
      journal = {\apj},
         year = 2020,
        month = may,
       volume = {895},
       number = {1},
          eid = {31},
        pages = {31},
          doi = {10.3847/1538-4357/ab8945},
archivePrefix = {arXiv},
       eprint = {1909.07304},
 primaryClass = {astro-ph.HE},
       adsurl = {https://ui.adsabs.harvard.edu/abs/2020ApJ...895...31B}
}

@ARTICLE{mutie2024,
       author = {{Mutie}, Isaac M. and {Williams-Baldwin}, David and {Beswick}, Robert J. and {Bempong-Manful}, Emmanuel K. and {Baki}, Paul O. and {Muxlow}, Tom W.~B. and {Gallimore}, Jack F. and {Aalto}, Susanne E. and {Dullo}, Bililign T. and {Baldi}, Ranieri D.},
        title = "{Radio jets in NGC 1068 with e-MERLIN and VLA: structure and morphology}",
      journal = {\mnras},
         year = 2024,
        month = feb,
       volume = {527},
       number = {4},
        pages = {11756-11765},
          doi = {10.1093/mnras/stad3864},
archivePrefix = {arXiv},
       eprint = {2312.09722},
 primaryClass = {astro-ph.GA},
       adsurl = {https://ui.adsabs.harvard.edu/abs/2024MNRAS.52711756M}
}

@ARTICLE{Rodrigues2020Active,
       author = {{Rodrigues}, Xavier and {Heinze}, Jonas and {Palladino}, Andrea and {van Vliet}, Arjen and {Winter}, Walter},
        title = "{Active Galactic Nuclei Jets as the Origin of Ultrahigh-Energy Cosmic Rays and Perspectives for the Detection of Astrophysical Source Neutrinos at EeV Energies}",
      journal = {\prl},
         year = 2021,
        month = may,
       volume = {126},
       number = {19},
          eid = {191101},
        pages = {191101},
          doi = {10.1103/PhysRevLett.126.191101},
archivePrefix = {arXiv},
       eprint = {2003.08392},
 primaryClass = {astro-ph.HE},
       adsurl = {https://ui.adsabs.harvard.edu/abs/2021PhRvL.126s1101R}
}

@ARTICLE{Collaboration2020Resolving,
       author = {{H.~E.~S.~S. Collaboration} and {Abdalla}, H. and {Adam}, R. and {Aharonian}, F. and {Ait Benkhali}, F. and {Ang{\"u}ner}, E.~O. and {Arakawa}, M. and {Arcaro}, C. and {Armand}, C. and {Ashkar}, H. and {Backes}, M. and {Barbosa Martins}, V. and {Barnard}, M. and {Becherini}, Y. and {Berge}, D. and {Bernl{\"o}hr}, K. and {Blackwell}, R. and {B{\"o}ttcher}, M. and {Boisson}, C. and {Bolmont}, J. and {Bonnefoy}, S. and {Bregeon}, J. and {Breuhaus}, M. and {Brun}, F. and {Brun}, P. and {Bryan}, M. and {B{\"u}chele}, M. and {Bulik}, T. and {Bylund}, T. and {Capasso}, M. and {Caroff}, S. and {Carosi}, A. and {Casanova}, S. and {Cerruti}, M. and {Chand}, T. and {Chandra}, S. and {Chen}, A. and {Colafrancesco}, S. and {Cury{\l}o}, M. and {Davids}, I.~D. and {Deil}, C. and {Devin}, J. and {deWilt}, P. and {Dirson}, L. and {Djannati-Ata{\"\i}}, A. and {Dmytriiev}, A. and {Donath}, A. and {Doroshenko}, V. and {Drury}, L. O'C. and {Dyks}, J. and {Egberts}, K. and {Emery}, G. and {Ernenwein}, J.-P. and {Eschbach}, S. and {Feijen}, K. and {Fegan}, S. and {Fiasson}, A. and {Fontaine}, G. and {Funk}, S. and {F{\"u}{\ss}ling}, M. and {Gabici}, S. and {Gallant}, Y.~A. and {Gat{\'e}}, F. and {Giavitto}, G. and {Glawion}, D. and {Glicenstein}, J.~F. and {Gottschall}, D. and {Grondin}, M.-H. and {Hahn}, J. and {Haupt}, M. and {Heinzelmann}, G. and {Henri}, G. and {Hermann}, G. and {Hinton}, J.~A. and {Hofmann}, W. and {Hoischen}, C. and {Holch}, T.~L. and {Holler}, M. and {Horns}, D. and {Huber}, D. and {Iwasaki}, H. and {Jamrozy}, M. and {Jankowsky}, D. and {Jankowsky}, F. and {Jardin-Blicq}, A. and {Jung-Richardt}, I. and {Kastendieck}, M.~A. and {Katarzy{\'n}ski}, K. and {Katsuragawa}, M. and {Katz}, U. and {Khangulyan}, D. and {Kh{\'e}lifi}, B. and {King}, J. and {Klepser}, S. and {Klu{\'z}niak}, W. and {Komin}, N. and {Kosack}, K. and {Kostunin}, D. and {Kraus}, M. and {Lamanna}, G. and {Lau}, J. and {Lemi{\`e}re}, A. and {Lemoine-Goumard}, M. and {Lenain}, J.-P. and {Leser}, E. and {Levy}, C. and {Lohse}, T. and {Lypova}, I. and {Mackey}, J. and {Majumdar}, J. and {Malyshev}, D. and {Marandon}, V. and {Marcowith}, A. and {Mares}, A. and {Mariaud}, C. and {Mart{\'\i}-Devesa}, G. and {Marx}, R. and {Maurin}, G. and {Meintjes}, P.~J. and {Mitchell}, A.~M.~W. and {Moderski}, R. and {Mohamed}, M. and {Mohrmann}, L. and {Moore}, C. and {Moulin}, E. and {Muller}, J. and {Murach}, T. and {Nakashima}, S. and {de Naurois}, M. and {Ndiyavala}, H. and {Niederwanger}, F. and {Niemiec}, J. and {Oakes}, L. and {O'Brien}, P. and {Odaka}, H. and {Ohm}, S. and {de Ona Wilhelmi}, E. and {Ostrowski}, M. and {Oya}, I. and {Panter}, M. and {Parsons}, R.~D. and {Perennes}, C. and {Petrucci}, P.-O. and {Peyaud}, B. and {Piel}, Q. and {Pita}, S. and {Poireau}, V. and {Priyana Noel}, A. and {Prokhorov}, D.~A. and {Prokoph}, H. and {P{\"u}hlhofer}, G. and {Punch}, M. and {Quirrenbach}, A. and {Raab}, S. and {Rauth}, R. and {Reimer}, A. and {Reimer}, O. and {Remy}, Q. and {Renaud}, M. and {Rieger}, F. and {Rinchiuso}, L. and {Romoli}, C. and {Rowell}, G. and {Rudak}, B. and {Ruiz-Velasco}, E. and {Sahakian}, V. and {Saito}, S. and {Sanchez}, D.~A. and {Santangelo}, A. and {Sasaki}, M. and {Schlickeiser}, R. and {Sch{\"u}ssler}, F. and {Schulz}, A. and {Schutte}, H.~M. and {Schwanke}, U. and {Schwemmer}, S. and {Seglar-Arroyo}, M. and {Senniappan}, M. and {Seyffert}, A.~S. and {Shafi}, N. and {Shiningayamwe}, K. and {Simoni}, R. and {Sinha}, A. and {Sol}, H. and {Specovius}, A. and {Spir-Jacob}, M. and {Stawarz}, {\L}. and {Steenkamp}, R. and {Stegmann}, C. and {Steppa}, C. and {Takahashi}, T. and {Tavernier}, T. and {Taylor}, A.~M. and {Terrier}, R. and {Tiziani}, D. and {Tluczykont}, M. and {Trichard}, C. and {Tsirou}, M. and {Tsuji}, N. and {Tuffs}, R.},
        title = "{Resolving acceleration to very high energies along the jet of Centaurus A}",
      journal = {\nat},
         year = 2020,
        month = jun,
       volume = {582},
       number = {7812},
        pages = {356-359},
          doi = {10.1038/s41586-020-2354-1},
archivePrefix = {arXiv},
       eprint = {2007.04823},
 primaryClass = {astro-ph.HE},
       adsurl = {https://ui.adsabs.harvard.edu/abs/2020Natur.582..356H}
}

@ARTICLE{Holdship2022,
       author = {{Holdship}, Jonathan and {Mangum}, Jeffrey G. and {Viti}, Serena and {Behrens}, Erica and {Harada}, Nanase and {Mart{\'\i}n}, Sergio and {Sakamoto}, Kazushi and {Muller}, Sebastien and {Tanaka}, Kunihiko and {Nakanishi}, Kouichiro and {Herrero-Illana}, Rub{\'e}n and {Yoshimura}, Yuki and {Aladro}, Rebeca and {Colzi}, Laura and {Emig}, Kimberly L. and {Henkel}, Christian and {Nishimura}, Yuri and {Rivilla}, V{\'\i}ctor M. and {van der Werf}, Paul P. and {Alma Comprehensive High-Resolution Extragalactic Molecular Inventory (Alchemi) Collaboration}},
        title = "{Energizing Star Formation: The Cosmic-Ray Ionization Rate in NGC 253 Derived from ALCHEMI Measurements of H$_{3}$O$^{+}$ and SO}",
      journal = {\apj},
         year = 2022,
        month = jun,
       volume = {931},
       number = {2},
          eid = {89},
        pages = {89},
          doi = {10.3847/1538-4357/ac6753},
archivePrefix = {arXiv},
       eprint = {2204.03668},
 primaryClass = {astro-ph.GA},
       adsurl = {https://ui.adsabs.harvard.edu/abs/2022ApJ...931...89H}
}

@article{Li2021Dense,title={Dense gas in local galaxies revealed by multiple tracers},journal = {\mnras},author={Fei Li and Junzhi Wang and F. Gao and Shu Liu and Zhi-Yu Zhang and Shanghuo Li and Y. Gong and Juan Li and Yong Shi},year={2021},doi={10.1093/mnras/stab745}}

@ARTICLE{Garcia2017,
       author = {{Garc{\'\i}a-Burillo}, S. and {Viti}, S. and {Combes}, F. and {Fuente}, A. and {Usero}, A. and {Hunt}, L.~K. and {Mart{\'\i}n}, S. and {Krips}, M. and {Aalto}, S. and {Aladro}, R. and {Ramos Almeida}, C. and {Alonso-Herrero}, A. and {Casasola}, V. and {Henkel}, C. and {Querejeta}, M. and {Neri}, R. and {Costagliola}, F. and {Tacconi}, L.~J. and {van der Werf}, P.~P.},
        title = "{ALMA imaging of C$_{2}$H emission in the disk of <ASTROBJ>NGC 1068</ASTROBJ>}",
      journal = {\aap},
         year = 2017,
        month = dec,
       volume = {608},
          eid = {A56},
        pages = {A56},
          doi = {10.1051/0004-6361/201731862},
archivePrefix = {arXiv},
       eprint = {1709.05895},
 primaryClass = {astro-ph.GA},
       adsurl = {https://ui.adsabs.harvard.edu/abs/2017A&A...608A..56G}
}

@ARTICLE{huang2023NGC253,
       author = {{Huang}, K. -Y. and {Viti}, S. and {Holdship}, J. and {Mangum}, J.~G. and {Mart{\'\i}n}, S. and {Harada}, N. and {Muller}, S. and {Sakamoto}, K. and {Tanaka}, K. and {Yoshimura}, Y. and {Herrero-Illana}, R. and {Meier}, D.~S. and {Behrens}, E. and {van der Werf}, P.~P. and {Henkel}, C. and {Garc{\'\i}a-Burillo}, S. and {Rivilla}, V.~M. and {Emig}, K.~L. and {Colzi}, L. and {Humire}, P.~K. and {Aladro}, R. and {Bouvier}, M.},
        title = "{Reconstructing the shock history in the CMZ of NGC 253 with ALCHEMI}",
      journal = {\aap},
         year = 2023,
        month = jul,
       volume = {675},
          eid = {A151},
        pages = {A151},
          doi = {10.1051/0004-6361/202245659},
archivePrefix = {arXiv},
       eprint = {2303.12685},
 primaryClass = {astro-ph.GA},
       adsurl = {https://ui.adsabs.harvard.edu/abs/2023A&A...675A.151H}
}

@ARTICLE{mijolla2019,
       author = {{de Mijolla}, D. and {Viti}, S. and {Holdship}, J. and {Manolopoulou}, I. and {Yates}, J.},
        title = "{Incorporating astrochemistry into molecular line modelling via emulation}",
      journal = {\aap},
         year = 2019,
        month = oct,
       volume = {630},
          eid = {A117},
        pages = {A117},
          doi = {10.1051/0004-6361/201935973},
archivePrefix = {arXiv},
       eprint = {1907.07472},
 primaryClass = {astro-ph.GA},
       adsurl = {https://ui.adsabs.harvard.edu/abs/2019A&A...630A.117D}
}

@ARTICLE{vollmer2022,
       author = {{Vollmer}, B. and {Davies}, R.~I. and {Gratier}, P. and {Liz{\'e}e}, Th. and {Imanishi}, M. and {Gallimore}, J.~F. and {Impellizzeri}, C.~M.~V. and {Garc{\'\i}a-Burillo}, S. and {Le Petit}, F.},
        title = "{From the Circumnuclear Disk in the Galactic Center to thick, obscuring tori of AGNs. Modeling the molecular emission of a parsec-scale torus as found in NGC 1068}",
      journal = {\aap},
         year = 2022,
        month = sep,
       volume = {665},
          eid = {A102},
        pages = {A102},
          doi = {10.1051/0004-6361/202141684},
archivePrefix = {arXiv},
       eprint = {2206.14513},
 primaryClass = {astro-ph.GA},
       adsurl = {https://ui.adsabs.harvard.edu/abs/2022A&A...665A.102V}
}

@ARTICLE{padovani2018,
       author = {{Padovani}, Marco and {Ivlev}, Alexei V. and {Galli}, Daniele and {Caselli}, Paola},
        title = "{Cosmic-ray ionisation in circumstellar discs}",
      journal = {\aap},
         year = 2018,
        month = jun,
       volume = {614},
          eid = {A111},
        pages = {A111},
          doi = {10.1051/0004-6361/201732202},
archivePrefix = {arXiv},
       eprint = {1803.09348},
 primaryClass = {astro-ph.HE},
       adsurl = {https://ui.adsabs.harvard.edu/abs/2018A&A...614A.111P}
}

@ARTICLE{Koudamani2022Two,
       author = {{Koudmani}, Sophie and {Sijacki}, Debora and {Smith}, Matthew C.},
        title = "{Two can play at that game: constraining the role of supernova and AGN feedback in dwarf galaxies with cosmological zoom-in simulations}",
      journal = {\mnras},
         year = 2022,
        month = oct,
       volume = {516},
       number = {2},
        pages = {2112-2141},
          doi = {10.1093/mnras/stac2252},
archivePrefix = {arXiv},
       eprint = {2206.11274},
 primaryClass = {astro-ph.GA},
       adsurl = {https://ui.adsabs.harvard.edu/abs/2022MNRAS.516.2112K}
}

@ARTICLE{Koudamani2019Fast,
       author = {{Koudmani}, Sophie and {Sijacki}, Debora and {Bourne}, Martin A. and {Smith}, Matthew C.},
        title = "{Fast and energetic AGN-driven outflows in simulated dwarf galaxies}",
      journal = {\mnras},
         year = 2019,
        month = apr,
       volume = {484},
       number = {2},
        pages = {2047-2066},
          doi = {10.1093/mnras/stz097},
archivePrefix = {arXiv},
       eprint = {1812.04629},
 primaryClass = {astro-ph.GA},
       adsurl = {https://ui.adsabs.harvard.edu/abs/2019MNRAS.484.2047K}
}

@ARTICLE{Leung2019The,
       author = {{Leung}, Gene C.~K. and {Coil}, Alison L. and {Aird}, James and {Azadi}, Mojegan and {Kriek}, Mariska and {Mobasher}, Bahram and {Reddy}, Naveen and {Shapley}, Alice and {Siana}, Brian and {Fetherolf}, Tara and {Fornasini}, Francesca M. and {Freeman}, William R. and {Price}, Sedona H. and {Sanders}, Ryan L. and {Shivaei}, Irene and {Zick}, Tom},
        title = "{The MOSDEF Survey: A Census of AGN-driven Ionized Outflows at z = 1.4-3.8}",
      journal = {\apj},
         year = 2019,
        month = nov,
       volume = {886},
       number = {1},
          eid = {11},
        pages = {11},
          doi = {10.3847/1538-4357/ab4a7c},
archivePrefix = {arXiv},
       eprint = {1905.13338},
 primaryClass = {astro-ph.GA},
       adsurl = {https://ui.adsabs.harvard.edu/abs/2019ApJ...886...11L}
}

@ARTICLE{Murthy2022Cold,
       author = {{Murthy}, Suma and {Morganti}, Raffaella and {Wagner}, Alexander Y. and {Oosterloo}, Tom and {Guillard}, Pierre and {Mukherjee}, Dipanjan and {Bicknell}, Geoffrey},
        title = "{Cold gas removal from the centre of a galaxy by a low-luminosity jet}",
      journal = {Nature Astronomy},
         year = 2022,
        month = feb,
       volume = {6},
        pages = {488-495},
          doi = {10.1038/s41550-021-01596-6},
archivePrefix = {arXiv},
       eprint = {2202.05222},
 primaryClass = {astro-ph.GA},
       adsurl = {https://ui.adsabs.harvard.edu/abs/2022NatAs...6..488M}
}

@ARTICLE{Saito2022,
       author = {{Saito}, Toshiki and {Takano}, Shuro and {Harada}, Nanase and {Nakajima}, Taku and {Schinnerer}, Eva and {Liu}, Daizhong and {Taniguchi}, Akio and {Izumi}, Takuma and {Watanabe}, Yumi and {Bamba}, Kazuharu and {Herbst}, Eric and {Kohno}, Kotaro and {Nishimura}, Yuri and {Stuber}, Sophia and {Tamura}, Yoichi and {Tosaki}, Tomoka},
        title = "{The Kiloparsec-scale Neutral Atomic Carbon Outflow in the Nearby Type 2 Seyfert Galaxy NGC 1068: Evidence for Negative AGN Feedback}",
      journal = {\apjl},
         year = 2022,
        month = mar,
       volume = {927},
       number = {2},
          eid = {L32},
        pages = {L32},
          doi = {10.3847/2041-8213/ac59ae},
archivePrefix = {arXiv},
       eprint = {2203.01355},
 primaryClass = {astro-ph.GA},
       adsurl = {https://ui.adsabs.harvard.edu/abs/2022ApJ...927L..32S}
}

@ARTICLE{Garcia2002,
       author = {{Garc{\'\i}a-Burillo}, S. and {Mart{\'\i}n-Pintado}, J. and {Fuente}, A. and {Usero}, A. and {Neri}, R.},
        title = "{Widespread HCO Emission in the Nuclear Starburst of M82}",
      journal = {\apjl},
         year = 2002,
        month = aug,
       volume = {575},
       number = {2},
        pages = {L55-L58},
          doi = {10.1086/342743},
archivePrefix = {arXiv},
       eprint = {astro-ph/0207313},
 primaryClass = {astro-ph},
       adsurl = {https://ui.adsabs.harvard.edu/abs/2002ApJ...575L..55G}
}

@ARTICLE{Martin2009,
       author = {{Mart{\'\i}n}, Sergio and {Mart{\'\i}n-Pintado}, J. and {Viti}, S.},
        title = "{Photodissociation Chemistry Footprints in the Starburst Galaxy NGC 253}",
      journal = {\apj},
         year = 2009,
        month = dec,
       volume = {706},
       number = {2},
        pages = {1323-1330},
          doi = {10.1088/0004-637X/706/2/1323},
archivePrefix = {arXiv},
       eprint = {0911.2673},
 primaryClass = {astro-ph.CO},
       adsurl = {https://ui.adsabs.harvard.edu/abs/2009ApJ...706.1323M}
}

@ARTICLE{Gerin2009,
       author = {{Gerin}, M. and {Goicoechea}, J.~R. and {Pety}, J. and {Hily-Blant}, P.},
        title = "{HCO mapping of the Horsehead: tracing the illuminated dense molecular cloud surfaces}",
      journal = {\aap},
         year = 2009,
        month = feb,
       volume = {494},
       number = {3},
        pages = {977-985},
          doi = {10.1051/0004-6361:200810933},
archivePrefix = {arXiv},
       eprint = {0811.1470},
 primaryClass = {astro-ph},
       adsurl = {https://ui.adsabs.harvard.edu/abs/2009A&A...494..977G}
}

@ARTICLE{Gao2004,
       author = {{Gao}, Yu and {Solomon}, Philip M.},
        title = "{The Star Formation Rate and Dense Molecular Gas in Galaxies}",
      journal = {\apj},
         year = 2004,
        month = may,
       volume = {606},
       number = {1},
        pages = {271-290},
          doi = {10.1086/382999},
archivePrefix = {arXiv},
       eprint = {astro-ph/0310339},
 primaryClass = {astro-ph},
       adsurl = {https://ui.adsabs.harvard.edu/abs/2004ApJ...606..271G}
}

@ARTICLE{Aladro2011,
       author = {{Aladro}, R. and {Mart{\'\i}n-Pintado}, J. and {Mart{\'\i}n}, S. and {Mauersberger}, R. and {Bayet}, E.},
        title = "{CS, HC$_{3}$N, and CH$_{3}$CCH multi-line analyses toward starburst galaxies. The evolution of cloud structures in the central regions of galaxies}",
      journal = {\aap},
         year = 2011,
        month = jan,
       volume = {525},
          eid = {A89},
        pages = {A89},
          doi = {10.1051/0004-6361/201014090},
archivePrefix = {arXiv},
       eprint = {1009.1831},
 primaryClass = {astro-ph.CO},
       adsurl = {https://ui.adsabs.harvard.edu/abs/2011A&A...525A..89A}
}

@ARTICLE{Holdship2021,
       author = {{Holdship}, J. and {Viti}, S. and {Mart{\'\i}n}, S. and {Harada}, N. and {Mangum}, J. and {Sakamoto}, K. and {Muller}, S. and {Tanaka}, K. and {Yoshimura}, Y. and {Nakanishi}, K. and {Herrero-Illana}, R. and {M{\"u}hle}, S. and {Aladro}, R. and {Colzi}, L. and {Emig}, K.~L. and {Garc{\'\i}a-Burillo}, S. and {Henkel}, C. and {Humire}, P. and {Meier}, D.~S. and {Rivilla}, V.~M. and {van der Werf}, P.},
        title = "{The distribution and origin of C$_{2}$H in NGC 253 from ALCHEMI}",
      journal = {\aap},
         year = 2021,
        month = oct,
       volume = {654},
          eid = {A55},
        pages = {A55},
          doi = {10.1051/0004-6361/202141233},
archivePrefix = {arXiv},
       eprint = {2107.04580},
 primaryClass = {astro-ph.GA},
       adsurl = {https://ui.adsabs.harvard.edu/abs/2021A&A...654A..55H}
}

@ARTICLE{Bland1997,
       author = {{Bland-Hawthorn}, J. and {Gallimore}, J.~F. and {Tacconi}, L.~J. and {Brinks}, E. and {Baum}, S.~A. and {Antonucci}, R.~R.~J. and {Cecil}, G.~N.},
        title = "{The Ringberg Standards for NGC 1068}",
      journal = {\apss},
         year = 1997,
        month = feb,
       volume = {248},
       number = {1-2},
        pages = {9-19},
          doi = {10.1023/A:1000567831370},
       adsurl = {https://ui.adsabs.harvard.edu/abs/1997Ap&SS.248....9B}
}

@ARTICLE{schinnerer2000,
       author = {{Schinnerer}, E. and {Eckart}, A. and {Tacconi}, L.~J. and {Genzel}, R. and {Downes}, D.},
        title = "{Bars and Warps Traced by the Molecular Gas in the Seyfert 2 Galaxy NGC 1068}",
      journal = {\apj},
         year = 2000,
        month = apr,
       volume = {533},
       number = {2},
        pages = {850-868},
          doi = {10.1086/308702},
archivePrefix = {arXiv},
       eprint = {astro-ph/9911488},
 primaryClass = {astro-ph},
       adsurl = {https://ui.adsabs.harvard.edu/abs/2000ApJ...533..850S}
}

@ARTICLE{Sanchez2022,
       author = {{S{\'a}nchez-Garc{\'\i}a}, M. and {Garc{\'\i}a-Burillo}, S. and {Pereira-Santaella}, M. and {Colina}, L. and {Usero}, A. and {Querejeta}, M. and {Alonso-Herrero}, A. and {Fuente}, A.},
        title = "{Spatially resolved star-formation relations of dense molecular gas in NGC 1068}",
      journal = {\aap},
         year = 2022,
        month = apr,
       volume = {660},
          eid = {A83},
        pages = {A83},
          doi = {10.1051/0004-6361/202142396},
archivePrefix = {arXiv},
       eprint = {2201.06552},
 primaryClass = {astro-ph.GA},
       adsurl = {https://ui.adsabs.harvard.edu/abs/2022A&A...660A..83S}
}

@ARTICLE{Scoville1988,
       author = {{Scoville}, N.~Z. and {Matthews}, K. and {Carico}, D.~P. and {Sanders}, D.~B.},
        title = "{The Stellar Bar in NGC 1068}",
      journal = {\apjl},
         year = 1988,
        month = apr,
       volume = {327},
        pages = {L61},
          doi = {10.1086/185140},
       adsurl = {https://ui.adsabs.harvard.edu/abs/1988ApJ...327L..61S}
}

@ARTICLE{Zhang2025,
       author = {{Zhang}, Y. and {Viti}, S. and {Garc{\'\i}a-Burillo}, S. and {Huang}, K. -Y.},
        title = "{ALMA uncovers optically thin and multi-component CO gas in the outflowing circumnuclear disk of NGC 1068}",
      journal = {\aap},
         year = 2025,
        month = jun,
       volume = {698},
          eid = {A17},
        pages = {A17},
          doi = {10.1051/0004-6361/202553704},
archivePrefix = {arXiv},
       eprint = {2504.02758},
 primaryClass = {astro-ph.GA},
       adsurl = {https://ui.adsabs.harvard.edu/abs/2025A&A...698A..17Z}
}

@ARTICLE{Fabian2012,
       author = {{Fabian}, A.~C.},
        title = "{Observational Evidence of Active Galactic Nuclei Feedback}",
      journal = {\araa},
         year = 2012,
        month = sep,
       volume = {50},
        pages = {455-489},
          doi = {10.1146/annurev-astro-081811-125521},
archivePrefix = {arXiv},
       eprint = {1204.4114},
 primaryClass = {astro-ph.CO},
       adsurl = {https://ui.adsabs.harvard.edu/abs/2012ARA&A..50..455F}
}

@ARTICLE{Harrison2017,
       author = {{Harrison}, C.~M.},
        title = "{Impact of supermassive black hole growth on star formation}",
      journal = {Nature Astronomy},
         year = 2017,
        month = jul,
       volume = {1},
          eid = {0165},
        pages = {0165},
          doi = {10.1038/s41550-017-0165},
archivePrefix = {arXiv},
       eprint = {1703.06889},
 primaryClass = {astro-ph.GA},
       adsurl = {https://ui.adsabs.harvard.edu/abs/2017NatAs...1E.165H}
}

@ARTICLE{Zubovas2013,
       author = {{Zubovas}, Kastytis and {Nayakshin}, Sergei and {King}, Andrew and {Wilkinson}, Mark},
        title = "{AGN outflows trigger starbursts in gas-rich galaxies}",
      journal = {\mnras},
         year = 2013,
        month = aug,
       volume = {433},
       number = {4},
        pages = {3079-3090},
          doi = {10.1093/mnras/stt952},
archivePrefix = {arXiv},
       eprint = {1306.0684},
 primaryClass = {astro-ph.GA},
       adsurl = {https://ui.adsabs.harvard.edu/abs/2013MNRAS.433.3079Z}
}

@ARTICLE{Silk2013,
       author = {{Silk}, Joseph},
        title = "{Unleashing Positive Feedback: Linking the Rates of Star Formation, Supermassive Black Hole Accretion, and Outflows in Distant Galaxies}",
      journal = {\apj},
         year = 2013,
        month = aug,
       volume = {772},
       number = {2},
          eid = {112},
        pages = {112},
          doi = {10.1088/0004-637X/772/2/112},
archivePrefix = {arXiv},
       eprint = {1305.5840},
 primaryClass = {astro-ph.CO},
       adsurl = {https://ui.adsabs.harvard.edu/abs/2013ApJ...772..112S}
}

@ARTICLE{holdship2017,
       author = {{Holdship}, J. and {Viti}, S. and {Jim{\'e}nez-Serra}, I. and {Makrymallis}, A. and {Priestley}, F.},
        title = "{UCLCHEM: A Gas-grain Chemical Code for Clouds, Cores, and C-Shocks}",
      journal = {\aj},
         year = 2017,
        month = jul,
       volume = {154},
       number = {1},
          eid = {38},
        pages = {38},
          doi = {10.3847/1538-3881/aa773f},
archivePrefix = {arXiv},
       eprint = {1705.10677},
 primaryClass = {astro-ph.GA},
       adsurl = {https://ui.adsabs.harvard.edu/abs/2017AJ....154...38H}
}

@ARTICLE{lauvikas2024,
       author = {{Lau{\v{z}}ikas}, M. and {Zubovas}, K.},
        title = "{Slow and steady does the trick: Slow outflows enhance the fragmentation of molecular clouds}",
      journal = {\aap},
         year = 2024,
        month = oct,
       volume = {690},
          eid = {A396},
        pages = {A396},
          doi = {10.1051/0004-6361/202450286},
archivePrefix = {arXiv},
       eprint = {2409.13234},
 primaryClass = {astro-ph.GA},
       adsurl = {https://ui.adsabs.harvard.edu/abs/2024A&A...690A.396L}
}

@ARTICLE{quillen2005,
       author = {{Quillen}, Alice C. and {Thorndike}, Stephen L. and {Cunningham}, Andy and {Frank}, Adam and {Gutermuth}, Robert A. and {Blackman}, Eric G. and {Pipher}, Judith L. and {Ridge}, Naomi},
        title = "{Turbulence Driven by Outflow-blown Cavities in the Molecular Cloud of NGC 1333}",
      journal = {\apj},
         year = 2005,
        month = oct,
       volume = {632},
       number = {2},
        pages = {941-955},
          doi = {10.1086/444410},
archivePrefix = {arXiv},
       eprint = {astro-ph/0503167},
 primaryClass = {astro-ph},
       adsurl = {https://ui.adsabs.harvard.edu/abs/2005ApJ...632..941Q}
}

@ARTICLE{Arce2010,
       author = {{Arce}, H{\'e}ctor G. and {Borkin}, Michelle A. and {Goodman}, Alyssa A. and {Pineda}, Jaime E. and {Halle}, Michael W.},
        title = "{The COMPLETE Survey of Outflows in Perseus}",
      journal = {\apj},
         year = 2010,
        month = jun,
       volume = {715},
       number = {2},
        pages = {1170-1190},
          doi = {10.1088/0004-637X/715/2/1170},
archivePrefix = {arXiv},
       eprint = {1005.1714},
 primaryClass = {astro-ph.SR},
       adsurl = {https://ui.adsabs.harvard.edu/abs/2010ApJ...715.1170A}
}

@ARTICLE{pabst2019,
       author = {{Pabst}, C. and {Higgins}, R. and {Goicoechea}, J.~R. and {Teyssier}, D. and {Berne}, O. and {Chambers}, E. and {Wolfire}, M. and {Suri}, S.~T. and {Guesten}, R. and {Stutzki}, J. and {Graf}, U.~U. and {Risacher}, C. and {Tielens}, A.~G.~G.~M.},
        title = "{Disruption of the Orion molecular core 1 by wind from the massive star {\ensuremath{\theta}}$^{1}$ Orionis C}",
      journal = {\nat},
         year = 2019,
        month = jan,
       volume = {565},
       number = {7741},
        pages = {618-621},
          doi = {10.1038/s41586-018-0844-1},
archivePrefix = {arXiv},
       eprint = {1901.04221},
 primaryClass = {astro-ph.GA},
       adsurl = {https://ui.adsabs.harvard.edu/abs/2019Natur.565..618P}
}

@ARTICLE{Bonne2022,
       author = {{Bonne}, L. and {Schneider}, N. and {Garc{\'\i}a}, P. and {Bij}, A. and {Broos}, P. and {Fissel}, L. and {Guesten}, R. and {Jackson}, J. and {Simon}, R. and {Townsley}, L. and {Zavagno}, A. and {Aladro}, R. and {Buchbender}, C. and {Guevara}, C. and {Higgins}, R. and {Jacob}, A.~M. and {Kabanovic}, S. and {Karim}, R. and {Soam}, A. and {Stutzki}, J. and {Tiwari}, M. and {Wyrowski}, F. and {Tielens}, A.~G.~G.~M.},
        title = "{The SOFIA FEEDBACK Legacy Survey Dynamics and Mass Ejection in the Bipolar H II Region RCW 36}",
      journal = {\apj},
         year = 2022,
        month = aug,
       volume = {935},
       number = {2},
          eid = {171},
        pages = {171},
          doi = {10.3847/1538-4357/ac8052},
archivePrefix = {arXiv},
       eprint = {2207.06479},
 primaryClass = {astro-ph.GA},
       adsurl = {https://ui.adsabs.harvard.edu/abs/2022ApJ...935..171B}
}

@ARTICLE{rosen2022,
       author = {{Rosen}, Anna L.},
        title = "{A Massive Star Is Born: How Feedback from Stellar Winds, Radiation Pressure, and Collimated Outflows Limits Accretion onto Massive Stars}",
      journal = {\apj},
         year = 2022,
        month = dec,
       volume = {941},
       number = {2},
          eid = {202},
        pages = {202},
          doi = {10.3847/1538-4357/ac9f3d},
archivePrefix = {arXiv},
       eprint = {2204.09700},
 primaryClass = {astro-ph.SR},
       adsurl = {https://ui.adsabs.harvard.edu/abs/2022ApJ...941..202R}
}

@ARTICLE{maeda2025,
       author = {{Maeda}, Fumiya and {Ohta}, Kouji and {Egusa}, Fumi and {Fujimoto}, Yusuke and {Kobayashi}, Masato I.~N. and {Inoue}, Shin and {Habe}, Asao},
        title = "{Galactic Structure Dependence of Cloud{\textendash}Cloud-collision-driven Star Formation in the Barred Galaxy NGC 3627}",
      journal = {\apj},
         year = 2025,
        month = mar,
       volume = {981},
       number = {2},
          eid = {156},
        pages = {156},
          doi = {10.3847/1538-4357/adb41e},
archivePrefix = {arXiv},
       eprint = {2502.06102},
 primaryClass = {astro-ph.GA},
       adsurl = {https://ui.adsabs.harvard.edu/abs/2025ApJ...981..156M}
}

@ARTICLE{choi2023,
       author = {{Choi}, Woorak and {Liu}, Lijie and {Bureau}, Martin and {Cappellari}, Michele and {Davis}, Timothy A. and {Gensior}, Jindra and {Liang}, Fu-Heng and {Lu}, Anan and {Williams}, Thomas G. and {Chung}, Aeree},
        title = "{WISDOM Project - XV. Giant molecular clouds in the central region of the barred spiral galaxy NGC 5806}",
      journal = {\mnras},
         year = 2023,
        month = jul,
       volume = {522},
       number = {3},
        pages = {4078-4097},
          doi = {10.1093/mnras/stad1211},
archivePrefix = {arXiv},
       eprint = {2304.10471},
 primaryClass = {astro-ph.GA},
       adsurl = {https://ui.adsabs.harvard.edu/abs/2023MNRAS.522.4078C}
}

@ARTICLE{maity2024,
       author = {{Maity}, A.~K. and {Inoue}, T. and {Fukui}, Y. and {Dewangan}, L.~K. and {Sano}, H. and {Yamada}, R.~I. and {Tachihara}, K. and {Bhadari}, N.~K. and {Jadhav}, O.~R.},
        title = "{Cloud{\textendash}Cloud Collision: Formation of Hub-filament Systems and Associated Gas Kinematics. Mass-collecting Cone{\textemdash}A New Signature of Cloud{\textendash}Cloud Collision}",
      journal = {\apj},
         year = 2024,
        month = oct,
       volume = {974},
       number = {2},
          eid = {229},
        pages = {229},
          doi = {10.3847/1538-4357/ad7098},
archivePrefix = {arXiv},
       eprint = {2408.06826},
 primaryClass = {astro-ph.GA},
       adsurl = {https://ui.adsabs.harvard.edu/abs/2024ApJ...974..229M}
}

@ARTICLE{wagner2012,
       author = {{Wagner}, A.~Y. and {Bicknell}, G.~V. and {Umemura}, M.},
        title = "{Driving Outflows with Relativistic Jets and the Dependence of Active Galactic Nucleus Feedback Efficiency on Interstellar Medium Inhomogeneity}",
      journal = {\apj},
         year = 2012,
        month = oct,
       volume = {757},
       number = {2},
          eid = {136},
        pages = {136},
          doi = {10.1088/0004-637X/757/2/136},
archivePrefix = {arXiv},
       eprint = {1205.0542},
 primaryClass = {astro-ph.CO},
       adsurl = {https://ui.adsabs.harvard.edu/abs/2012ApJ...757..136W}
}

@ARTICLE{Combes2019,
       author = {{Combes}, F. and {Gupta}, N. and {Jozsa}, G.~I.~G. and {Momjian}, E.},
        title = "{Discovery of CO absorption at z = 0.05 in G0248+430}",
      journal = {\aap},
         year = 2019,
        month = mar,
       volume = {623},
          eid = {A133},
        pages = {A133},
          doi = {10.1051/0004-6361/201935057},
archivePrefix = {arXiv},
       eprint = {1901.04683},
 primaryClass = {astro-ph.GA},
       adsurl = {https://ui.adsabs.harvard.edu/abs/2019A&A...623A.133C}
}

@ARTICLE{Rose2020,
       author = {{Rose}, Tom and {Edge}, A.~C. and {Combes}, F. and {Hamer}, S. and {McNamara}, B.~R. and {Russell}, H. and {Gaspari}, M. and {Salom{\'e}}, P. and {Sarazin}, C. and {Tremblay}, G.~R. and {Baum}, S.~A. and {Bremer}, M.~N. and {Donahue}, M. and {Fabian}, A.~C. and {Ferland}, G. and {Nesvadba}, N. and {O'Dea}, C. and {Oonk}, J.~B.~R. and {Peck}, A.~B.},
        title = "{A molecular absorption line survey towards the AGN of Hydra-A}",
      journal = {\mnras},
         year = 2020,
        month = jul,
       volume = {496},
       number = {1},
        pages = {364-380},
          doi = {10.1093/mnras/staa1474},
archivePrefix = {arXiv},
       eprint = {2005.10252},
 primaryClass = {astro-ph.GA},
       adsurl = {https://ui.adsabs.harvard.edu/abs/2020MNRAS.496..364R}
}

@ARTICLE{Rose2019,
       author = {{Rose}, Tom and {Edge}, A.~C. and {Combes}, F. and {Gaspari}, M. and {Hamer}, S. and {Nesvadba}, N. and {Peck}, A.~B. and {Sarazin}, C. and {Tremblay}, G.~R. and {Baum}, S.~A. and {Bremer}, M.~N. and {McNamara}, B.~R. and {O'Dea}, C. and {Oonk}, J.~B.~R. and {Russell}, H. and {Salom{\'e}}, P. and {Donahue}, M. and {Fabian}, A.~C. and {Ferland}, G. and {Mittal}, R. and {Vantyghem}, A.},
        title = "{Constraining cold accretion on to supermassive black holes: molecular gas in the cores of eight brightest cluster galaxies revealed by joint CO and CN absorption}",
      journal = {\mnras},
         year = 2019,
        month = oct,
       volume = {489},
       number = {1},
        pages = {349-365},
          doi = {10.1093/mnras/stz2138},
archivePrefix = {arXiv},
       eprint = {1907.13526},
 primaryClass = {astro-ph.GA},
       adsurl = {https://ui.adsabs.harvard.edu/abs/2019MNRAS.489..349R}
}

@ARTICLE{Bellocchi2023,
       author = {{Bellocchi}, E. and {Mart{\'\i}n-Pintado}, J. and {Rico-Villas}, F. and {Mart{\'\i}n}, S. and {Jim{\'e}nez-Sierra}, I.},
        title = "{Positive feedback, quenching, and sequential super star cluster (SSC) formation in NGC 4945}",
      journal = {\mnras},
         year = 2023,
        month = feb,
       volume = {519},
       number = {1},
        pages = {L68-L73},
          doi = {10.1093/mnrasl/slac154},
archivePrefix = {arXiv},
       eprint = {2211.17268},
 primaryClass = {astro-ph.GA},
       adsurl = {https://ui.adsabs.harvard.edu/abs/2023MNRAS.519L..68B}
}

@ARTICLE{indriolo2015,
       author = {{Indriolo}, Nick and {Neufeld}, D.~A. and {Gerin}, M. and {Schilke}, P. and {Benz}, A.~O. and {Winkel}, B. and {Menten}, K.~M. and {Chambers}, E.~T. and {Black}, John H. and {Bruderer}, S. and {Falgarone}, E. and {Godard}, B. and {Goicoechea}, J.~R. and {Gupta}, H. and {Lis}, D.~C. and {Ossenkopf}, V. and {Persson}, C.~M. and {Sonnentrucker}, P. and {van der Tak}, F.~F.~S. and {van Dishoeck}, E.~F. and {Wolfire}, Mark G. and {Wyrowski}, F.},
        title = "{Herschel Survey of Galactic OH$^{+}$, H$_{2}$O$^{+}$, and H$_{3}$O$^{+}$: Probing the Molecular Hydrogen Fraction and Cosmic-Ray Ionization Rate}",
      journal = {\apj},
         year = 2015,
        month = feb,
       volume = {800},
       number = {1},
          eid = {40},
        pages = {40},
          doi = {10.1088/0004-637X/800/1/40},
archivePrefix = {arXiv},
       eprint = {1412.1106},
 primaryClass = {astro-ph.GA},
       adsurl = {https://ui.adsabs.harvard.edu/abs/2015ApJ...800...40I}
}

@ARTICLE{Nagai2007,
       author = {{Nagai}, Makoto and {Tanaka}, Kunihiko and {Kamegai}, Kazuhisa and {Oka}, Tomoharu},
        title = "{Physical Conditions of Molecular Gas in the Galactic Center}",
      journal = {\pasj},
         year = 2007,
        month = feb,
       volume = {59},
        pages = {25-31},
          doi = {10.1093/pasj/59.1.25},
archivePrefix = {arXiv},
       eprint = {astro-ph/0701845},
 primaryClass = {astro-ph},
       adsurl = {https://ui.adsabs.harvard.edu/abs/2007PASJ...59...25N}
}

@ARTICLE{Tanaka2018,
       author = {{Tanaka}, Kunihiko and {Nagai}, Makoto and {Kamegai}, Kazuhisa and {Iino}, Takahiro and {Sakai}, Takeshi},
        title = "{HCN J = 4-3, HNC J = 1-0, H$^{13}$CN J = 1-0, and HC$_{3}$N J = 10-9 Maps of the Galactic Center Region. I. Spatially Resolved Measurements of Physical Conditions and Chemical Composition}",
      journal = {\apjs},
         year = 2018,
        month = jun,
       volume = {236},
       number = {2},
          eid = {40},
        pages = {40},
          doi = {10.3847/1538-4365/aab9a5},
archivePrefix = {arXiv},
       eprint = {1804.00666},
 primaryClass = {astro-ph.GA},
       adsurl = {https://ui.adsabs.harvard.edu/abs/2018ApJS..236...40T}
}

@ARTICLE{Miura2018,
       author = {{Miura}, Rie E. and {Espada}, Daniel and {Hirota}, Akihiko and {Nakanishi}, Kouichiro and {Bendo}, George J. and {Sugai}, Hajime},
        title = "{ALMA Observations toward the Starburst Dwarf Galaxy NGC 5253. I. Molecular Cloud Properties and Scaling Relations}",
      journal = {\apj},
         year = 2018,
        month = sep,
       volume = {864},
       number = {2},
          eid = {120},
        pages = {120},
          doi = {10.3847/1538-4357/aad69f},
archivePrefix = {arXiv},
       eprint = {1808.10089},
 primaryClass = {astro-ph.GA},
       adsurl = {https://ui.adsabs.harvard.edu/abs/2018ApJ...864..120M}
}

@ARTICLE{Mills2021,
       author = {{Mills}, E.~A.~C. and {Gorski}, M. and {Emig}, K.~L. and {Bolatto}, A.~D. and {Levy}, R.~C. and {Leroy}, A.~K. and {Ginsburg}, A. and {Henshaw}, J.~D. and {Zschaechner}, L.~K. and {Veilleux}, S. and {Tanaka}, K. and {Meier}, D.~S. and {Walter}, F. and {Krieger}, N. and {Ott}, J.},
        title = "{Clustered Star Formation in the Center of NGC 253 Contributes to Driving the Ionized Nuclear Wind}",
      journal = {\apj},
         year = 2021,
        month = oct,
       volume = {919},
       number = {2},
          eid = {105},
        pages = {105},
          doi = {10.3847/1538-4357/ac0fe8},
archivePrefix = {arXiv},
       eprint = {2106.14970},
 primaryClass = {astro-ph.GA},
       adsurl = {https://ui.adsabs.harvard.edu/abs/2021ApJ...919..105M}
}

@ARTICLE{Sun2024,
       author = {{Sun}, Jiayi and {He}, Hao and {Batschkun}, Kyle and {Levy}, Rebecca C. and {Emig}, Kimberly and {Rodr{\'\i}guez}, M. Jimena and {Hassani}, Hamid and {Leroy}, Adam K. and {Schinnerer}, Eva and {Ostriker}, Eve C. and {Wilson}, Christine D. and {Bolatto}, Alberto D. and {Mills}, Elisabeth A.~C. and {Rosolowsky}, Erik and {Lee}, Janice C. and {Dale}, Daniel A. and {Larson}, Kirsten L. and {Thilker}, David A. and {Ubeda}, Leonardo and {Whitmore}, Bradley C. and {Williams}, Thomas G. and {Barnes}, Ashley T. and {Bigiel}, Frank and {Chevance}, M{\'e}lanie and {Glover}, Simon C.~O. and {Grasha}, Kathryn and {Groves}, Brent and {Henshaw}, Jonathan D. and {Indebetouw}, R{\'e}my and {Jim{\'e}nez-Donaire}, Mar{\'\i}a J. and {Klessen}, Ralf S. and {Koch}, Eric W. and {Liu}, Daizhong and {Mathur}, Smita and {Meidt}, Sharon and {Menon}, Shyam H. and {Neumann}, Justus and {Pinna}, Francesca and {Querejeta}, Miguel and {Sormani}, Mattia C. and {Tress}, Robin G.},
        title = "{Hidden Gems on a Ring: Infant Massive Clusters and Their Formation Timeline Unveiled by ALMA, HST, and JWST in NGC 3351}",
      journal = {\apj},
         year = 2024,
        month = jun,
       volume = {967},
       number = {2},
          eid = {133},
        pages = {133},
          doi = {10.3847/1538-4357/ad3de6},
archivePrefix = {arXiv},
       eprint = {2401.14453},
 primaryClass = {astro-ph.GA},
       adsurl = {https://ui.adsabs.harvard.edu/abs/2024ApJ...967..133S}
}

@ARTICLE{Lambda2005,
       author = {{Sch{\"o}ier}, F.~L. and {van der Tak}, F.~F.~S. and {van Dishoeck}, E.~F. and {Black}, J.~H.},
        title = "{An atomic and molecular database for analysis of submillimetre line observations}",
      journal = {\aap},
         year = 2005,
        month = mar,
       volume = {432},
       number = {1},
        pages = {369-379},
          doi = {10.1051/0004-6361:20041729},
archivePrefix = {arXiv},
       eprint = {astro-ph/0411110},
 primaryClass = {astro-ph},
       adsurl = {https://ui.adsabs.harvard.edu/abs/2005A&A...432..369S}
}

@ARTICLE{Harada2019,
       author = {{Harada}, Nanase and {Sakamoto}, Kazushi and {Mart{\'\i}n}, Sergio and {Watanabe}, Yoshimasa and {Aladro}, Rebeca and {Riquelme}, Denise and {Hirota}, Akihiko},
        title = "{Chemical Evolution along the Circumnuclear Ring of M83}",
      journal = {\apj},
         year = 2019,
        month = oct,
       volume = {884},
       number = {2},
          eid = {100},
        pages = {100},
          doi = {10.3847/1538-4357/ab41ff},
archivePrefix = {arXiv},
       eprint = {1909.01549},
 primaryClass = {astro-ph.GA},
       adsurl = {https://ui.adsabs.harvard.edu/abs/2019ApJ...884..100H}
}

@ARTICLE{Viti2017,
       author = {{Viti}, Serena},
        title = "{Molecular transitions as probes of the physical conditions of extragalactic environments}",
      journal = {\aap},
         year = 2017,
        month = nov,
       volume = {607},
          eid = {A118},
        pages = {A118},
          doi = {10.1051/0004-6361/201628877},
archivePrefix = {arXiv},
       eprint = {1708.09188},
 primaryClass = {astro-ph.GA},
       adsurl = {https://ui.adsabs.harvard.edu/abs/2017A&A...607A.118V}
}

@ARTICLE{Leroy2017,
       author = {{Leroy}, Adam K. and {Schinnerer}, Eva and {Hughes}, Annie and {Kruijssen}, J.~M. Diederik and {Meidt}, Sharon and {Schruba}, Andreas and {Sun}, Jiayi and {Bigiel}, Frank and {Aniano}, Gonzalo and {Blanc}, Guillermo A. and {Bolatto}, Alberto and {Chevance}, M{\'e}lanie and {Colombo}, Dario and {Gallagher}, Molly and {Garcia-Burillo}, Santiago and {Kramer}, Carsten and {Querejeta}, Miguel and {Pety}, Jerome and {Thompson}, Todd A. and {Usero}, Antonio},
        title = "{Cloud-scale ISM Structure and Star Formation in M51}",
      journal = {\apj},
         year = 2017,
        month = sep,
       volume = {846},
       number = {1},
          eid = {71},
        pages = {71},
          doi = {10.3847/1538-4357/aa7fef},
archivePrefix = {arXiv},
       eprint = {1706.08540},
 primaryClass = {astro-ph.GA},
       adsurl = {https://ui.adsabs.harvard.edu/abs/2017ApJ...846...71L}
}

@BOOK{Dyson1997,
       author = {{Dyson}, J.~E. and {Williams}, D.~A.},
        title = "{Physics of the interstellar medium}",
         year = 1980,
       adsurl = {https://ui.adsabs.harvard.edu/abs/1980pim..book.....D}
}

@ARTICLE{Holdship2022b,
       author = {{Holdship}, J. and {Viti}, S.},
        title = "{History-independent tracers. Forgetful molecular probes of the physical conditions of the dense interstellar medium}",
      journal = {\aap},
         year = 2022,
        month = feb,
       volume = {658},
          eid = {A103},
        pages = {A103},
          doi = {10.1051/0004-6361/202142398},
archivePrefix = {arXiv},
       eprint = {2201.01312},
 primaryClass = {astro-ph.GA},
       adsurl = {https://ui.adsabs.harvard.edu/abs/2022A&A...658A.103H}
}

\begin{appendix}

\section{Neural Network performance}
\label{sec:NN-performance}

\begin{figure*}
\begin{center}
\includegraphics[width=\textwidth]{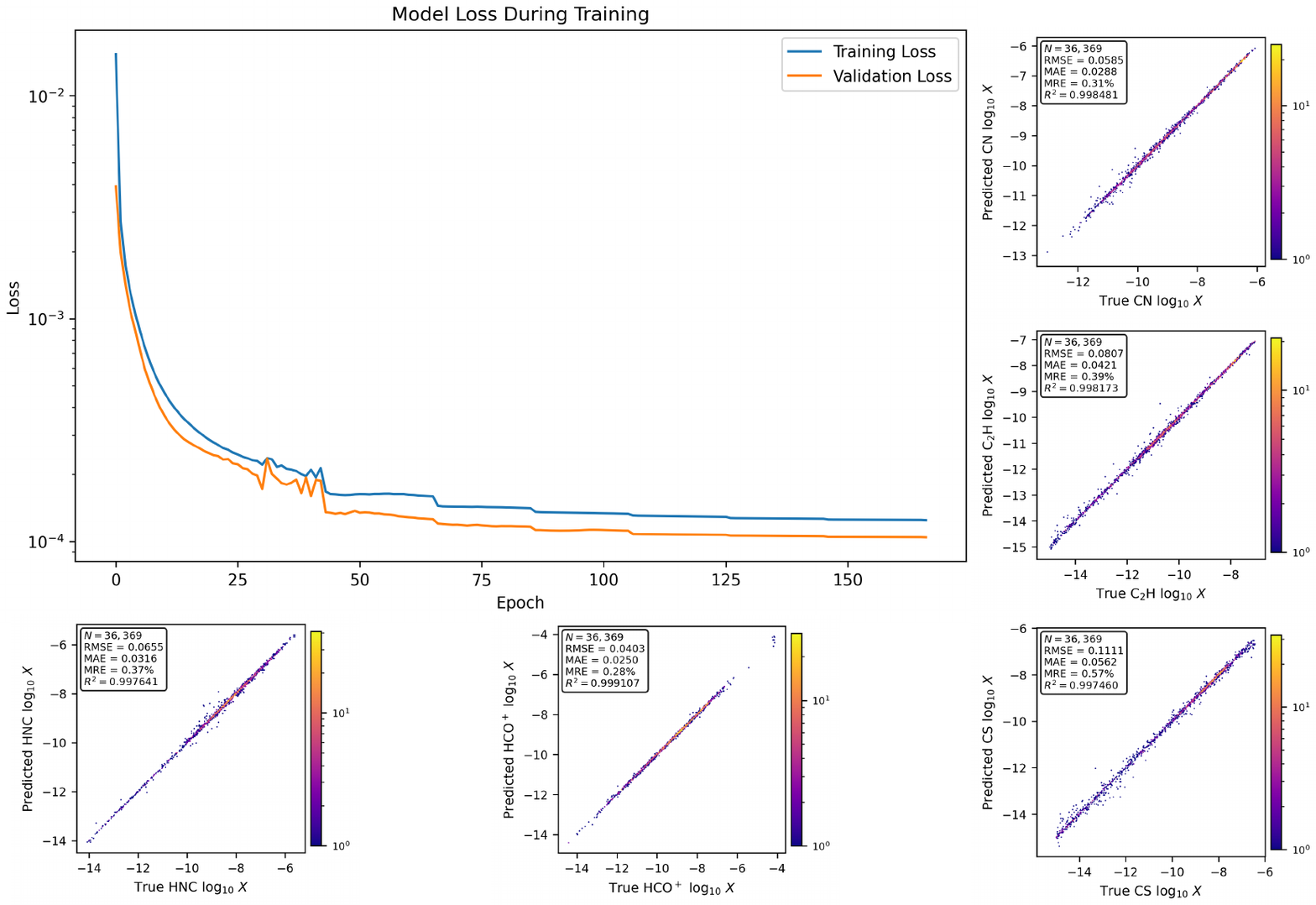}
\captionof{figure}{Performance of the neural network during training and testing.
The top left panel shows the evolution of the training and validation loss as a function of epoch. The remaining panels compare the predicted and \texttt{UCLCHEM}-derived abundances for the molecules used in this work, illustrating the accuracy of the trained model across the full dynamic range of abundances.}
\label{fig:nn_performance}
\end{center}
\end{figure*}

Figure~\ref{fig:nn_performance} summarizes the performance of the trained neural network for all species used in this work. The top-left panel shows the training and validation loss as a function of epoch. Both curves decrease rapidly and then flatten, and the validation loss closely follows the training loss, which indicates that the network generalizes well and does not overfit. The remaining panels compare the abundances predicted by the network with those computed by \texttt{UCLCHEM} for the test set. For every species the predictions lie close to the one-to-one relation across the full range of abundances, with only a small number of outliers near the edges of the parameter space. This confirms that the emulator is accurate and reliable.

\section{Corner plots for the CND and SSCs in NGC~1068}
\begin{figure*}[htbp]
    \centering
    \includegraphics[width=\textwidth]{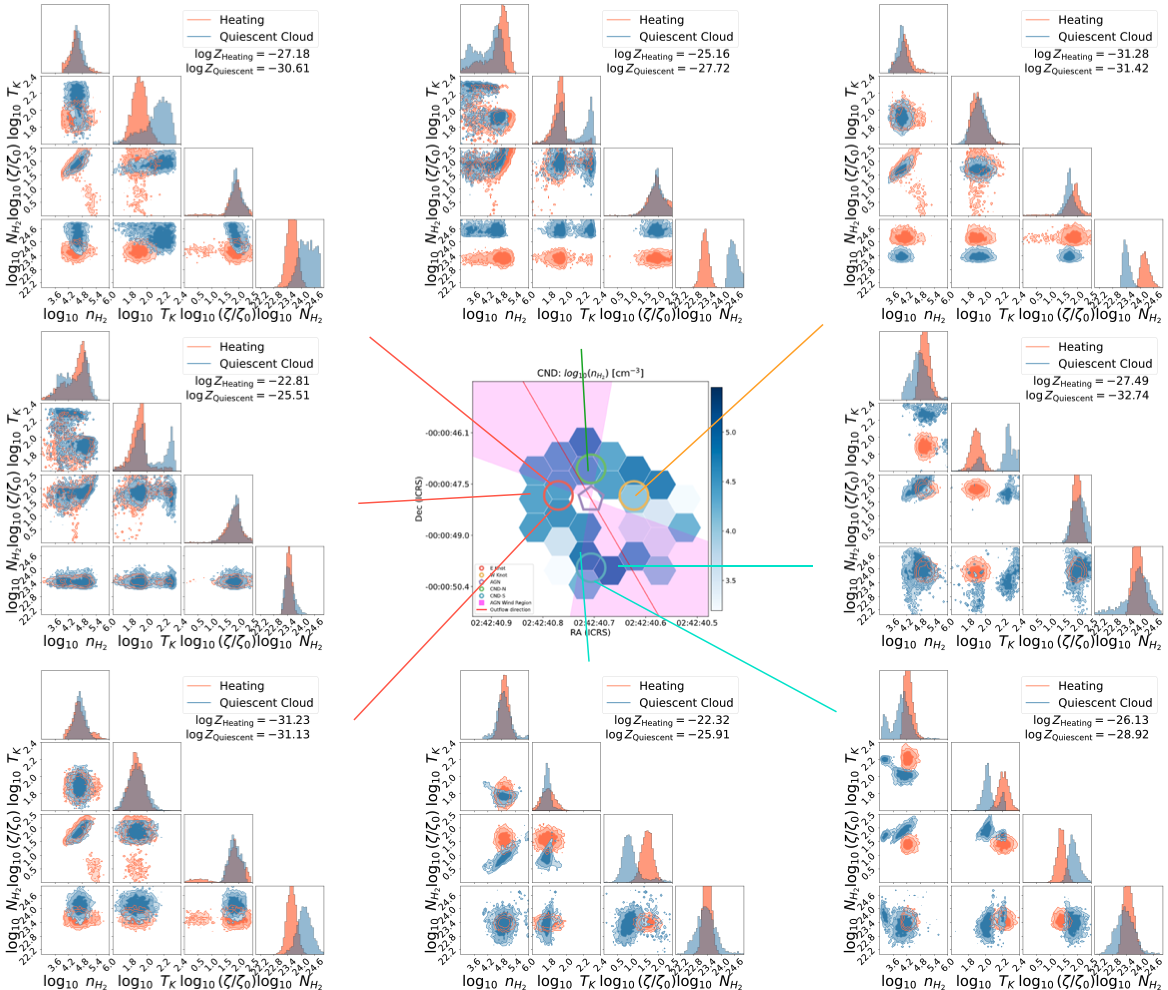} 
    \caption{
    Corner plots of the posterior distributions of the physical parameters in each hexagonal region of the CND. 
    The central panel shows the hexagonal sampling grid, with the coloured circles marking the regions for which the corner plots are displayed. 
    Orange points and histograms correspond to the heating model and blue ones to the quiescent cloud model.
    }
    \label{fig:cnd_corner_hex}
\end{figure*}

\begin{figure*}[htbp]
    \centering
    \includegraphics[width=\textwidth]{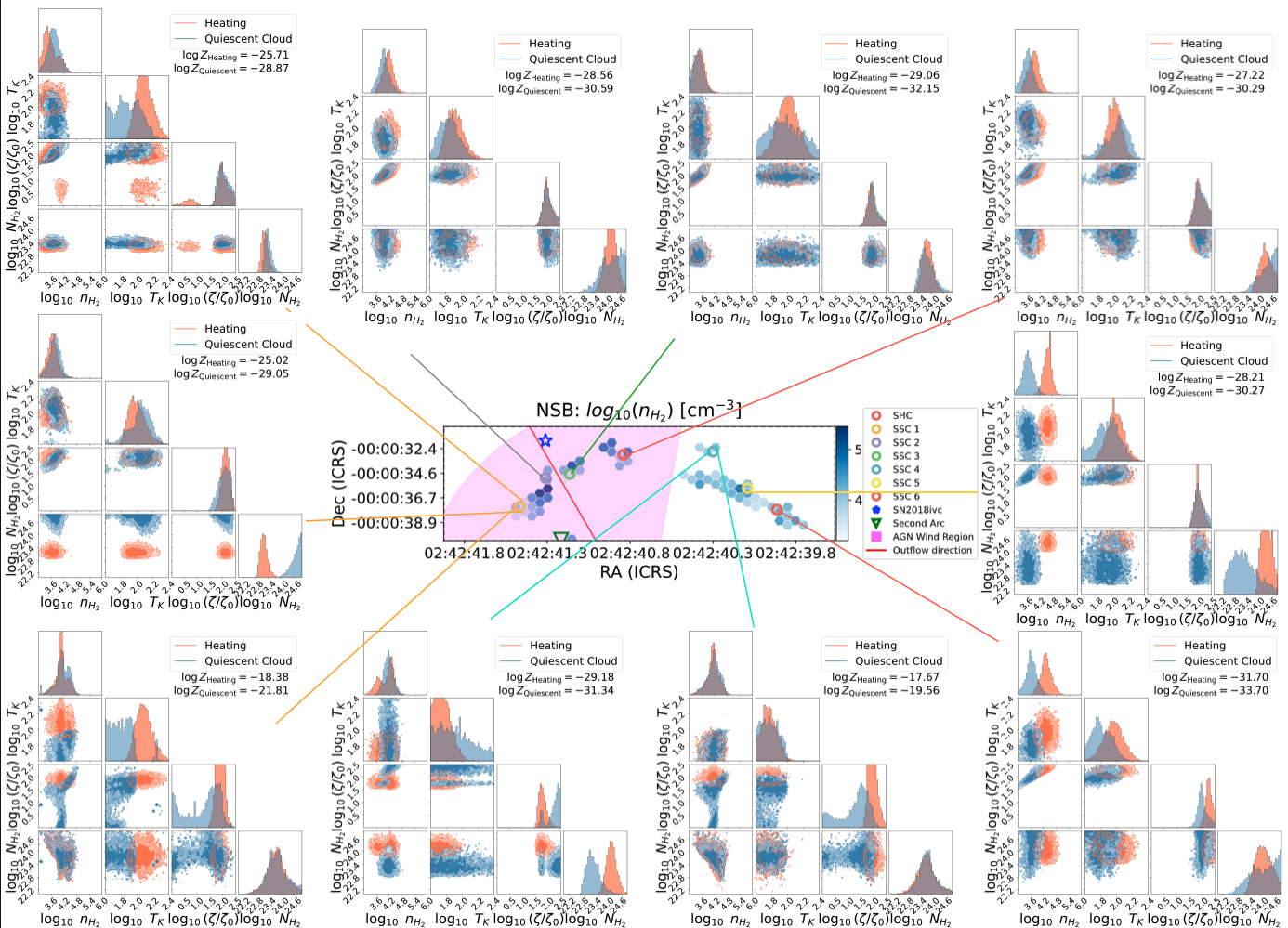} 
    \caption{Same as Fig~.\ref{fig:cnd_corner_hex}, but for the northern starburst ring.
    }
    \label{fig:nsb_corner_hex}
\end{figure*}

\begin{figure*}[htbp]
    \centering
    \includegraphics[width=\textwidth]{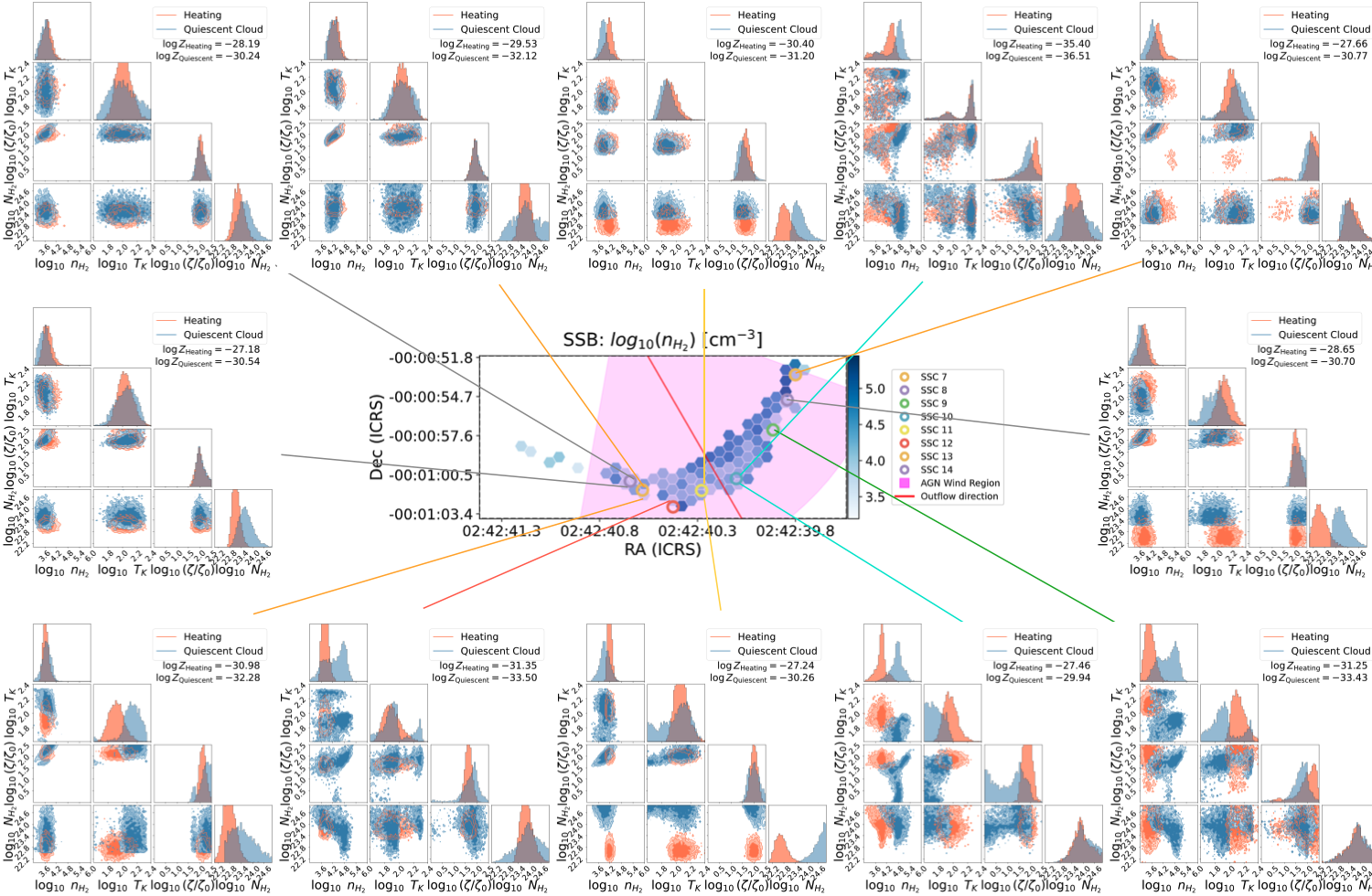} 
    \caption{Same as Fig~.\ref{fig:cnd_corner_hex}, but for the southern starburst ring.
    }
    \label{fig:ssb_corner_hex}
\end{figure*}

Fig.~\ref{fig:cnd_corner_hex}-Fig.~\ref{fig:ssb_corner_hex} presents the posterior distributions of the four physical parameters (gas density, kinetic temperature, column density, and cosmic ray ionization rate) for representative regions in the CND and the starburst ring. The diagonal panels show the marginalized distribution of each parameter, while the off-diagonal panels show the pairwise correlations.

\section{CS abundance as function of $\zeta$}
\begin{figure}
    \centering
    \includegraphics[width=\hsize]{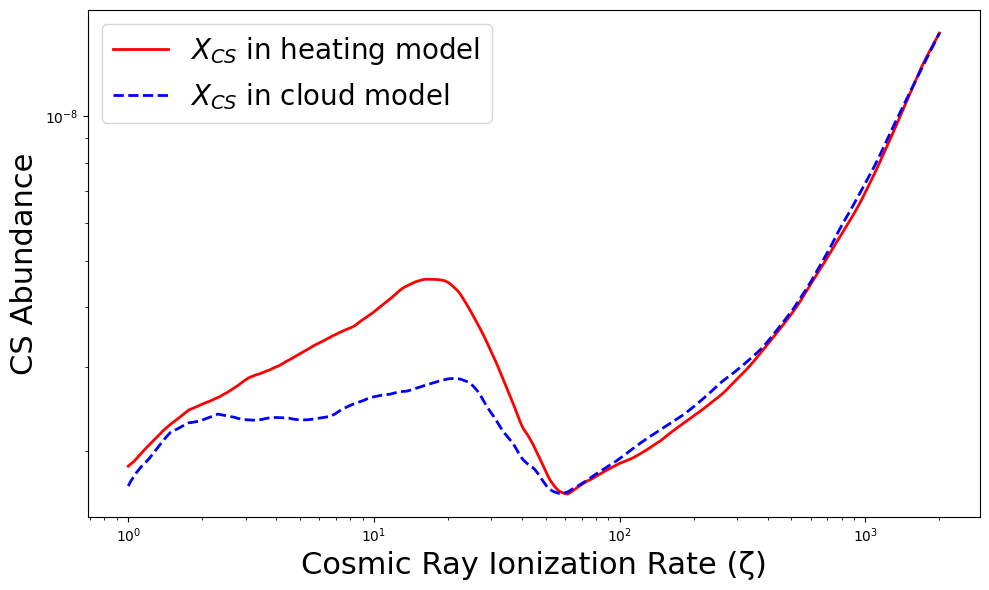}
    \caption{CS abundance as a function of cosmic ray ionization rate, using $n_{H_{2}}$=10$^{4}$ cm$^{-3}$ and temperature equals to 70 K, which represents an average physical condition in NGC 1068. The solid red line represents the CS fractional abundance ($X_\mathrm{CS}$) in the heating model, while the dashed blue line shows $X_\mathrm{CS}$in the quiescent cloud model. Both models show linear relation at high cosmic ray ionization rates ($\zeta$>100$\zeta_{0}$), but for lower CRIR regime, $X_\mathrm{CS}$ variation suggests that it's not a good tracer alone.}
    \label{fig:CS_abundance}
    \end{figure}

Figure~\ref{fig:CS_abundance} shows the predicted CS fractional abundance as a function of the cosmic ray ionization rate for the heating and quiescent models. The relation between $X_{\rm CS}$ and $\zeta$ becomes linear only at $\zeta/\zeta_{0} \gtrsim 100$. Since the regions in NGC\,1068 lie below this value, CS falls in the nonlinear regime and on its own does not provide a reliable constraint on $\zeta$.

\section{Posterior predictive check}
To assess the goodness of fit of the Bayesian inference, we perform posterior predictive checks (PPCs) for all the sub-regions. A PPC generates synthetic data from the posterior distributions of the inferred parameters and compares them with the observations. If the model could capture the data, the observed values should fall within the range spanned by the predicted distributions.

\begin{figure*}

\centering
\includegraphics[width=0.33\textwidth]{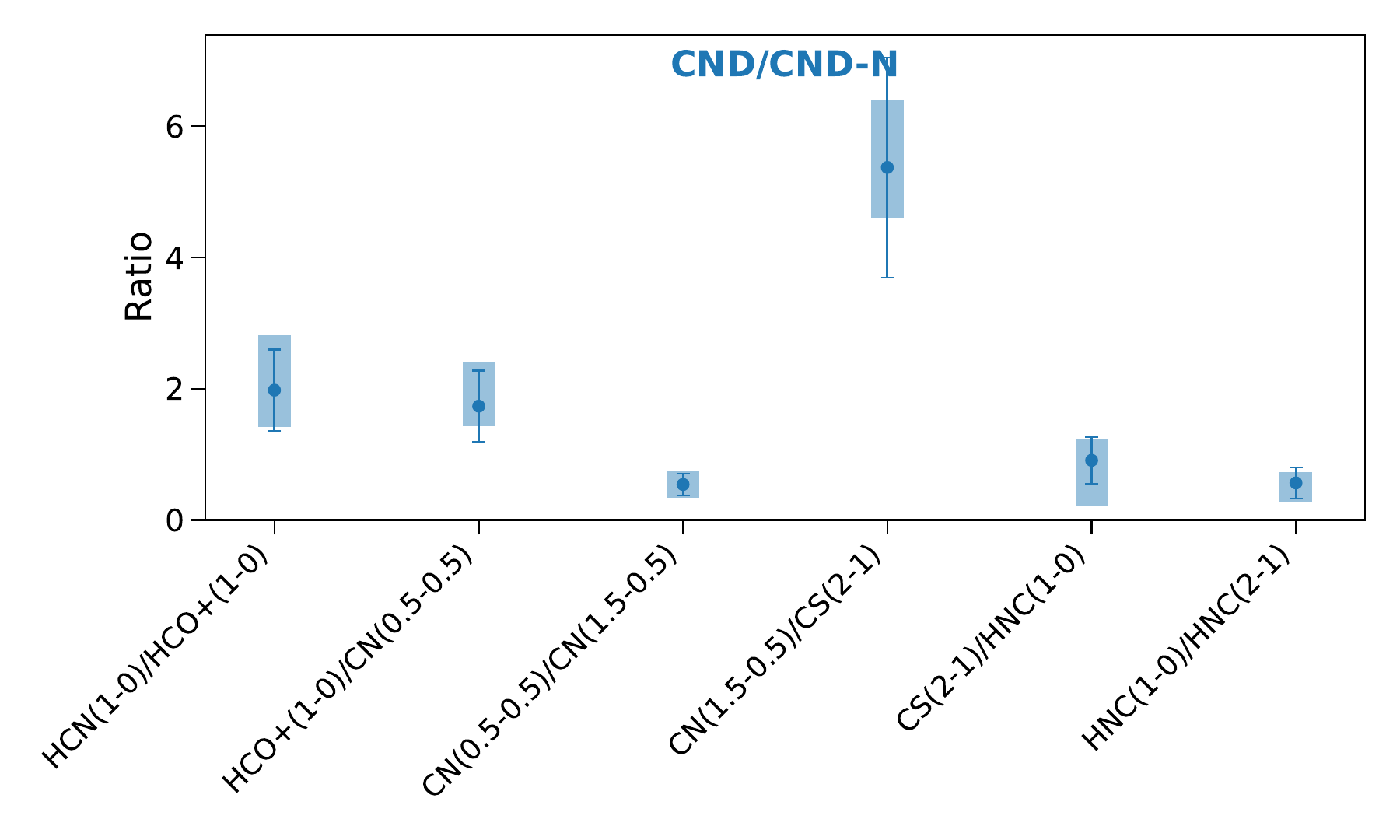}
\includegraphics[width=0.33\textwidth]{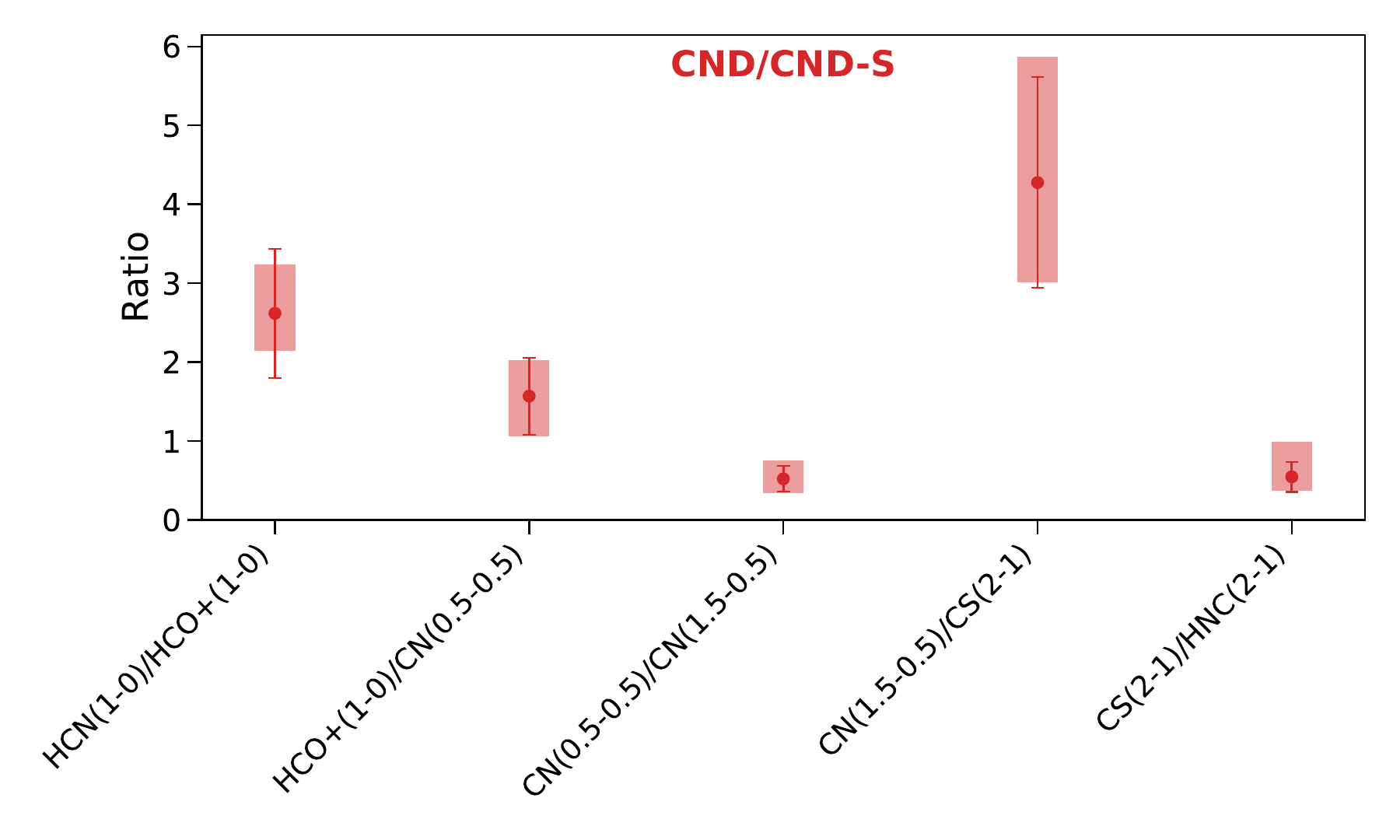}
\includegraphics[width=0.33\textwidth]{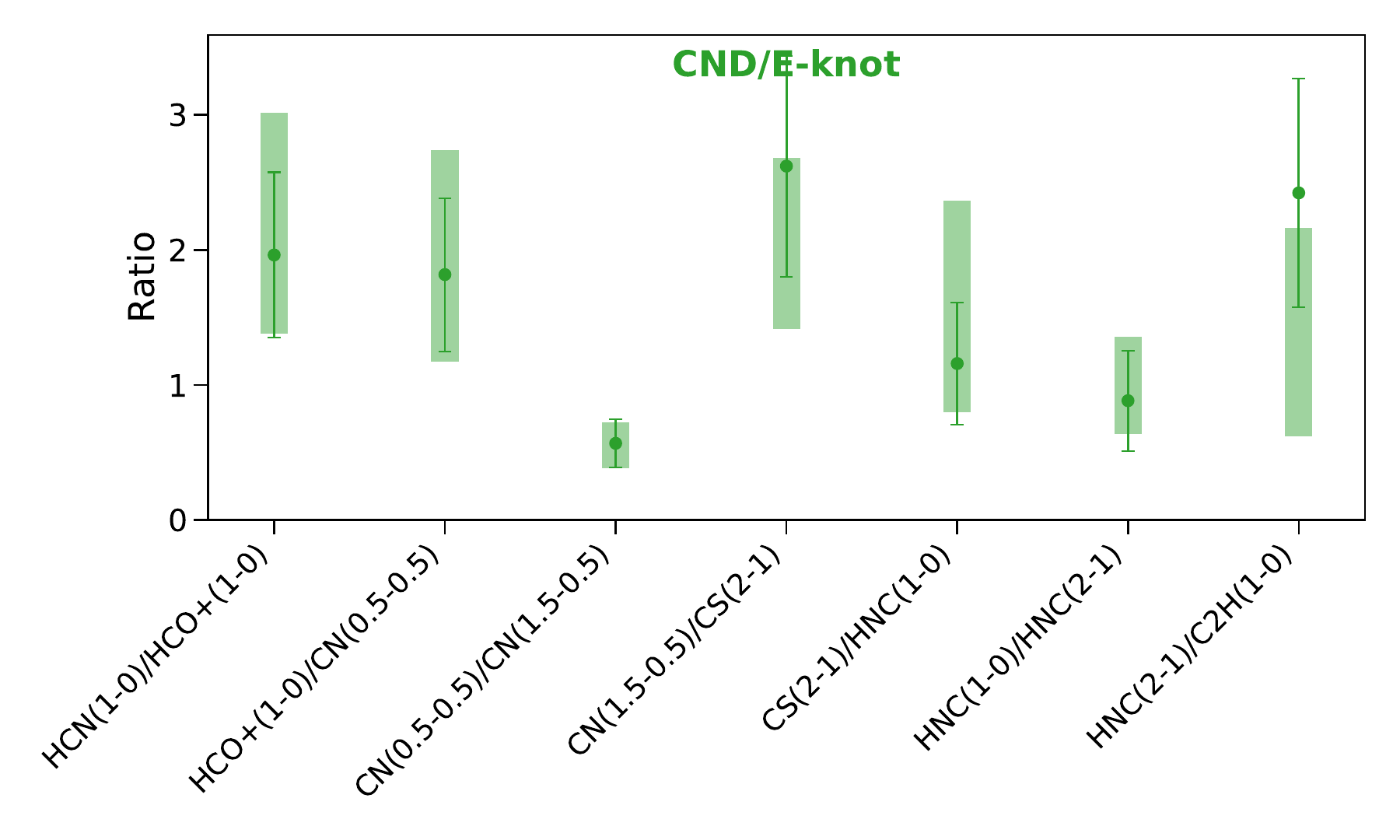}
\includegraphics[width=0.33\textwidth]{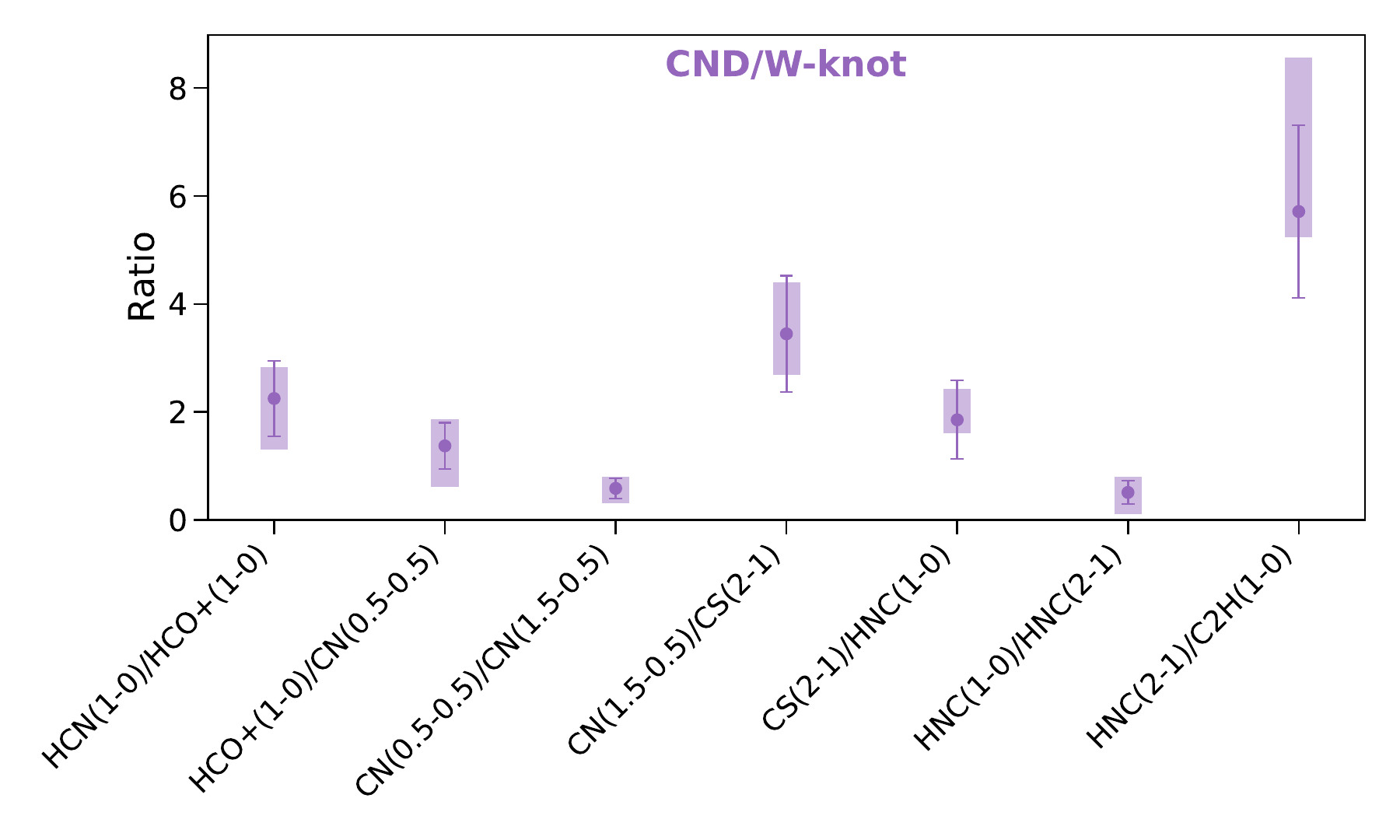}
\includegraphics[width=0.33\textwidth]{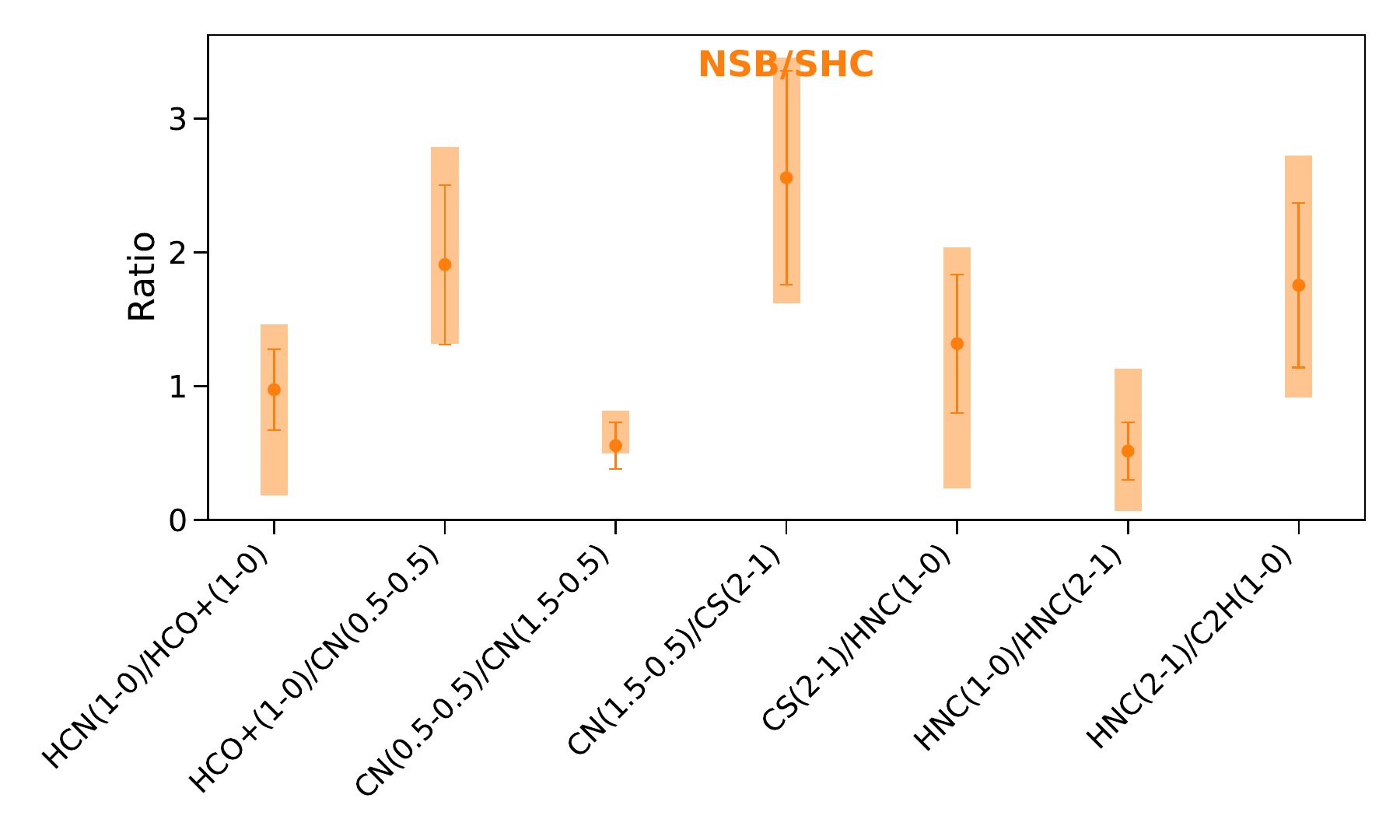}
\includegraphics[width=0.33\textwidth]{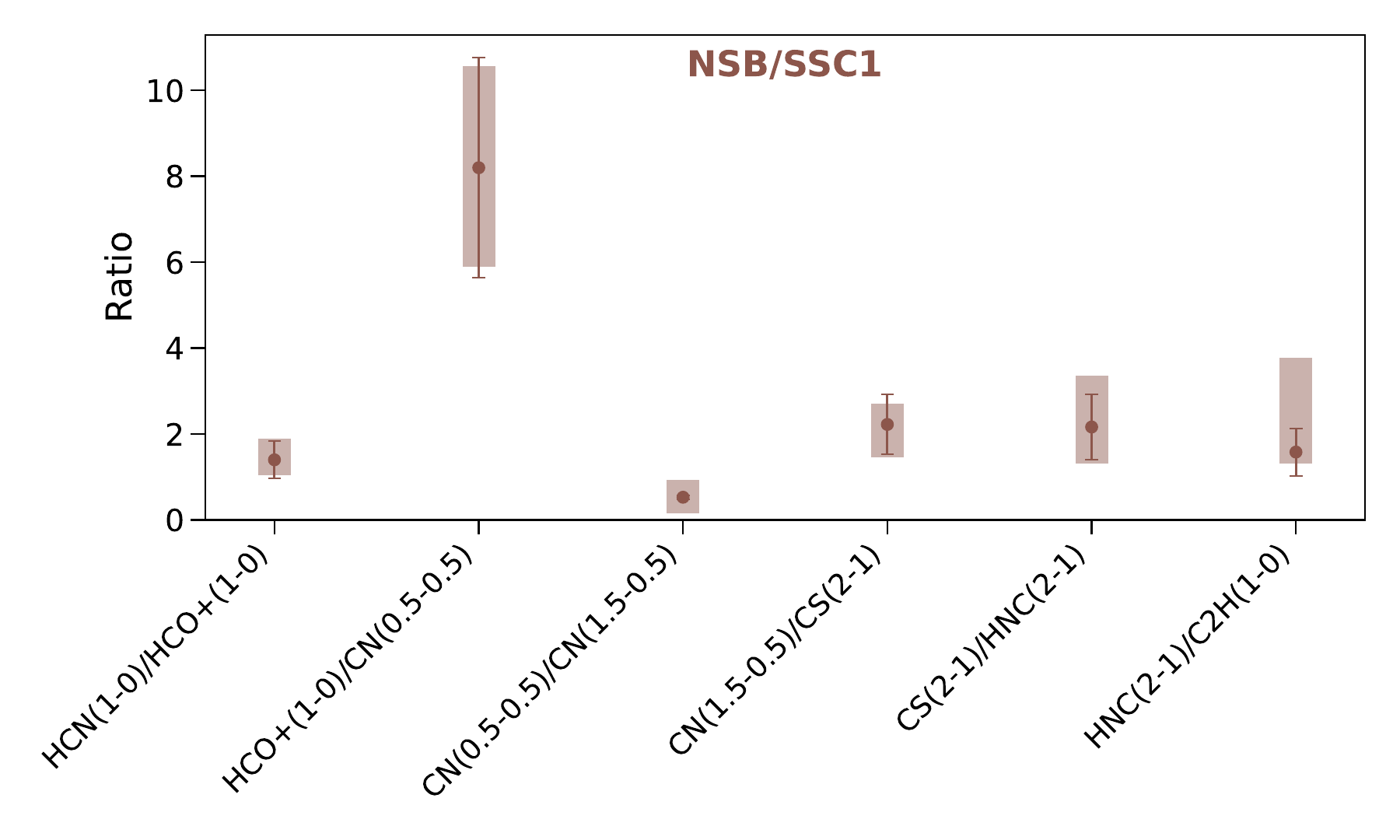}
\includegraphics[width=0.33\textwidth]{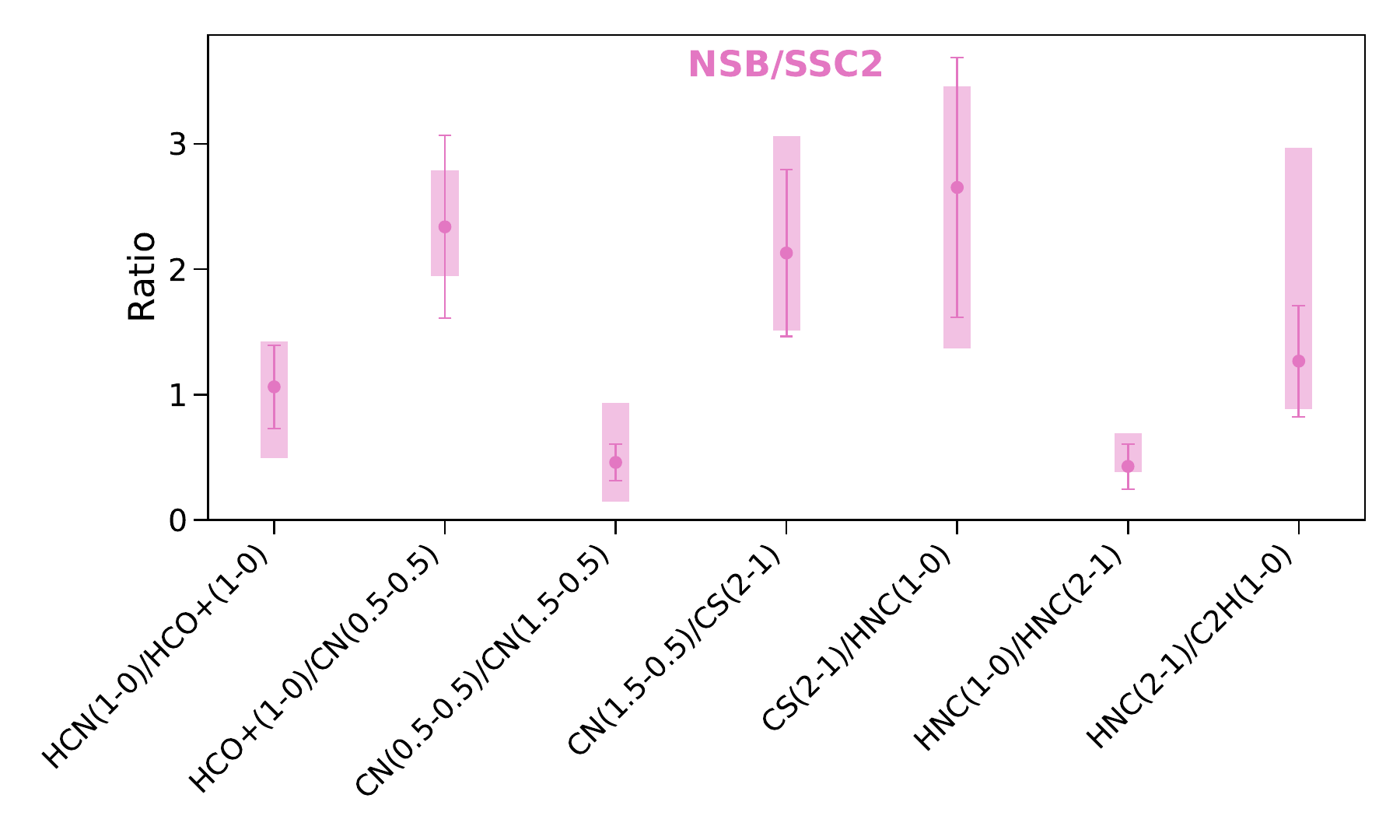}
\includegraphics[width=0.33\textwidth]{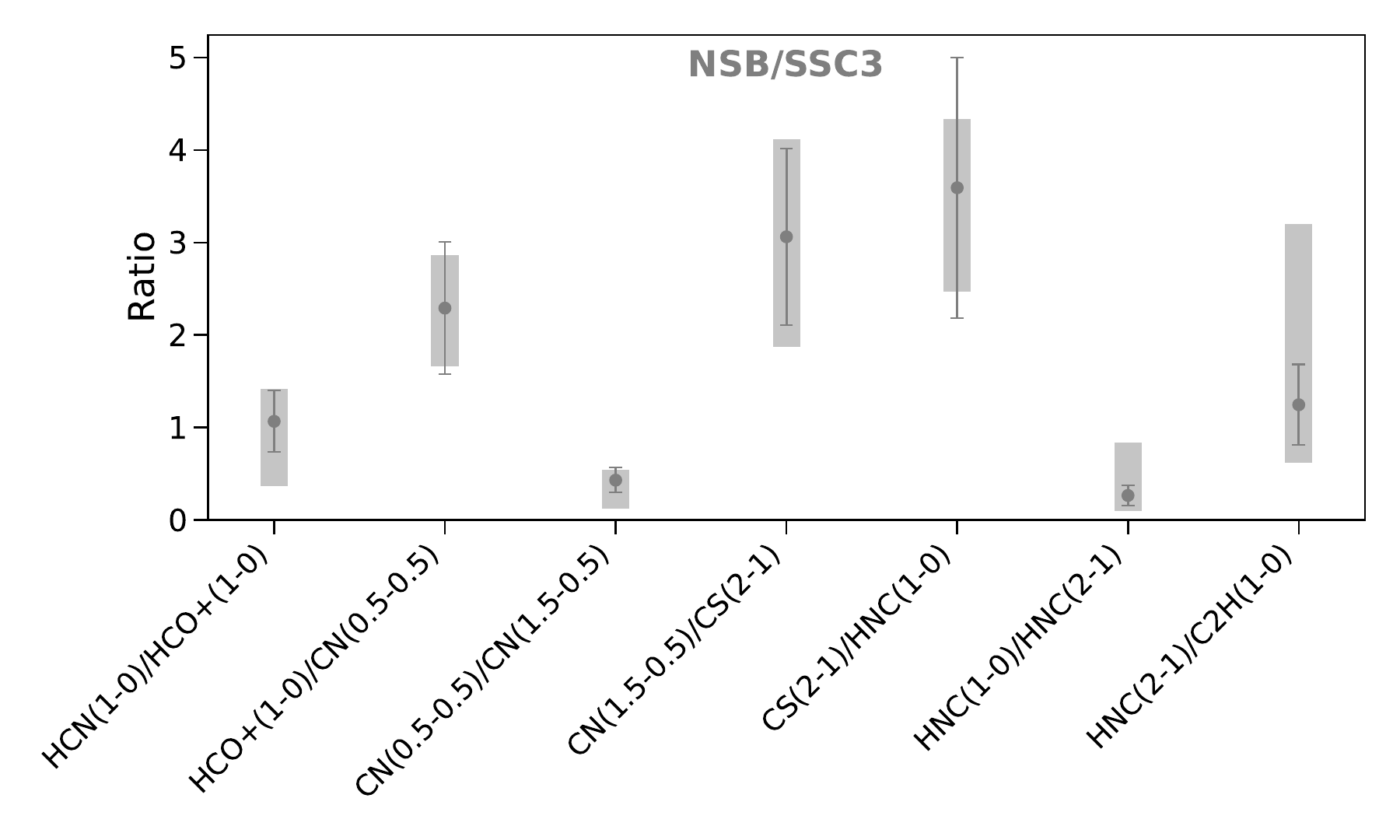}
\includegraphics[width=0.33\textwidth]{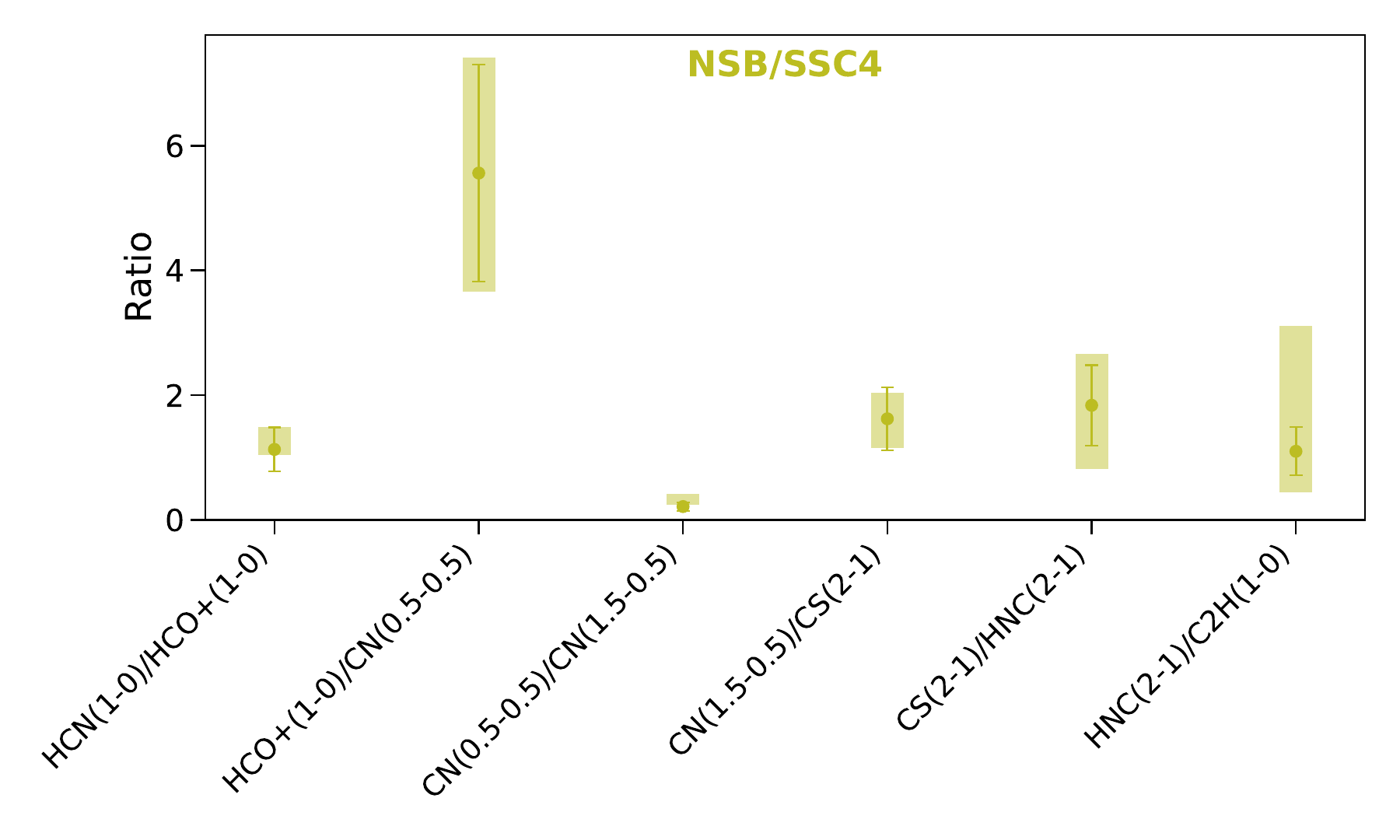}
\includegraphics[width=0.33\textwidth]{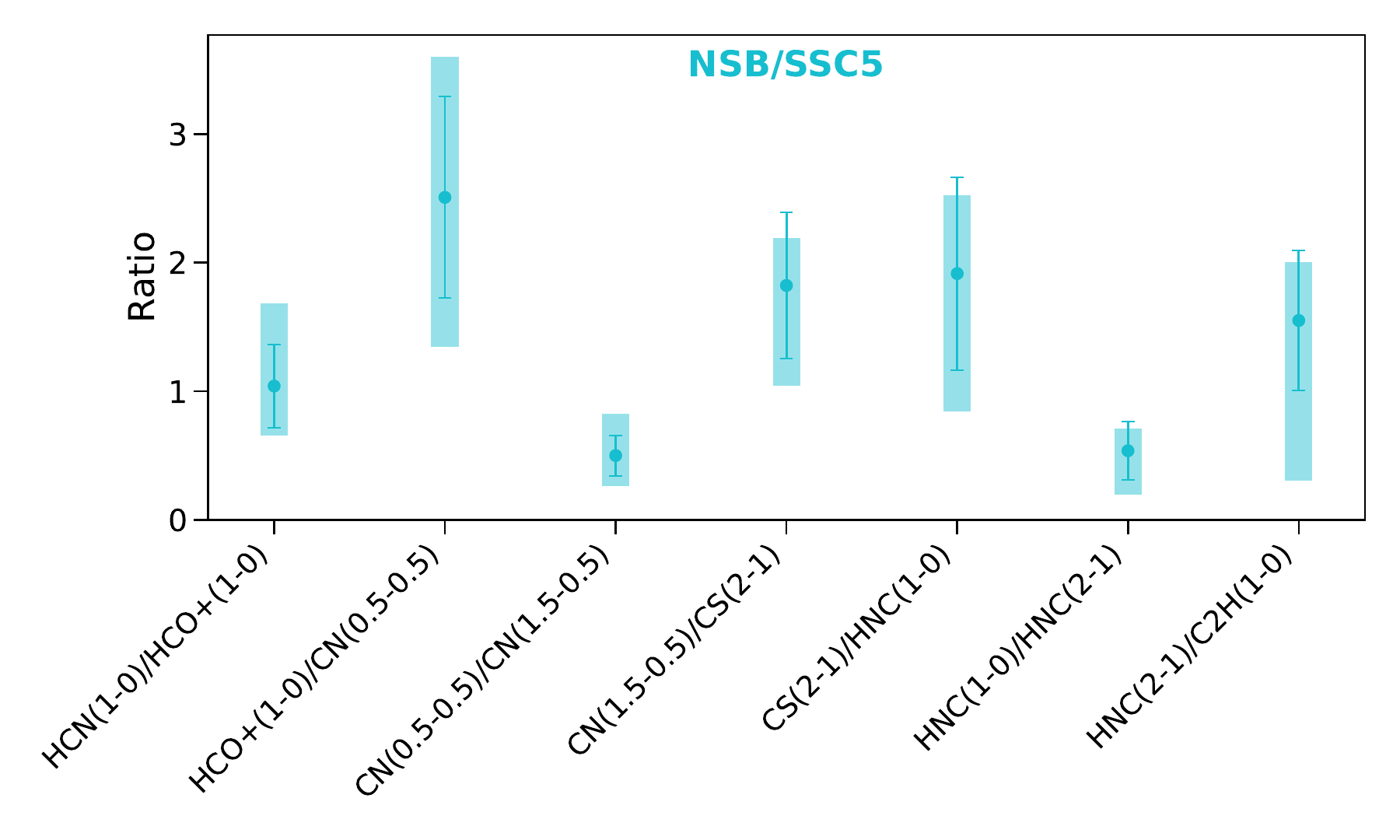}
\includegraphics[width=0.33\textwidth]{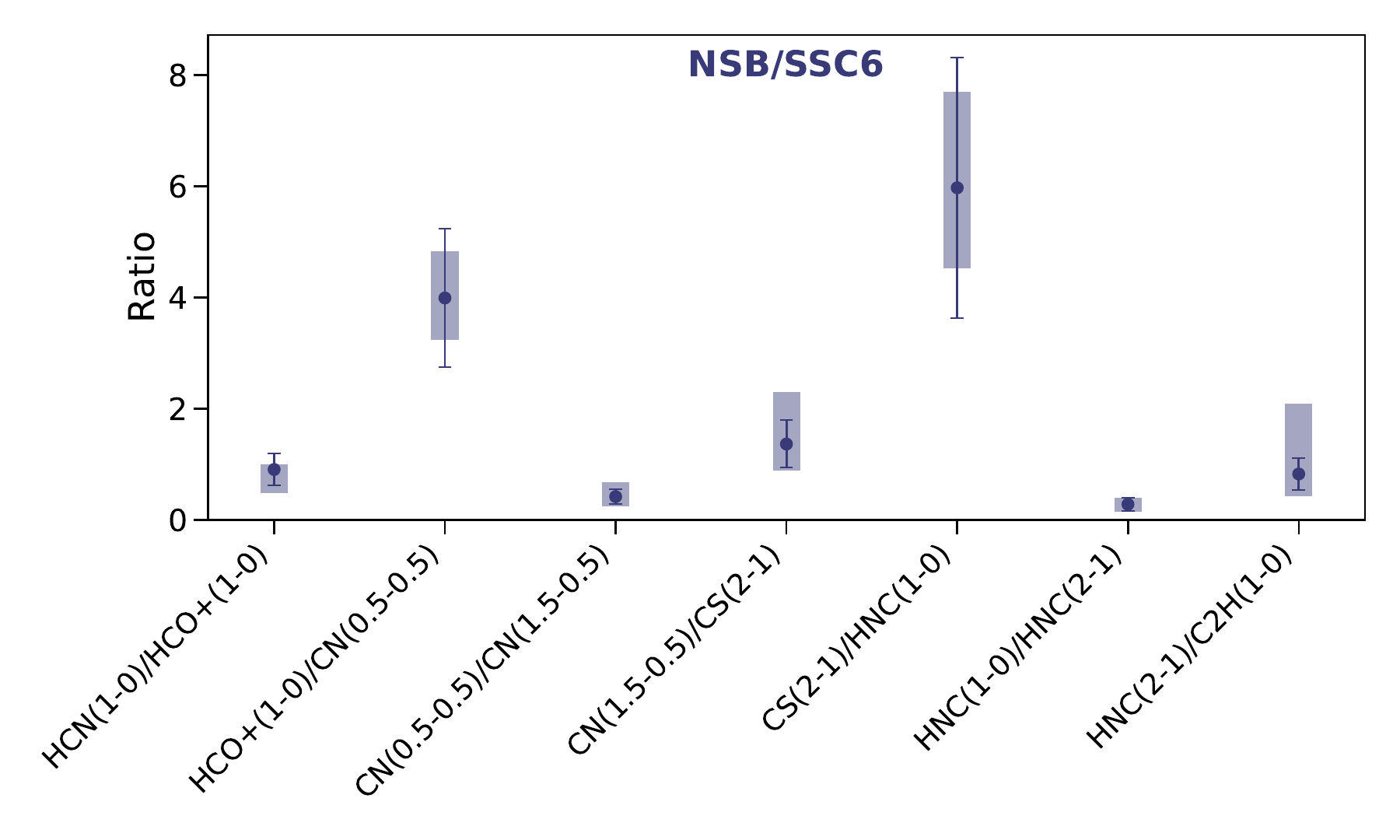}
\includegraphics[width=0.33\textwidth]{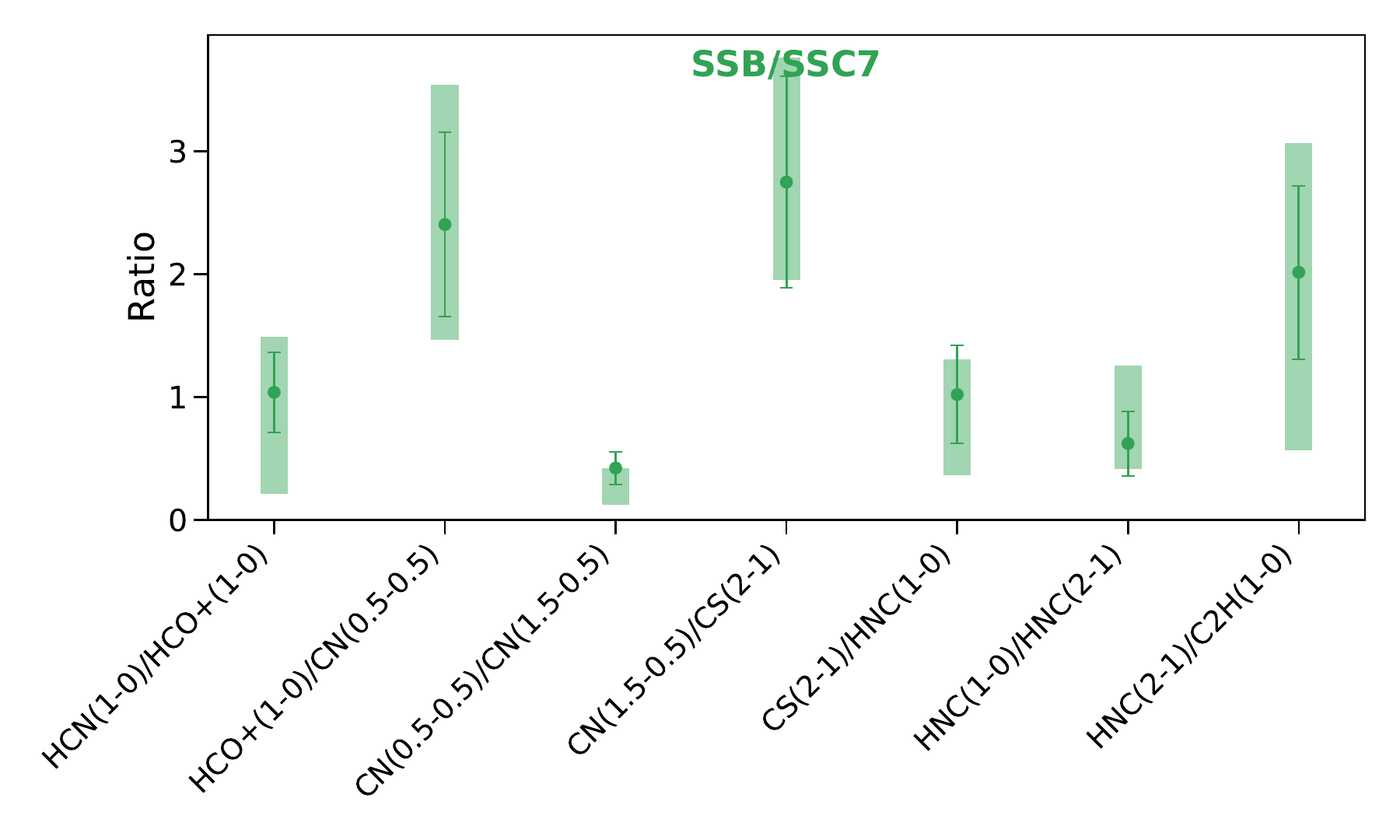}
\includegraphics[width=0.33\textwidth]{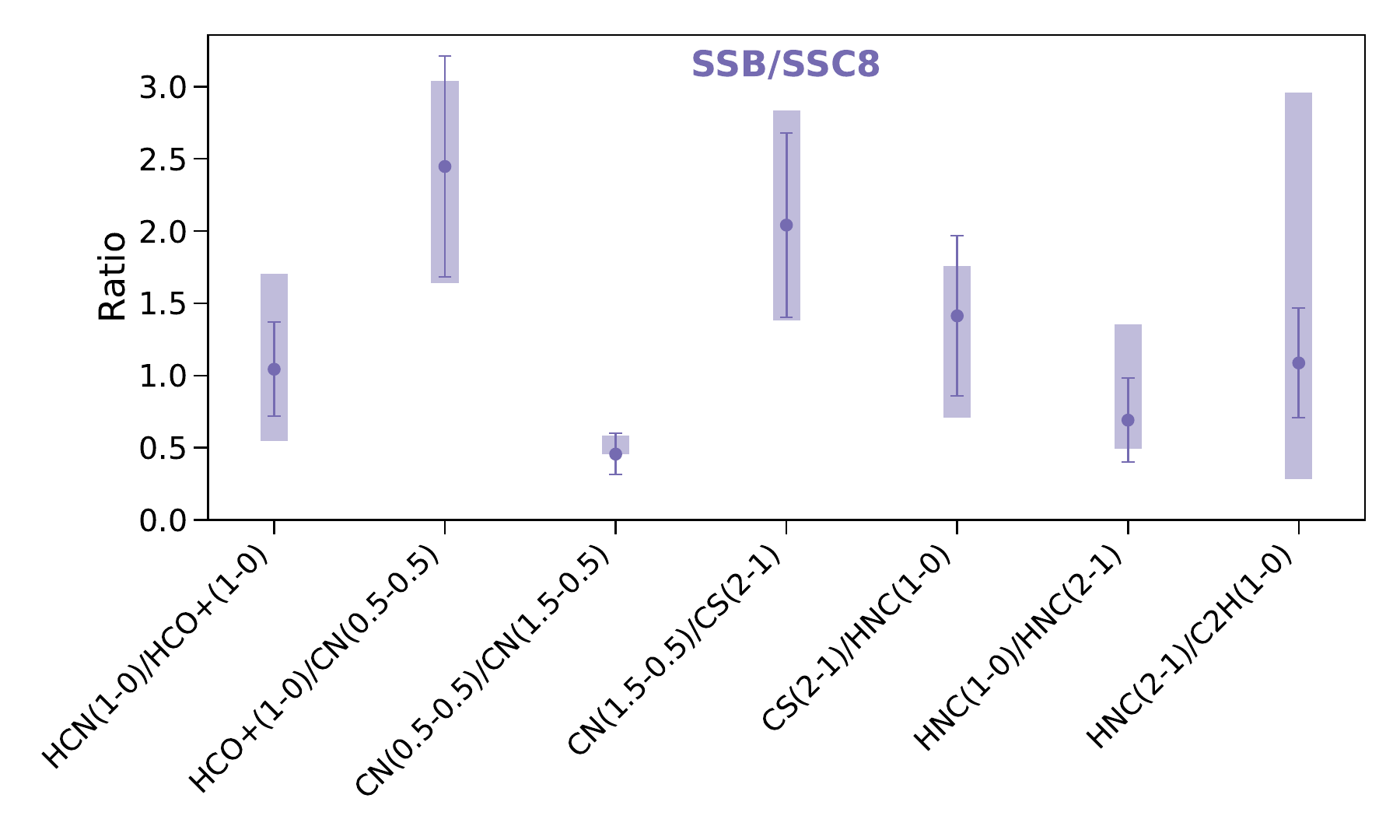}
\includegraphics[width=0.33\textwidth]{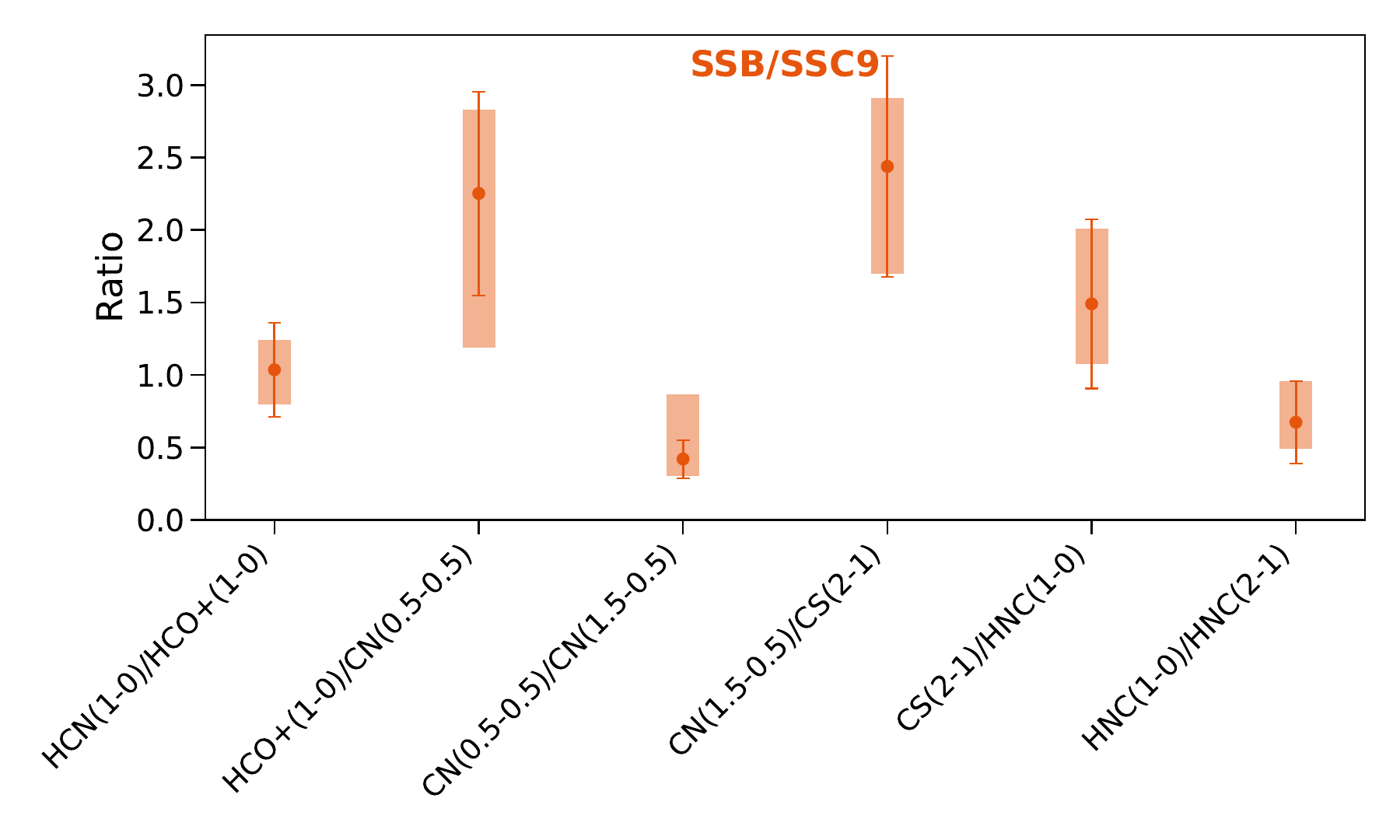}
\includegraphics[width=0.33\textwidth]{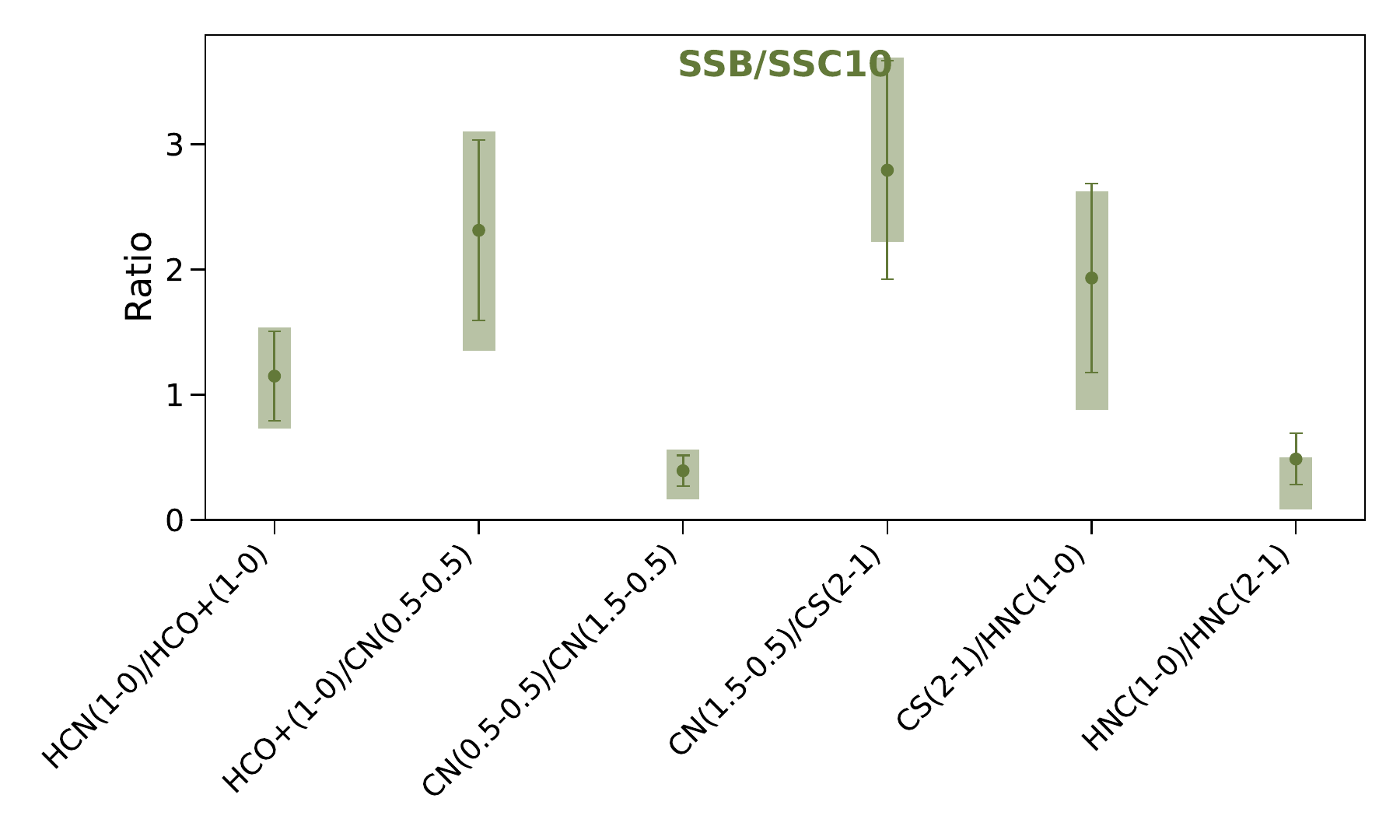}
\includegraphics[width=0.33\textwidth]{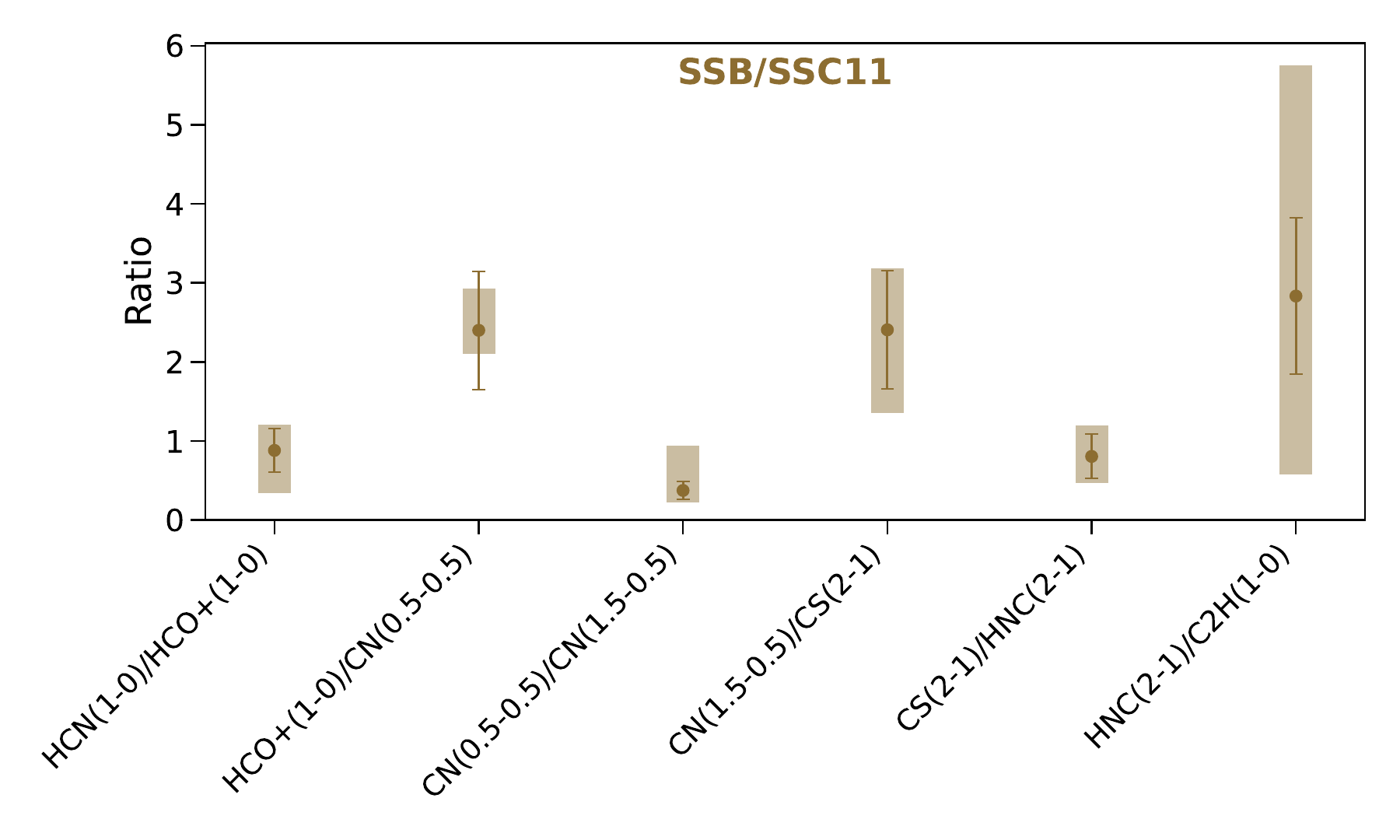}
\includegraphics[width=0.33\textwidth]{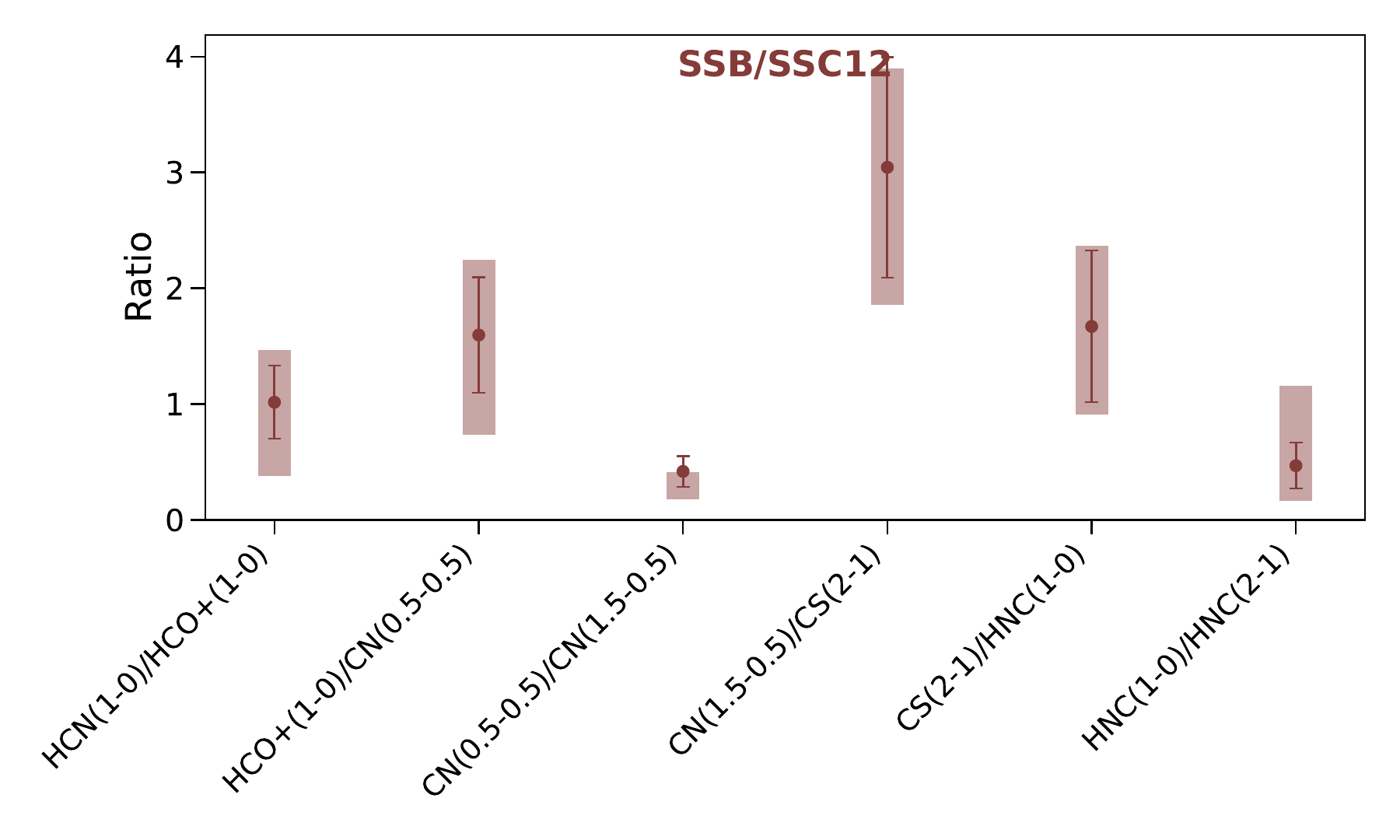}
\includegraphics[width=0.33\textwidth]{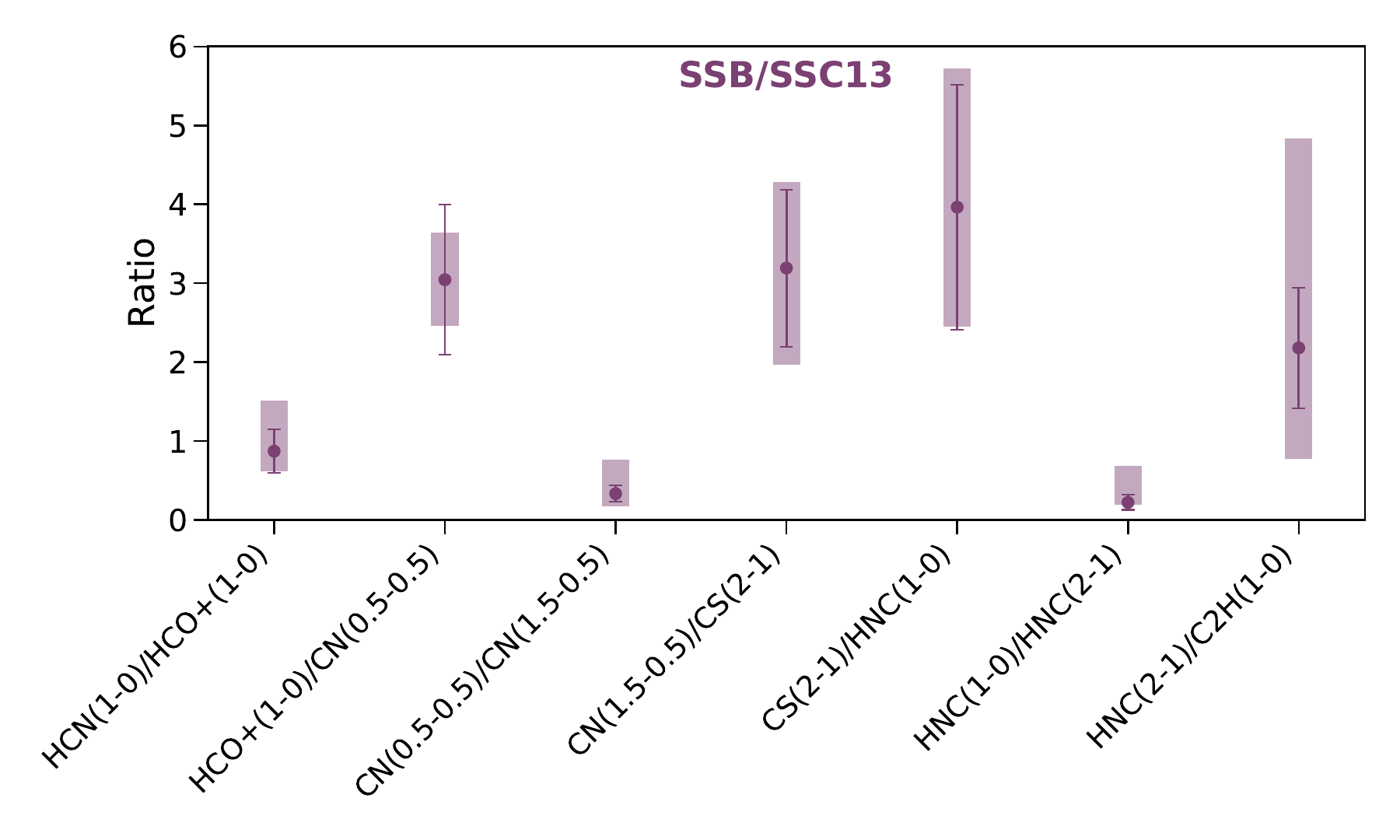}
\includegraphics[width=0.33\textwidth]{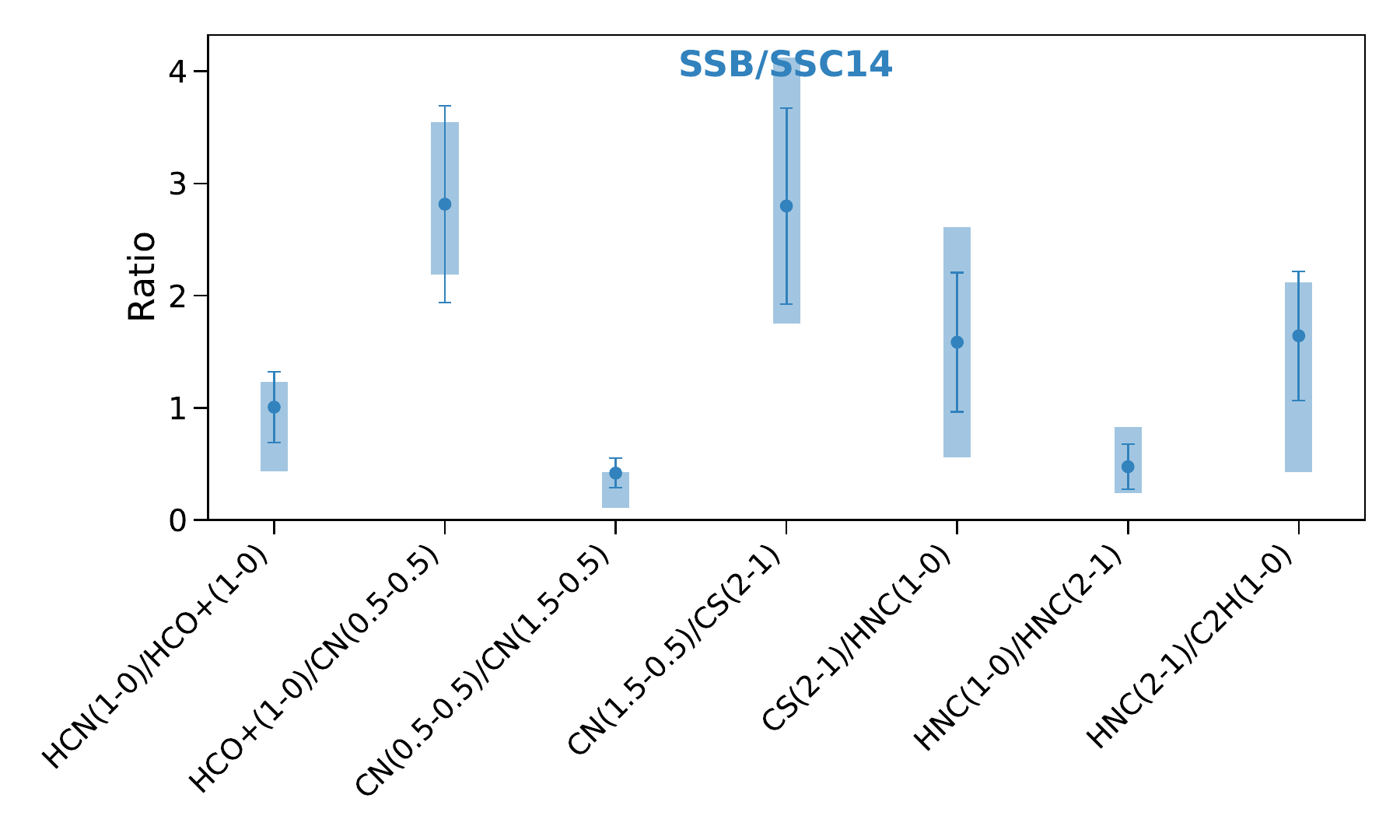}

\caption{Posterior predictive checks for the subregions of the CND,  the SSCs and the SHC on the starburst ring.}
\label{fig:PPCs}
\end{figure*}

For each region, we compute the predicted line ratios from the posterior distribution between the 16th and 84th percentiles. Figure~\ref{fig:PPCs} presents the PPCs for the sub-regions of the CND, the SSCs, and the SHC on the starburst ring. In each panel, the points indicate the observed line ratios, the shaded bars show the 68\% credible intervals of the posterior predictive distribution.
 
For most line ratios, the observed values fall within the predicted credible intervals, confirming that the model adequately reproduces the observed emission across the different environments in NGC~1068. The C$_2$H related ratio is noticeably less well reproduced compared to the other ratios. The posterior predictive distributions for this ratio show larger scatter and, in several regions, the observed values lie near the boundary of the error bars. This is mainly because the C$_2$H($N = 1$--0) transition consists of six hyperfine components grouped into two fine structure groups that are blended at our velocity resolution (Sect.~\ref{sec:regionselection}; \citealt{Garcia2017}). The integrated intensity therefore encompasses the full $N = 1$--0 multiplet. In the radiative transfer calculation, we model C$_2$H using the $N = 1$--0 transition as a whole, which does not fully account for the relative excitation of individual hyperfine components. This simplified treatment introduces additional uncertainty in the predicted C$_2$H intensity that propagates into the line ratio. Despite this limitation, the C$_2$H related ratio does not drive the inference toward systematically biased physical parameters, as demonstrated by the consistency of our results across regions where C$_2$H is and is not detected.

While the posterior predictive checks confirm that the model reproduces the observed line ratios within the measurement uncertainties, this does not guarantee that all physical parameters are tightly constrained by the available data, since some ratios stay nearly constant over a wide range of conditions and can be matched by many parameter combinations\citep{viti2014}. We also use fewer species than \citet{Holdship2022b}, which limits the constraining power and leaves some degeneracy between density, temperature, and column density. The hierarchical Bayesian framework can compensate for this limitation, but compared to the rich ALCHEMI dataset \citep{tanaka2024}, the dataset used in this work is less adequate to fully overcome the potential degeneracy. Nevertheless, the derived parameter distributions should be interpreted as the best constraints obtainable with the current molecular species set rather than as uniquely determined values. Future observations that cover the starburst ring and include additional transitions of the same species, which would provide excitation constraints, are essential to better constrain the parameters.

\section{Comparison with previous work}
\label{sec:appxE}
Table~\ref{tab:prev_studies} summarizes the molecular transitions, spatial resolutions, and methods used in previous multi-molecular studies of NGC\,1068 that we compare with in Sect.~\ref{subsection:Comparsion}. The studies differ in the set of transitions analyzed and in whether chemical modelling was included, which helps to interpret the differences in the derived physical parameters shown in Fig.~\ref{fig:parameter_comparison}. \cite{viti2014} and \cite{Scourfield2020} coupled radiative transfer with \texttt{UCLCHEM}, whereas \cite{Josh2022} used radiative transfer alone. These differences in tracers and methods, together with the range of spatial resolutions, account for much of the scatter seen when comparing the inferred densities, temperatures, and ionization rates.

\begin{table*}
\caption{Summary of molecular transitions and methods used in previous multi-molecular studies of NGC\,1068.}
\label{tab:prev_studies}
\centering
\begin{tabular}{llll}
\hline\hline
Study & Transitions used & Resolution & Methods \\
\hline
Viti et al. (2014) & CO(1--0), CO(2--1), CO(3--2), CO(6--5) & $\sim$100\,pc & Non-LTE RADEX grids + UCLCHEM \\
 & HCN(1--0), HCN(4--3) &  & \\
 & HCO$^+$(1--0), HCO$^+$(4--3) &  & \\
 & CS(2--1), CS(7--6) &  &  \\
\hline
Scourfield et al. (2020) & CS(2--1), CS(3--2), CS(6--5), CS(7--6) & $\sim$35\,pc & RADEX $\chi^2$ fitting coupled with UCLCHEM\\
\hline
Butterworth et al. (2022) & HCN(1--0), HCN(4--3) & $\sim$40\,pc & Non-LTE RADEX with MCMC \\
 & HCO$^+$(1--0), HCO$^+$(4--3) &  & sampling. No chemical modelling. \\
 & CS(2--1), CS(7--6) &  & \\
\hline
\end{tabular}
\end{table*}

\end{appendix}
\end{document}